\documentclass[12pt]{article}
 
\pdfoutput=1

\usepackage{amsmath, amssymb, amscd, amsthm, amsfonts}
\usepackage{physics}

\usepackage{verbatim}

\usepackage{amsmath,amssymb,amsfonts}
\usepackage{psfrag}
\usepackage{footmisc}
\usepackage{url}
\usepackage{mathtools}
\usepackage{color}
\usepackage[dvipsnames]{xcolor}

\usepackage{tikz}
    \usepackage{amssymb,amsfonts,amsmath}
    \usepackage{tkz-euclide}
        \usetikzlibrary{arrows,calc,patterns}
\usepackage{pgfplots}

\definecolor{darkblue}{rgb}{0.1,0.1,.7}
\definecolor{darkgreen}{rgb}{0.1,0.6,.1}
\definecolor{purple}{rgb}{0.6,0,0.6}
\definecolor{orange}{rgb}{0.9,0.6,0}
\definecolor{lightred}{rgb}{0.85,0.35,0.35}
\usepackage[colorlinks, linkcolor=darkblue, citecolor=darkblue, urlcolor=darkblue, linktocpage]{hyperref} 
\usepackage[square, comma, compress,numbers]{natbib}
\usepackage[]{amsmath}
\usepackage[]{graphicx}
\usepackage[]{latexsym}
\usepackage[utf8]{inputenc}
\usepackage{geometry}
\usepackage{amscd}
\usepackage[all,cmtip]{xy}
\usepackage{mathrsfs}

\usepackage[framemethod=TikZ]{mdframed}

\usepackage[margin=10pt,font=small,labelfont=bf]{caption}
\usepackage{changepage}
\usepackage{setspace}
\usepackage{upgreek}

\numberwithin{equation}{section}

\pgfplotsset{compat=1.18}

\newcommand{\address}[1]{\vbox{\center\em#1}}
\newcommand{\gt}{{\tilde{g}}}
\newcommand{\Khighlow}{{\mathbb{K}}}
\newcommand{\highElabel}{{+}}
\newcommand{\lowElabel}{{-}}

\usepackage{tikz}
\newcommand*{\boxcolor}{black}
\makeatletter
\renewcommand{\boxed}[1]{\textcolor{\boxcolor}{%
\tikz[baseline={([yshift=-1ex]current bounding box.center)}] \node [rectangle, minimum width=1ex,rounded corners,draw] {\normalcolor\m@th$\;\,\displaystyle#1\;\,$};}}
 \makeatother

\begin{document}

\thispagestyle{empty}


\begin{center}

{\LARGE \bf {Mind the crosscap:\\ $\uptau$-scaling in non-orientable gravity and time-reversal-invariant systems}}
\end{center}

\bigskip \noindent

\bigskip

\begin{center}

Gabriele Di Ubaldo,${}^{a,b}$ Altay Etkin,${}^c\,$ Felix M.\ Haehl,${}^c\,$ Moshe Rozali${}^d$ 

\address{
a) Leinweber Institute for Theoretical Physics and Department of Physics\\
University of California, Berkeley, CA 94720, USA\\\vspace{.3cm}
b) RIKEN iTHEMS Center for Interdisciplinary Theoretical and Mathematical Sciences,\\ 2-1 Hirosawa, Wako, Saitama 351-0198 Japan\\\vspace{.3cm}
c) School of Mathematical Sciences \& STAG Research Centre,\\
University of Southampton, SO17 1BJ, U.K.\\\vspace{.3cm}
d) Department of Physics and Astronomy, University of British Columbia,\\ Vancouver, V6T 1Z1, Canada}

\vspace{0.15in}
    
{\tt gdiubaldo@berkeley.edu, a.etkin@soton.ac.uk,\\ f.m.haehl@soton.ac.uk, rozali@phas.ubc.ca}

\bigskip

\vspace{1cm}

\end{center}

Spectral statistics of quantum chaotic systems are governed by random matrix universality. In many cases of interest, time-reversal symmetry selects the Gaussian Orthogonal Ensemble (GOE) as the relevant universality class. In holographic CFTs, this is mirrored by the presence of non-orientable geometries in the dual gravitational path integral. In this work, we analyze general properties of these matrix models and their gravitational counterparts. First, we develop a formalism to express the universal level statistics in the canonical ensemble for arbitrary spectral curves, leading to a topological expansion with finite radius of convergence in the late-time $\uptau$-scaling limit. Then, we focus on topological gravity and study topological recursion on the moduli space of non-orientable surfaces. We find that the Weil–Petersson volumes display non-analytic behaviour multiplying polynomials in the boundary lengths. The volumes give rise to wormholes with late-time divergences, in contrast with the orientable case, which is finite. We identify systematic cancellations among WP volumes implied by the consistency and finiteness of the $\uptau$-scaling limit. In particular, the cancellation of  late-time divergences requires a nontrivial genus resummation. Working in the gravitational microcanonical ensemble, we derive and resum all orders of the topological expansion matching the GOE matrix model in the high-energy regime.

\newpage

\setcounter{tocdepth}{2}
{}
\tableofcontents

\newpage


\section{Introduction and summary of results}
\label{sec:intro}

\subsection{Motivation}

The gravitational path integral has proven to be a very successful tool in  the study of quantum gravity and black holes. A lot of progress has come from understanding the correct rules according to which geometries should be included in the path integral. This has led to breakthrough results on the late-time physics of black holes \cite{Almheiri:2019psf,Penington:2019npb}, the statistical realization of random matrix universality in semi-classical gravity \cite{Saad:2018bqo,Saad:2019lba,Cotler:2020ugk}, and many other problems. At the heart of these advances is the realization that spacetimes with nontrivial topologies must be included in the path integral to solve this problem.

\paragraph{The necessity of non-orientable geometries in AdS/CFT.}
One class of spacetimes that has often been neglected consists of non-orientable geometries. It has only recently been appreciated that these should  be included in the path integral on very general symmetry grounds. In quantum gravity, global symmetries are expected to be absent, whether continuous or discrete. In the context of AdS/CFT, this statement has been proven to hold in \cite{Harlow:2018tng}, and we can explore its consequences by examining the symmetries of the dual CFT. Since $\mathcal{CRT}$ is a symmetry of any relativistic quantum field theory \cite{Streater:1989vi}, including Conformal Field Theories, it is also a symmetry of the dual AdS quantum gravity. Since no global symmetries can exist in gravity, $\mathcal{CRT}$ (and any other  symmetries like $\mathcal{R}$ and $\mathcal{CT}$) must be a bulk gauge symmetry.

The consequences of gauging such a discrete spacetime symmetry in gravity were recently studied in \cite{Harlow:2023hjb}. It was shown that gauging spacetime inversions implies  the necessary inclusion of non-orientable spacetimes in the gravitational path integral. Furthermore, these geometries are not necessarily subleading  and there is a priori no reason to exclude them from the path integral. In fact, in \cite{Harlow:2023hjb} a specific example was exhibited, where a \textit{$\mathcal{CRT}$-twisted} BTZ black hole needs to be included to match a boundary CFT prediction. It was pointed out \cite{Grabovsky:2024vnb,Chen:2023mbc} that in Euclidean signature, these simply belong to the class of $SL(2,\mathbb{Z})$ black holes, which in Lorentzian signature may be rendered non-orientable. Thus, inadvertently, non-orientable geometries were already included in the partition function of pure three-dimensional gravity which sums over all $SL(2,\mathbb{Z})$ black holes \cite{Maloney:2007ud,Keller:2014xba}. In particular, failing to include these non-orientable contributions would lead to a violation of modular invariance. On the other hand, summing over all the $SL(2,\mathbb{Z})$ black holes leads to the well-known problem of the Maloney-Witten density of states being negative in certain regions of the spectrum \cite{Benjamin:2019stq}. This has led to several resolutions and proposals for what else to include in the path integral (e.g., \cite{Benjamin:2020mfz,Alday:2019vdr,DiUbaldo:2023hkc,Maxfield:2020ale}).

In two-dimensional Jackiw–Teitelboim (JT) gravity, dual to an ensemble average, bulk gauge symmetries are realized as global symmetries in each member of the ensemble \cite{Hsin:2020mfa,Stanford:2019vob}. If the dual quantum mechanical system is time reversal $\mathcal{T}$ symmetric, such as the SYK model with $q=0 \,{\rm mod}\, 4$ \cite{You_2017,Cotler:2016fpe}, then we must gauge $\mathcal{T}$ in the bulk theory by summing over both orientable and non-orientable surfaces in the gravitational path integral. 
Stanford and Witten \cite{Stanford:2019vob} showed that JT gravity, when non-orientable geometries are included, is dual to an ensemble of random Hamiltonians with time-reversal symmetry $\mathcal{T}$, belonging to the Gaussian Orthogonal Ensemble (GOE) universality class. As an application, crosscap contributions to correlation functions were considered in \cite{Yan:2022nod} and found to modify the late-time plateau. Recently, a version of topological recursion for non-orientable surfaces was developed in \cite{Stanford:2023dtm}, which we will discuss  in detail in the main text. 

The connection between random matrices and discrete symmetries is an old subject which goes far beyond two-dimensional gravity. In fact, any quantum mechanical system which exhibits chaos in the form of random matrix statistics will fall into one of several universality classes according to its discrete symmetries \cite{haake,Altland_1997,zirnbauer2010symmetryclasses}. The specific symmetry class will change qualitatively and quantitatively the random matrix ensemble that describes the statistics of the energy levels of the quantum system under consideration. The GUE universality class of Hermitian random matrices, corresponding to systems with no anti-unitary symmetries such as time reversal, is extremely useful and practical due to is technical simplicity. However, since relativistic quantum field theories are $\mathcal{CRT}$-invariant, there is always an anti-unitary symmetry acting on the Hilbert space and consequently the Gaussian Unitary Ensemble (GUE) is not the relevant symmetry class. This motivates us to study other symmetry classes, particularly the GOE and Gaussian Symplectic Ensemble (GSE), since these are relevant for chaotic QFTs. From the holographic perspective, this provides another reason why contributions from non-orientable geometries are not only necessary but essential: their inclusion is required to correctly capture the chaotic behaviour expected of dual CFTs.

\paragraph{Quantum chaos and the case of AdS${}_3$/CFT$_2$.}
Let us consider the case of AdS$_3$/CFT$_2$ in some more detail. Consider a unitary 2d CFT with $c_L=c_R$, not assuming that it is parity invariant. The Hamiltonian decomposes into blocks with fixed spin $H=\bigoplus_{j\in \mathbb{Z}} H_j$. The CFT is $\mathcal{RT}$-invariant and the $\mathcal{RT}$ symmetry preserves a spin-$j$ block, namely it sends $j\rightarrow j$. Since there is an antiunitary symmetry acting within a block $H_j$ which squares to $(\mathcal{RT})^2=(-1)^F$, the universality class of a block $H_j$ is GOE for bosonic states and GSE for fermionic states, including $j=0$.\footnote{In Lorentzian cylinder  quantization, $\mathcal{T}$ and $\mathcal{R}$ both send $j\rightarrow -j$ so $\mathcal{RT}$ preserves $H_j$. In Euclidean radial quantization, the Osterwalder-Schrader time-reversal $\Theta$ acts as conjugation $\Theta L_n\Theta^{-1}=L^{\dagger}_n=L_{-n}$ \cite{Simmons-Duffin:2016gjk,Ginsparg:1988ui}, leaving $j$ invariant. In particular, $\Theta$ and $\mathcal{RT}$ correspond to each other, since they have the same action on Virasoro generators. Further assuming that the CFT is parity $\mathcal{R}$-invariant only implies that the spectra of $H_j$ and $H_{-j}$ coincide.}

 On the  AdS$_3$ side, this means we should include contributions from non-orientable geometries.  This reasoning was applied  in \cite{Yan:2023rjh} where it was pointed out that one should also include a time-reversal of the Cotler-Jensen torus wormhole \cite{Cotler:2020ugk}, to account for the correct slope of the linear ramp in the spectral form factor (SFF). Another class of geometries which can be obtained by crosscap-type identifications in 3D gravity is given by so-called twisted I-bundles, considered in \cite{Collier:2024mgv} in the framework of Virasoro TQFT \cite{Collier:2023fwi} and in \cite{Yan:2023rjh} in the semiclassical approximation. Simply put, these are $\mathbb{Z}_2$-quotients of two-boundary wormholes along a two-dimensional slice which results in a single-boundary contribution to the path integral. 
 
In the near-extremal limit of AdS$_3$ black holes and wormholes, we expect JT gravity to correctly capture the relevant physics \cite{Maxfield:2020ale}. The gauged $\mathcal{RT}$ symmetry implies that the correct effective theory in the near-extremal limit of AdS$_3$ gravity is non-orientable JT gravity. Similar conclusions should be valid in higher-dimensional AdS/CFT, since JT gravity has been shown to be the correct effective theory for near-extremal black holes in several contexts \cite{Iliesiu:2020qvm,Iliesiu:2021are,Heydeman:2020hhw}. 
In particular, the contribution of non-orientable geometries was not included in the work of Maxfield and Turiaci \cite{Maxfield:2020ale}, which considered only ordinary orientable JT gravity. In light of this, it would be very interesting to revisit their calculation in non-orientable JT gravity with defects. The interplay between crosscaps and defects has not been discussed before. From the matrix integral side, the conclusion of \cite{Maxfield:2020ale} was simply that one has to consider a matrix model with a shifted edge $E_0\ll1$ with respect to the semiclassical black hole threshold, set to $E=0$ by convention. A shifted edge GOE matrix model is the obvious candidate that should correspond to non-orientable JT with defects, where the sum over defects is simply the gravitational counterpart of the $E_0\ll 1$ expansion.  Matching the effective 2D theory to 3D gravity sets $E_0= \mathcal{O}\qty(e^{-S_0/2})$ and similarly for the defect coupling $\lambda=\mathcal{O}\qty(e^{-S_0/2})$. A possible subtlety lies in the fact that a $4k$ defect contribution is of the same order as the contribution of a genus $g=4k$ surface, where the genus can now take half-integer values due to crosscaps: $g=0,\frac{1}{2},1, \frac{3}{2},\dots$.

Going beyond the density of states, to correctly describe the RMT statistics of chaotic 2D CFTs and their AdS$_3$ holographic duals, it is crucial to consider the constraints coming from Virasoro symmetry and $\sl$ modular invariance. 
Recently, a subset of the authors constructed RMT$_2$, a fully modular-invariant random matrix ensemble which captures the nontrivial interplay between random matrix statistics in CFT$_2$ and modular invariance \cite{DiUbaldo:2023qli,Boruch:2025ilr} (see also \cite{Haehl:2023tkr,Haehl:2023xys,Haehl:2023mhf}).  RMT$_2$ serves as a benchmark for chaos in CFT$_2$ and produces candidate partition functions for off-shell wormholes in AdS$_3$ pure gravity. RMT$_2$ can be developed for any of the GUE, GOE and GSE symmetry classes (and in principle for the other seven classes of the Altland-Zirnbauer classification \cite{Altland_1997}), but to apply it consistently to CFT$_2$, we should use the GOE ensemble.\footnote{See also  \cite{Jafferis:2025vyp,deBoer:2025rct} for further comments on the importance of considering the GOE class in this context.}

The necessity and importance of including non-orientable manifolds are transparent, particularly for understanding chaos in the form of random matrix universality in gravity and CFT. This motivates us to consider the SFF in GOE matrix integrals and correspondingly the topological expansion in non-orientable JT gravity. This physically relevant case is unfortunately not a simple generalization of the GUE case. There are several both qualitative and quantitative differences between orientable and non-orientable cases, among which are the presence of two types of divergences and an overall significantly more complex analytic structure, as explored in section \ref{sec:gravitySummary}.

\paragraph{Objectives.}
We will focus on the SFF as a diagnostic of random matrix universality which has a characteristic ramp-plateau structure in all symmetry classes. It was shown in \cite{Saad:2022kfe} that in orientable JT gravity it is possible to capture the ramp-plateau transition by considering the topological $e^{-S_0}$ expansion of the SFF in the $\uptau$-scaling limit, where $\uptau=Te^{-S_0}$ is held fixed as $T,e^{S_0}\rightarrow \infty$. This is an extremely non-trivial property since the topological expansion is asymptotic and the plateau is usually understood to be a doubly non-perturbative effect in $e^{ie^{S_0}}$ from the GUE matrix integral. The $\uptau$-scaling limit reorganizes the topological expansion into a series with a finite radius of convergence, which upon analytic continuation describes the full ramp-plateau transition in canonical variables. 
From the gravity side, this shows that the plateau, which indicates the discreteness of black hole microstates, is obtained by resumming infinitely many wormhole geometries and does not necessarily require new non-perturbative effects.

In this work, we extend and generalize these lessons to other symmetry classes, with a particular focus on time-reversal invariant systems described by GOE matrix models. As we will see, the $\uptau$-scaling limit is considerably more subtle in this case and it involves qualitatively new ingredients and challenges. Nevertheless, ultimately it still leads to a convergent topological expansion, allowing for a perturbative study of the plateau. 

In the remainder of this section we give a light introduction and an overview of our results.

\subsection{Random matrix universality and $\uptau$-scaling}
\label{sec:RMTsummary}

In this paper, we study spectral correlators in random matrix theory with particular focus on two-point correlators. We denote the connected SFF as
\begin{equation}
\label{eq:Ky1y2Def}
    K(\beta_1,\beta_2) = \big\langle \text{Tr}(e^{-\beta_1 H}) \text{Tr}(e^{-\beta_2 H}) \big\rangle_c\,,
\end{equation}
where the expectation value indicates a connected matrix integral of the form 
\begin{equation}
    \langle \; \cdot \; \rangle_c \equiv \frac{1}{Z} \int dH \, (\; \cdot \;) \,e^{-N \, \text{Tr}V(H)} \,.
\end{equation}
The characteristics of the matrices $H$, as well as the matrix potential, depend on the discrete symmetries. We will consider cases where they belong to one of the following universality classes:
\begin{itemize}
    \item {\bf GUE:} $H$ are Hermitian and $V_0(H) = \frac{1}{2} H^2$; a model for systems without time-reversal symmetry.
    \item {\bf GOE:} $H$ are real-symmetric and $V_0(H) = \frac{1}{4} H^2$; a model for systems with time-reversal symmetry in which ${\cal T}^2 =1$.
    \item {\bf GSE:} $H$ are Hermitian quaternionic and $V_0(H) = H^2$; a model for systems with time-reversal symmetry in which ${\cal T}^2 =-1$ (e.g., fermionic systems).
\end{itemize}
While we review the GUE for illustration and discuss the GSE as an interesting generalization, our main focus will be on the GOE. The latter is often the most realistic universality class for systems of interest in holography.\footnote{See \cite{Stanford:2019vob,Yan:2022nod,Stanford:2023dtm,Yan:2023rjh,Tall:2024hgo} for some recent discussions.} We note that further generalizations taking into account additional symmetries exist and would be interesting to study further \cite{Altland_1997}.

In the double-scaling limit, we take $N\rightarrow \infty$ and suitably tune the model parameters to zoom in on the edge of the spectrum. The matrix model can then be characterized by a continuous spectral curve, which we assume to be of the form 
\begin{equation}
\label{eq:rhoGeneral}
    \rho(E) = \rho_0(E) \, e^{S_0} = \frac{\sqrt{E}}{2\pi}\left( 1 + a_1 E + a_2 E^2 + \ldots \right) e^{S_0} \,, 
\end{equation}
where $S_0$ is a large parameter that controls the double-scaling limit.
The SFF admits a topological expansion in powers of $e^{-S_0}$:
\begin{equation}
    K(\beta_1,\beta_2) = \sum_{g=0,\frac{1}{2},1,\frac{3}{2},\ldots}
    K_{g,2}(\beta_1,\beta_2)\, e^{-2gS_0}\,,
\end{equation}
where half-integer terms are only relevant for the GOE and the GSE. Much of this paper is concerned with the analysis of this expansion, such as finding the coefficients $K_{g,2}$ as a function of the coefficients $a_i$ of the spectral curve. For example, a simple universal result is the genus-0 contribution:
\begin{equation}
    K_{0,2}(\beta_1,\beta_2) = \frac{\text{\tt C}}{2\pi} \frac{\sqrt{\beta_1\beta_2}}{\beta_1+\beta_2} \,,
    \label{eq:Kg0Universal}
\end{equation}
with $\text{\tt C}_\text{GUE} = 1$, $\text{\tt C}_\text{GOE} = 2$, $\text{\tt C}_\text{GSE} = \frac{1}{2}$.

We will use the following standard analytic continuation of \eqref{eq:Ky1y2Def}. In canonical expressions, we write
\begin{equation}
\beta_1 = \beta + i T \,,\qquad \beta_2 = \beta - i T \,,
\end{equation}
and denote the corresponding SFF as $K_\beta(T)$. The corresponding microcanonical expression is related by a Laplace-Fourier transform: 
\begin{equation}
\label{eq:KEomega}
    K_E(\omega) \equiv \int_{-\infty}^\infty dE d\omega \; e^{-2\beta E -i T \omega} \, K_\beta(T) \,.
\end{equation}
More conveniently, we will often consider the microcanonical SFF in the time domain, i.e.
\begin{equation}
    K_E(T) \equiv \int_{-\infty}^\infty dE \; e^{-2\beta E} \, K_\beta(T) \,.
\end{equation}


\paragraph{The $\uptau$-scaling limit.}
It has been suggested in \cite{Saad:2022kfe} (see also \cite{Haake:2009scd,Okuyama:2020ncd}) that a simple scaling limit of $K_\beta(T)$ can be used to focus on the physics at the scale of the average level spacings and thereby extract the universal RMT behavior.  Furthermore, it turns out that the canonical SFF is best suited for this: it leads to a convergent topological expansion that displays a smooth transition to the plateau.
We define this $\uptau$-scaling limit of the SFF as follows:
\begin{equation}
\label{eq:tauDef}
  \uptau\text{-scaling:}\qquad  T\rightarrow\infty \,,\qquad S_0 \rightarrow \infty \,,\qquad \uptau \equiv T \,e^{-S_0} \text{ fix}.
\end{equation}
In this limit, the topological expansion simplifies, and we write:
\begin{equation}
\label{eq:topExp}
 {\cal K}_\beta(\uptau) \equiv \; \lim_{\substack{T,S_0 \rightarrow \infty \\ \uptau \text{ fixed}}} \; e^{-S_0}\, K_\beta(T=\uptau e^{S_0}) 
  \,.
\end{equation}
For instance, the genus-0 term \eqref{eq:Kg0Universal} implies a universal contribution ${\cal K}_\beta(\uptau) \supset \frac{\text{\tt C}\,\uptau}{4\pi \beta}$.

A conjecture due to \cite{Saad:2022kfe} (see also \cite{Muller2005,Cotler:2016fpe,Blommaert:2022lbh} and \cite{Weber:2024ieq,Tall:2024hgo} for related recent discussions in the context of the GOE) is that the $\uptau$-scaled SFF agrees exactly with the universal expression (``sine kernel'') derived in the corresponding Gaussian random matrix model.
\begin{figure}
\begin{center}\includegraphics[width=.48\textwidth]{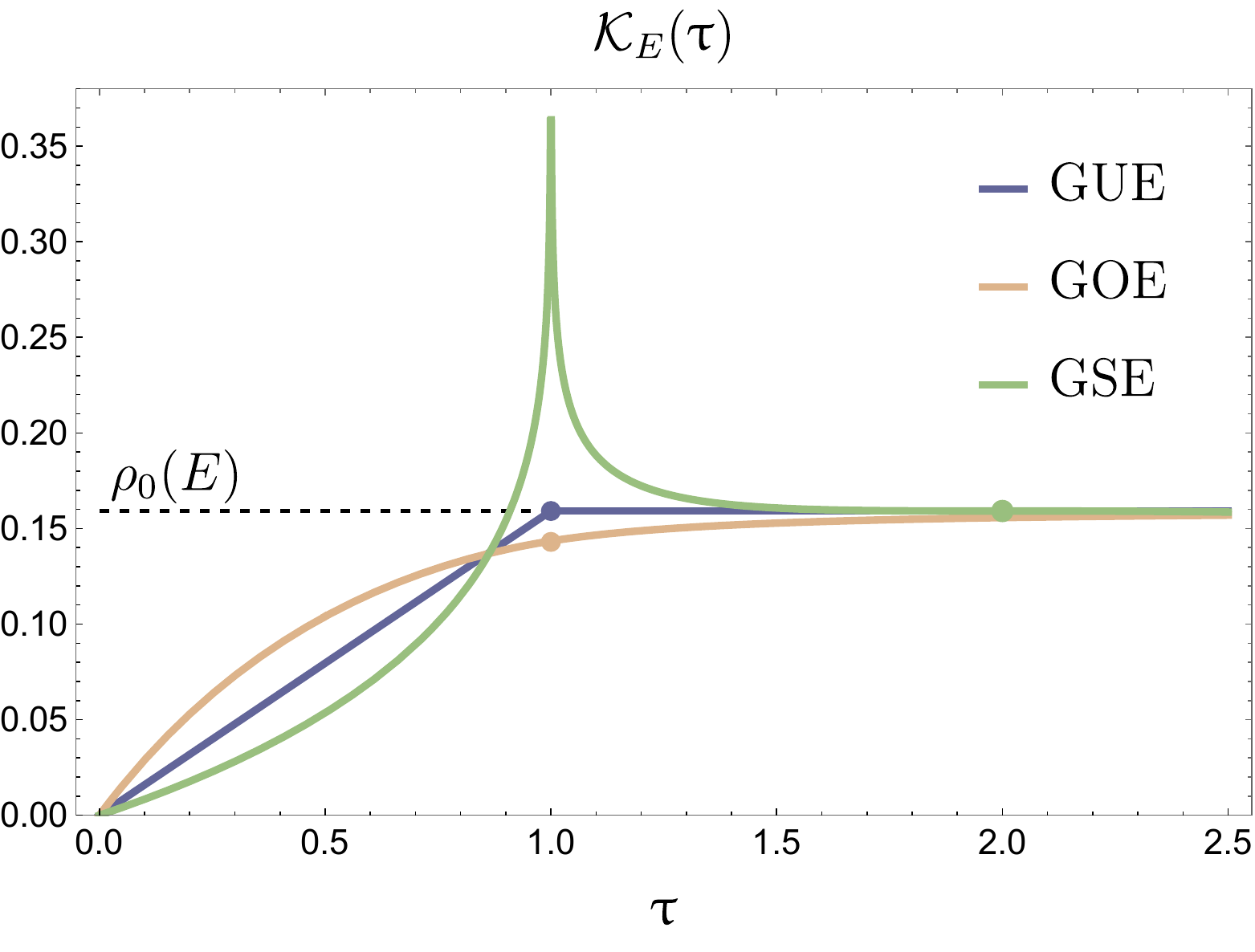}
$\;\;$
\includegraphics[width=.48\textwidth]{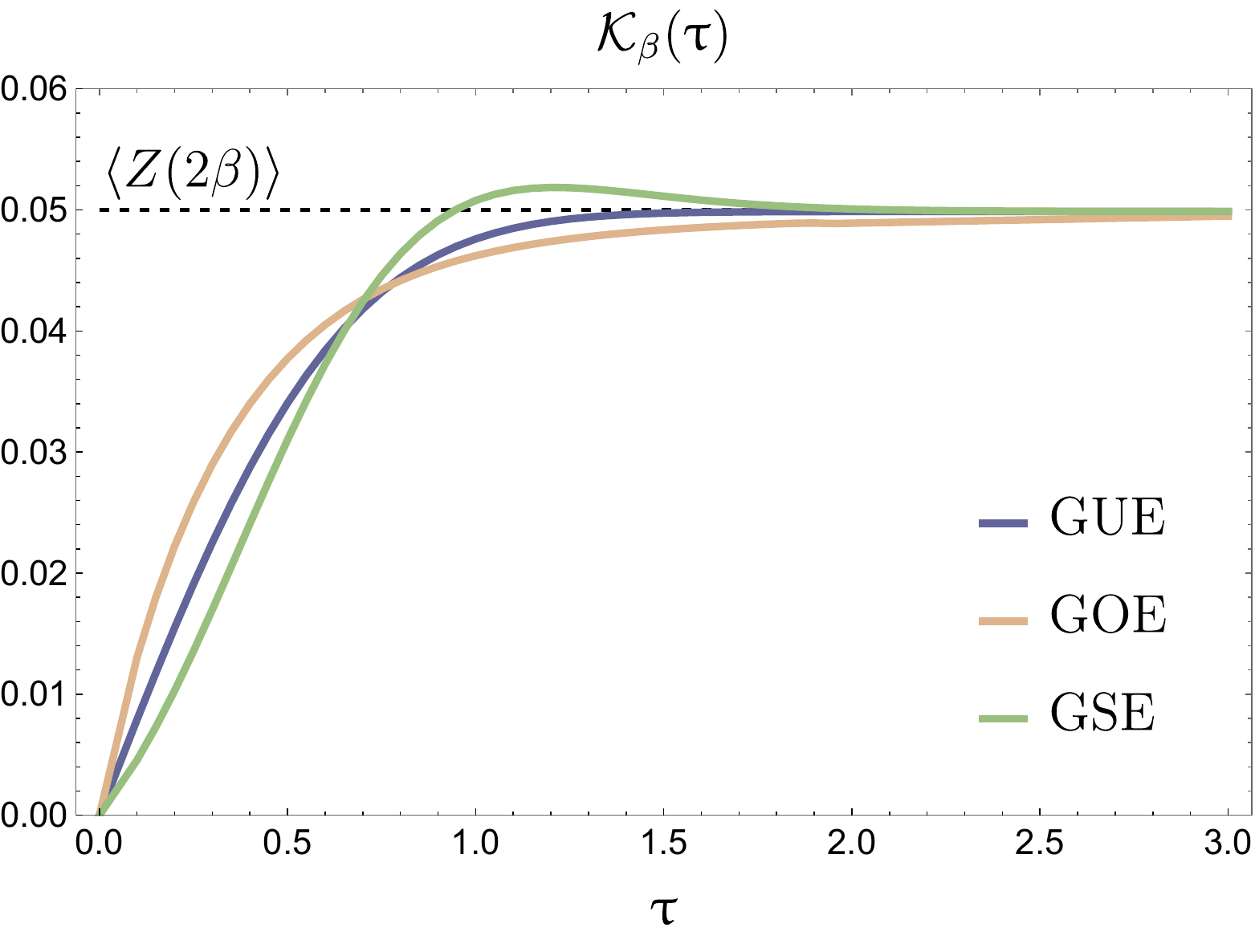}
    \end{center}
    \caption{Universal spectral form factors in three universality classes, for the Airy model. {\it Left:} Microcanonical variables (fixed $E=1$); the non-smooth transition between ramp and plateau is indicated by dots. {\it Right:} Canonical variables (fixed $\beta=1$); the transition to the plateau is smooth.}
\label{fig:RMTplots}
\end{figure}
That is, one expects:
\begin{equation}
\label{eq:LaplaceGeneral}
    {\cal K}_\beta(\uptau)\; \stackrel{?}{=} \; \int_0^\infty dE \, e^{-2\beta E} \; {\cal K}_E(\uptau)\,,
\end{equation}
where the universal expressions for spectral correlators in RMT  are \cite{Liu:2018hlr}:
\begin{equation}
\begin{split}
    \text{GUE RMT:} \quad {\cal K}_E(\uptau) &= \text{min} \left\{ \frac{\uptau}{2\pi} \,,\; \rho_0(E) \right\}  \,,\\
    \text{GOE RMT:} \quad {\cal K}_E(\uptau) &= \text{min} \left\{ \frac{\uptau}{\pi} - \frac{\uptau}{2\pi} \, \log \left( 1 + \frac{\uptau}{\pi\rho_0(E)} \right) \,,\; 2\rho_0(E) - \frac{\uptau}{2\pi} \, \log \left( \frac{\uptau+\pi \rho_0(E)}{\uptau-\pi \rho_0(E)}\right) \right\} \,,\\
    \text{GSE RMT:} \quad {\cal K}_E(\uptau) &= \text{min} \left\{ \frac{\uptau}{4\pi} - \frac{\uptau}{8\pi} \, \log \left| 1 - \frac{\uptau}{2\pi\rho_0(E)} \right| \,,\; \rho_0(E) \right\} \,.
\end{split}
\label{eq:KEtauGeneral}
\end{equation}
See figure \ref{fig:RMTplots} (left) for illustration. The GUE expression is continuous but not differentiable across the ``ramp-to-plateau transition'' at energy $E_*$ determined by $\rho_0(E_*) = \frac{\uptau}{2\pi}$. The universal GOE SFF is twice differentiable across the same transition. In the GSE case, the transition happens at $\rho_0(E_*) = \frac{\uptau}{4\pi}$ and is twice differentiable; in addition there is a discontinuity in the ``ramp'' at $\rho_0(E_*) = \frac{\uptau}{2\pi}$.

In section \ref{sec:RMT}, we study in detail the Laplace transform of the universal $\uptau$-scaled RMT correlators, see \eqref{eq:LaplaceGeneral}. Our main result is a derivation of the coefficients in the $\uptau$-scaled topological expansion of ${\cal K}_\beta(\uptau)$ for the three universality classes for an arbitrary spectral curve \eqref{eq:rhoGeneral}. We find that all of them lead to a convergent topological expansion of the form
\begin{equation}
\begin{split}
 {\cal K}_{\beta}(\uptau) = 
  \frac{\text{\tt C}}{4\pi \beta}\,\uptau + \sum_{g=1,2,...}^\infty \left[ A_{g}(\rho_0;\beta)+ B_{g}(\rho_0;\beta) \, \log(\uptau) \right] \, \uptau^{2g+1} + \sum_{\gt=\frac{1}{2},\frac{3}{2},...}^\infty C_{\gt}(\rho_0;\beta) \, \uptau^{2\gt+1}\,.
  \end{split}
 \label{eq:SFFtaugeneralStructure}
\end{equation}
The coefficients in this expansion for general spectral curves can be found for GUE, GOE, and GSE in \eqref{eq:PgDef}, \eqref{eq:generalA}, and \eqref{eq:GSEgeneralResult}, respectively. While the GUE only features the $A_g$ coefficients, the GOE and GSE attain the general structure shown above, where $g\in \mathbb{Z}$ indicates integer-genus contributions and $\tilde{g} \in \mathbb{Z}+\frac{1}{2}$ labels half-integer-genus contributions. The latter are associated in gravity models with non-orientable crosscap geometries. See figure \ref{fig:RMTplots} (right) for an example of ${\cal K}_\beta(\uptau)$.

To derive these results, we develop tools to perform the Laplace transform of the universal expressions \eqref{eq:KEtauGeneral} for arbitrary $\rho_0(E)$. We also give evidence that these universal expressions do indeed capture the $\uptau$-scaling limit of SFF's computed in various models of interest. Furthermore, we confirm in examples that the topological expansion in the $\uptau$-scaling limit is convergent.

\subsection{Non-orientable gravity and late-time divergences}
\label{sec:gravitySummary}

Besides universal RMT correlators, we also study the $\uptau$-scaling limit of spectral correlators in toy models of gravity with a known matrix model description.

The paradigmatic example is the non-perturbative JT gravity path integral, which is known to admit a {\it genus expansion} \cite{Saad:2019lba}: 
\begin{equation}
    Z^{\text{JT}}\bigl(\beta_1,...,\beta_n\bigr) \equiv \Bigl\langle Z\bigl(\beta_1\bigr)...Z\bigl(\beta_n\bigr)\Bigr \rangle= \sum_{g \geq 0} e^{\chi S_0} \,Z_{g,n}^\text{JT}\bigl(\beta_1,...,\beta_n\bigr),
\end{equation}
where $\chi := 2 - 2g - n$ is the Euler characteristic of the hyperbolic surface, which the path integral is evaluated on. The genus $g$ partition function is computed by gluing external throats (``trumpets'') to internal geometries with $g$ handles and $n$ punctures: 
\begin{equation}
    Z_{g,n}^\text{JT}\bigl(\beta_1,...,\beta_n\bigr) = \prod_{i=1}^n \int_0^{\infty} b_i db_i \ Z_{\text{tr}}\bigl(\beta_i, b_i\bigr) V_{g,n}^\text{JT}\bigl(b_1,...,b_n\bigr),
    \label{eq:ZgGeneral}
\end{equation}
with associated trumpet wavefunction
\begin{equation}
    Z_{\text{tr}}\bigl(\beta_i,b_i\bigr) = \frac{e^{-\frac{b_i^2}{4 \beta_i}}}{\sqrt{4 \pi \beta_i}}\,,
\end{equation}
and $V^\text{JT}_{g,n}(b_1,...,b_n)$ being the Weil-Petersson (WP) volumes for the genus $g$ moduli space with $n$ geodesic boundaries. Geodesic lengths are denoted by $b_i$, and they serve as gluing boundaries between the trumpets and the hyperbolic surfaces.

In order to compute the gravitational path integral, one requires a systematic way of calculating the WP volumes. This is achieved by Mirzakhani's topological recursion \cite{Mirzakhani:2006fta} and its non-orientable generalization \cite{Stanford:2023dtm}. The geometric intuition behind the recursion is a cutting and pasting procedure, which describes the moduli space of surfaces $\Sigma_{g,n}$ in terms of the moduli space of surfaces either with lower genus or fewer boundaries.

Stanford and Witten \cite{Stanford:2019vob} first examined the JT gravity path integral on non-orientable surfaces, showing that it diverges due to contributions from small crosscaps. In the moduli space integral, the measure for a crosscap of size $a$ is \cite{norbury2007lengthsgeodesicsnonorientablehyperbolic,Stanford:2019vob}: 
\begin{equation}
    \frac{da}{2\tanh(\frac{a}{4})}\,,
\end{equation}
which diverges near $a=0$. This has long posed an obstacle to formulating a version of Mirzakhani’s topological recursion for non-orientable surfaces and relating it to a dual matrix integral in the GOE symmetry class. Recently, Stanford \cite{Stanford:2023dtm} overcame this by demonstrating an analog of Mirzakhani’s recursion for suitably regularized WP volumes of non-orientable surfaces, and showed its correspondence with the loop equations of a GOE matrix integral.

The recursion relation, together with its integral kernels, is finite; however, the resulting volumes diverge. This led to a definition of $\epsilon$-regularized volumes $V^{\epsilon}_{g,n}$ of the general form: 
\begin{equation}
    V^{\text{JT},\epsilon}_{g,n}(b_1,\dots,b_n)= \sum_{k=0}^{2g} \log(\epsilon^{-1})^k \,v_{g,k}(b_1,\dots,b_n) +\mathcal{O}(\epsilon).
\end{equation}
In practice, however, performing  the necessary integrals and computing  $v_{g,k}$ is challenging. One way of practically computing $V^{\epsilon}_{g,n}$ is by considering instead the spectral curve of the  $(2,p)$ minimal string and identifying $p=\tfrac{1}{\epsilon}$. On the matrix model side, this method was used  in \cite{Tall:2024hgo} to compute the $v_{g,k}$ up to genus $g=1$ and $n=1,2$. The structure of the volumes $v_{g,k}$ becomes significantly more complex as they are no longer simply given by symmetric polynomials in the geodesic lengths $b_i$. Instead, they feature  multiple polylogarithms of exponentials of the lengths $b_i$ of the form $\text{Li}_{\{k_i\}}(e^{L(\{b_i\})})$, where $L(\{b_i\})$ are linear functions of $b_i$. This functional form is based on a limited set of examples, and higher topologies could give rise to new functions.

Remarkably, the $\uptau$-scaling limit is well defined in the GOE class and yields a convergent genus expansion that captures the late-time ramp–plateau structure of the spectral form factor. To appreciate the nontrivial nature of this result from the gravitational perspective, it is instructive to compare it with the $\uptau$-scaling limit in conventional orientable JT gravity. There are three main reasons why the $\uptau$-scaling limit in non-orientable gravity is significantly more challenging than the orientable one:
\begin{itemize}
    \item \textbf{Crosscap divergences:} Due to the UV divergences generated by small crosscaps, the regularized WP volumes depend explicitly on the regulator $\epsilon$. In contrast, the $\uptau$-scaled SFF is manifestly regulator-independent, raising the question of how the $\epsilon$-dependent volumes $V^{\text{JT},\epsilon}_{g,2}$ can reproduce it. For the cases $g=\tfrac{1}{2}$ and $g=1$, and up to $\mathcal{O}(\uptau^4)$, taking the $\uptau$-scaling limit prior to sending $\epsilon \rightarrow 0$ yields the correct $\epsilon$-independent result \cite{Weber:2024ieq}. However, the general mechanism behind this remains unclear. 
    \item \textbf{Volume cancellations and integrable structure:} In the GUE case, the $\uptau$-scaling limit was traced to special cancellations in the WP volumes \cite{Saad:2022kfe,Blommaert:2022lbh}, first observed in \cite{Eynard:2021zcj}, which reduce the degree of divergence of the topological expansion. These cancellations ensure that the $\uptau$-scaling limit yields a topological expansion with a finite radius of convergence. As explained in \cite{Blommaert:2022lbh}, this phenomenon is rooted in the integrable KdV hierarchy structure underlying matrix models and intersection numbers, and the cancellations were later proven rigorously in \cite{Eynard:2023qdr}. Thus, for the GUE case, the $\uptau$-scaling limit is understood as a manifestation of a rich integrable mathematical structure. The fact that the GOE class also admits such a limit suggests the presence of an analogous—albeit more intricate—integrable structure governing it. In the GUE analysis, a key role was played by the fact that the volumes are symmetric polynomials, making the cancellations manifest in the basis of elementary symmetric polynomials. By contrast, the non-orientable volumes involve a significantly more complicated functional form, featuring polylogarithms, rendering the existence of the $\uptau$-scaling limit even more nontrivial than in the already rich GUE case.
    \item \textbf{Late-time divergences:} The way the $\uptau$-scaling limit is realized in GUE JT gravity is particularly simple: the limit can be taken directly at the level of a single geometry of fixed-genus. Each fixed-genus contribution scales as $\uptau^{2g+1}$, and these terms can be resummed into a convergent series.  In the GOE case, however, taking the $\uptau$-scaling limit for a single geometry leads to divergences, as the limit retains dependence on both $T$ and $\uptau$. For instance, at genus $g=1$ one encounters a logarithmic divergence of the form $\sim \uptau^3 \log T$ \cite{Tall:2024hgo}. The fact that the full SFF nevertheless admits a well-defined $\uptau$-dependent limit indicates that the sum over geometries must involve a nontrivial resummation canceling these divergences.
\end{itemize}
Taken together—the crosscap divergences, the cancellations and underlying integrable structure, and the late-time divergences—these features make the $\uptau$-scaling limit in non-orientable gravity particularly challenging to analyze, underscoring the nontriviality of its very existence. Remarkably, the limit not only exists, but can also be computed analytically in closed form, as derived in section \ref{sec:RMT}.

In section \ref{sec:toprec}, we review the non-orientable topological recursion, and derive its simpler version for topological gravity (Airy model). A central result is \eqref{eq:AiryTopRecNew}, a manifestly finite topological recursion for non-orientable geometries (or ribbon graphs) in the Airy model. We work out many examples (see table \ref{tab:NOairyWP}) and describe the general structure of non-orientable WP volumes in this model. Generally, they consist of a piece which is a symmetric polynomial in the boundary geodesic lengths $\{b_i\}$, as well as a non-analytic piece consisting of step functions multiplying polynomials. This simple yet non-trivial appearance of non-analyticities serves as a valuable toy model for the more challenging case of non-orientable JT gravity. We also describe how the topological recursion for the Airy model is connected to a corresponding set of loop equations, see section \ref{sec:loopEqs}.

In section \ref{sec:cancellations}, we consider the recursively obtained Airy WP volumes for $n=2$ boundaries in more detail, adding to the recent analysis of \cite{Weber:2024ieq,Tall:2024hgo}. We use the WP volumes to compute the gravitational path integral \eqref{eq:ZgGeneral} in the $\uptau$-scaling limit. We observe a large number of potential divergences in the $\uptau$-scaling limit. We give evidence that many of these divergences are absent due to fortuitous cancellations between the coefficients defining the WP volumes. We conjecture the general form of these constraints for any genus, for example \eqref{eq:conjecture1}. Some divergences do not cancel and their removal requires a non-trivial resummation of the topological expansion.

\noindent{\bf Note:} Ref.\ \cite{Weber:2025mow} appeared while this work was nearing completion. It overlaps with section \ref{sec:cancellations}.


\newpage

\section{The $\uptau$-scaled spectral form factor in GUE/GOE/GSE}
\label{sec:RMT}

In this section, we derive the $\uptau$-scaling limit of the SFF in the GOE and GSE universality classes for arbitrary spectral densities $\rho_0(E)$. We start by reviewing the derivation for the GUE class (following appendix D of \cite{Saad:2022kfe}), in a way that will be useful in preparation for the non-orientable cases.

\subsection{Review of the GUE $\uptau$-scaled SFF}
\label{sec:GUEreview}

Our task is to compute the following Laplace transform of the microcanonical SFF, see \eqref{eq:tauDef} and \eqref{eq:KEtauGeneral}:
\begin{equation}
\label{eq:GUElaplaceStart}
    {\cal K}_\beta^\text{GUE}(\uptau) = \int_0^\infty dE \, e^{-2\beta E} \; \text{min} \left\{ \frac{\uptau}{2\pi} \,,\; \rho_0(E) \right\} \,.
\end{equation}
The resulting SFF is written as a topological expansion that corresponds to a power series in $\uptau$:
\begin{equation}
 \boxed{  {\cal K}_\beta^\text{GUE}(\uptau) = \frac{\uptau}{4\pi \beta} + \sum_{g=1}^\infty  A_g^\text{GUE}(\rho_0;\beta) \,  \uptau^{2g+1} }
 \label{eq:GUESFF}
\end{equation}
with coefficients given in terms of the following contour integral over the spectral density:
\begin{equation}
\label{eq:PgDef}
A_g^\text{GUE}(\rho_0;\beta) \equiv -\frac{1}{2\pi g(2g+1)}  \, c_{2g}(\rho_0;\beta)\,,\qquad c_{2g}(\rho_0;\beta) \equiv 
\frac{1}{(2\pi)^{2g}} \oint_0 \frac{dE}{2\pi i} \, \frac{e^{-2 \beta E}}{\rho_0(E)^{{2g}}}\,.
\end{equation}
There are two remarkable properties of this expression: $(i)$ the canonical ensemble has smoothed the SFF from a non-analytic function into an analytic one (c.f., figure \ref{fig:RMTplots}), and $(ii)$ the topological (or small $\uptau$) expansion has become convergent, with a finite radius of convergence. Thanks to these properties, the transition from the ramp to the plateau is accessible by resumming this convergent series.

We will now review how to derive this formula. The reader may want to jump to section \ref{sec:GUEexamples} for examples.

\paragraph{Derivation.} It is convenient to treat the genus-0 term separately from the rest. That is, we write \eqref{eq:GUElaplaceStart} as follows:
\begin{equation}
 {\cal K}_\beta^\text{GUE}(\uptau) = \frac{\uptau}{4\pi\beta} + \underbrace{\int_0^{E_*} dE \, e^{-2\beta E} \, \left( \rho_0(E) - \frac{\uptau}{2\pi} \right)}_{ {\equiv \Khighlow_{\lowElabel,\beta}^{{\text{GUE}}}}} \,,
\end{equation}
where $E_*$ is the energy where the two terms in \eqref{eq:GUElaplaceStart} exchange dominance, $\rho_0(E_*) = \frac{\uptau}{2\pi}$. The notation $\Khighlow_{\lowElabel,\beta}^{{\text{GUE}}}$ indicates that we are considering a low-energy integral ($E<E_*$). In the GOE and GSE cases, there will also be non-trivial high-energy integrals for $E>E_*$.

Consider now $\Khighlow_{\lowElabel,\beta}^{{\text{GUE}}}$. We review the general idea used in \cite{Saad:2022kfe} and show how it can be further refined. We first write the expression as an integral in the complex $\rho_0$ plane:
\begin{equation}
\Khighlow_{\lowElabel,\beta}^{{\text{GUE}}} = \int_0^{\frac{\uptau}{2\pi}} d\rho_0 \, E'(\rho_0)\; e^{-2\beta E(\rho_0)} \, \left( \rho_0 - \frac{\uptau}{2\pi} \right) \,.
\end{equation}
We wish to find a
function $f(\rho_0)$, which has a branch cut in the complex $\rho_0$ plane along $\rho_0 \in [0,\frac{\uptau}{2\pi}]$ (i.e., for $E \in [0,E_*]$), and whose discontinuity across the cut is the integrand $\rho_0 - \frac{\uptau}{2\pi}$. We can then replace the integral with a contour integration for a contour that wraps the interval $[0,\frac{\uptau}{2\pi}]$. In this case the integrand is simple and this can be done exactly, but for later sections it will prove useful to find a way of obtaining the expanded result systematically.

We use the general way to recover a function from its cuts, the dispersion relation (or Stieltjes transform),
\begin{equation}\label{eq:stieltjes transform}
f(\rho_0)= -\int_0^{\tfrac{\uptau}{2\pi}} \frac{ d\rho'}{2 \pi i}\,\frac{\rho'-\frac{\uptau}{2\pi}}{\rho'-\rho_0} \qquad\quad \left(\rho_0 \in \mathbb{C} \smallsetminus \Bigl[0,\frac{\uptau}{2\pi}\Bigr] \right)\,,
\end{equation}
which gives a function whose discontinuity is the prescribed
\begin{equation}
\text{Disc}_{\rho_0}\bigl[f(\rho_0)\bigr] \equiv f(\rho_0-i0) - f(\rho_0+i0) =\rho_0 - \frac{\uptau}{2\pi}\,, \qquad \rho_0 \in \left[ 0,\frac{\uptau}{2\pi} \right]\,.
\end{equation}
We find the following function:\footnote{ The dispersive expression does not require any corrections by ``subtractions" as the function of interest decays sufficiently fast at infinity.}
\begin{equation}\label{eq:GUE Stiltjes transform low integral answer}
    f(\rho_0) = \frac{1}{2\pi i} \left[ \left( \frac{\uptau}{2\pi} -\rho_0\right) \log\left( 1- \frac{\uptau}{2\pi\rho_0}\right)-\frac{\uptau}{2\pi}\right]\,.
\end{equation}
We can then write:
\begin{align}
    \Khighlow_{\lowElabel,\beta}^{{\text{GUE}}} &= \int_0^{\frac{\tau}{2\pi}} d\rho_0\,E'(\rho_0) \, e^{-2\beta E(\rho_0)} \text{Disc}_{\rho_0} \bigl[f(\rho_0)\bigr] 
    = \oint_{\bigl[0,\tfrac{\uptau}{2\pi}\bigr]} d\rho_0 \,E'(\rho_0) \, e^{-2\beta E(\rho_0)} f(\rho_0)\,,
\end{align}
where the notation indicates a contour integral along a contour that wraps the interval $[0,\tfrac{\tau}{2\pi}]$ counterclockwise.

\begin{figure}
\begin{center}\includegraphics[width=.9\textwidth]{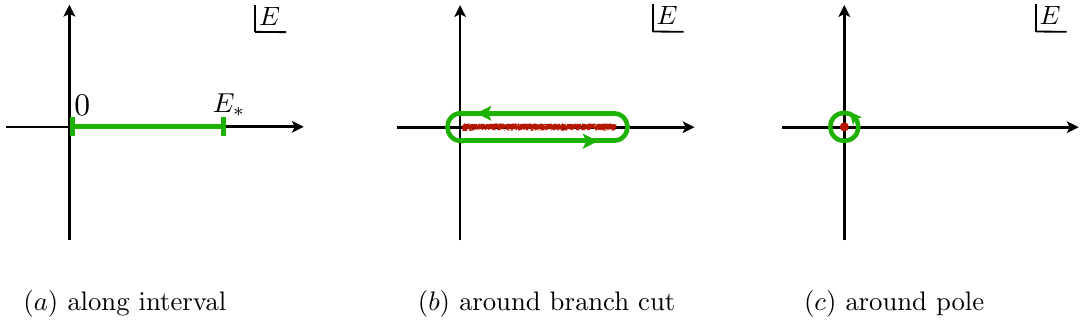}
    \end{center}
    \caption{Illustration of the contour deformation: the initial integral along $E\in [0,E_*]$ is first turned into a contour integral around a branch cut, and subsequently into a sum over contour integral around a pole at $E=0$.}
\label{fig:contours}
\end{figure}

The expression \eqref{eq:stieltjes transform} constructs a function which has a cut as a function of $\rho_0$, by using an integral in the $\rho'$-plane. This is distinguished from the cut as a function of $E$: The density is a double-valued function of $E$, and from the function $f(\rho_0)$ given in \eqref{eq:GUE Stiltjes transform low integral answer}, we form the symmetric combination; this gives the discontinuity as a function of $\rho_0$ in terms of the discontinuity as a function of $E$, which is what is needed to evaluate the integral in the $E$-plane. We find for the functions of interest:
\begin{equation}
    \text{Disc}_{\rho_0}\bigl[f(\rho_0)\bigr]= \text{Disc}_{E} \bigl[f(\rho_0(E))\bigr] + \text{Disc}_{E} \bigl[f(-\rho_0(E))\bigr].
\end{equation}
Note that in the $E$-plane, the logarithm contains a branch cut for $E<0$. This contribution is zero since the symmetric combination for the discontinuity as a function of $E$ vanishes:
\begin{equation}
    \text{Disc}_{E<0} \bigl[f(\rho_0(E))\bigr] + \text{Disc}_{E<0} \bigl[f(-\rho_0(E))\bigr] = 0.
\end{equation}
Therefore, the only contribution to the discontinuity comes from the interval $[0,E_*]$. Thus, converting the $\rho_0$ contour integral into an $E$ contour integral, we obtain:
\begin{align}
\Khighlow_{\lowElabel,\beta}^{{\text{GUE}}}
    &= \int_0^{E_*} dE \ e^{-2\beta E} \left\{ \text{Disc}_{E} \bigl[f(\rho_0(E))\bigr] + \text{Disc}_{E} \bigl[f(-\rho_0(E))\bigr]\right\} \nonumber\\
    &= \oint_{[0,E_*]} dE \ e^{-2\beta E} \bigl(f(\rho_0(E)) + f(-\rho_0(E))\bigr)\,.
\end{align}
This is illustrated in figure \ref{fig:contours}, (a) and (b).

The contour integrals are suited for extracting the topological expansion. To this end, we need to expand the function $f(\rho_0)$ in powers of $\uptau$, or equivalently in inverse powers of $\rho_0$. More precisely, we deform the $E$ contour to enclose the original interval but taking $|\rho_0|$ arbitrarily large. Since $|\rho'|$ is bounded we can expand $\frac{1}{\rho'-\rho_0}$ as a geometric series inside the integral:
\begin{equation}
f(\rho_0)=\int_0^{\frac{\uptau}{2\pi}}\frac{ d\rho'}{2 \pi i} \, \left(\rho'-\frac{\uptau}{2\pi}\right)\sum_{n=1}^\infty \frac{(\rho')^{n-1}}{\rho_0^n}= -\frac{1}{2 \pi i} \sum_{n=1}^\infty  \frac{\left(\frac{\uptau}{2\pi}\right)^{n+1}}{n(n+1)} \frac{1}{\rho_0^n}\,.
\end{equation}
Forming the symmetric combination $f(\rho_0) + f(-\rho_0)$, this gives
\begin{equation}
\begin{split}
\Khighlow_{\lowElabel,\beta}^{{\text{GUE}}}
&=- \sum_{n=1}^\infty \frac{\uptau^{n+1}}{2\pi n(n+1)} \frac{1}{(2\pi)^n}\oint_{[0,E_*]} \frac{dE}{2\pi i} \, e^{-2\beta E} \,\left(  \frac{1}{\rho_0(E)^n} +  \frac{1}{(-\rho_0(E))^n} \right) \,.
\end{split}
\end{equation}
Note that only the even powers $n=2g$ survive in the symmetric combination:
\begin{equation}
\begin{split}
\Khighlow_{\lowElabel,\beta}^{{\text{GUE}}}
&=- \sum_{g=1}^\infty \frac{\uptau^{2g+1}}{2\pi g(2g+1)} \frac{1}{(2\pi)^{2g}}\oint_{[0,E_*]} \frac{dE}{2\pi i} \, e^{-2\beta E} \,  \frac{1}{\rho_0(E)^{2g}}  \,.
\end{split}
\end{equation}
 For even powers, the integrand is then a meromorphic function of $E$ (recall \eqref{eq:rhoGeneral}). While the branch cut disappears, a pole at $E=0$ remains. The last step is therefore to deform the $E$ contour for individual terms in the sum to a small circle around $E=0$, see figure \ref{fig:contours}(c). Contracting the contour in this manner yields the final expression for the $\uptau$-scaled SFF, \eqref{eq:GUESFF}.

\subsubsection{Examples}
\label{sec:GUEexamples}

For illustration of the general formula \eqref{eq:GUESFF}, we discuss a few examples.

\paragraph{Example I: Airy model.}
The simplest example of a double-scaled random matrix model with a spectral curve of the form \eqref{eq:rhoGeneral} is the Airy model:
\begin{equation}
\rho_0^\text{Airy}(E) = \frac{\sqrt{E}}{2\pi}\,,
\end{equation}
 we have
\begin{equation}
\label{eq:AGUEairy}
c_{2g}\big(\rho_0^\text{Airy};\beta\big) = \frac{(-2\beta)^{g-1}}{(g-1)!} \qquad \Rightarrow \qquad 
A_g^\text{GUE}\big(\rho_0^\text{Airy};\beta\big)= - \frac{(-2\beta)^{g-1}}{2\pi(2g+1)g!} \,.
\end{equation}
Plugging these coefficients into the general formula \eqref{eq:GUESFF}, we can perform the sum over $g$ explicitly and find the known expression:
\begin{equation}
\label{eq:KGUEAiry}
 {\cal K}_\beta^{{\text{GUE,Airy}}}(\uptau) = \frac{1}{8\sqrt{2\pi}\beta^{3/2}} \, \text{Erf} \left( \sqrt{2\beta}\, \uptau \right) \,.
\end{equation}

\paragraph{Example II: $(2,p)$ minimal string for $p=3$.}
The $(2,p)$ minimal string with $p=3$ has a matrix model description with the spectral density \cite{Maldacena:2004sn,Saad:2019lba}
\begin{equation}
\label{eq:rho023}
\rho_0^{(2,p=3)}(E) = \frac{\sqrt{E}}{2\pi} \left( 1 + \frac{2}{3\kappa} \, E \right) \,.
\end{equation}
In this case we compute the coefficients \eqref{eq:PgDef} using the following expansion:
\begin{equation}
\label{eq:23expansion}
    \frac{1}{\left[\rho_0^{(2,p=3)}(E)\right]^{2g}} = (2\pi)^{2g} \sum_{k\geq 0} \frac{(2g+k-1)!}{(2g-1)!\,k!} \left( - \frac{2}{3\kappa} \right)^k E^{k-g} \,.
\end{equation}
    To compute the residue at $E=0$, we note that the integrand in \eqref{eq:PgDef} has a pole of order $g$ at $E=0$. This lets us compute the following residue:
\begin{equation}
\begin{split}
c_{2g}\big(\rho_0^{(2,p=3)};\beta\big) 
&= 
\frac{1}{(g-1)!} \, \lim_{E \rightarrow 0} \left(\frac{d}{dE}\right)^{g-1} \left[  \sum_{k,\ell\geq 0} \frac{(2g+k-1)!}{(2g-1)!\,k!\,\ell!} \left( - \frac{2}{3\kappa} \right)^k (-2\beta)^\ell E^{k+\ell} \right]
\\
&= \frac{1}{(g-1)!} \, \left( -\frac{2}{3\kappa} \right)^{g-1} (3\kappa\beta)^{3g-1} \, U\left(2g,3g;3\kappa\beta \right)
\,.
\end{split}
\label{eq:23ccoeff}
\end{equation}
This yields the coefficients in the genus expansion of ${\cal K}_\beta^\text{GUE}(\uptau)$:
\begin{equation}  A_g^\text{GUE}\big(\rho_0^{(2,p=3)};\beta\big) = \frac{(-2)^{g}}{4\pi (2g+1) g!} \,  \beta^{3g-1}(3\kappa)^{2g} \, U\left(2g,3g;3\kappa\beta \right) \,.
\end{equation}
Note that the confluent hypergeometric function appearing here truncates for any integer $g$, i.e., it is a finite polynomial in $\beta$.

\paragraph{Example III: JT gravity.}
Next, we compute $A_g^\text{GUE}$ for JT gravity with density of states given by
\begin{equation}
\label{eq:rho0JT}
  \rho_0^\text{JT}(E) = \frac{\sinh(2\pi \sqrt{E})}{4\pi^2}.
\end{equation}
Using the series expansion of $\sinh(x)$, we can work out the expansion required for the computation of the integrals $c_{2g}(\rho_0^\text{JT};\beta)$; this involves a sum over integer partitions to account for all terms appearing in the coefficient of any given power of $E$:
\begin{equation}
\label{eq:rho0JTexpanded}
\begin{split}
 \frac{1}{\left[\rho_0^\text{JT}(E)\right]^{2g}} &= (2\pi)^{2g} \sum_{p\geq 0} \left[ (2\pi)^{2p} \sum_{\{m_r\}_p} \frac{\Gamma\left(2g + \sum_r m_r\right)}{\Gamma(2g)} \prod_r \frac{1}{m_r!} \left(- \frac{1}{(2r+1)!} \right)^{m_r} \right] \, E^{p-g}
\end{split}
\end{equation}
where each set $\{m_r\}_p = \{m_1,\ldots, m_p\}$ denotes an integer partition of $p$ of the form $p = \sum_r r m_r$.
Performing the residue \eqref{eq:PgDef} term by term, we find the following coefficients:
\begin{equation}
\begin{split}
  c_{2g}\big(\rho_0^\text{JT};\beta\big) 
  &=  \frac{1}{\Gamma(2g)} \sum_{m_1,m_2,\ldots =0}^\infty \left[\frac{\Gamma\left(2g + \sum_r m_r\right)}{\Gamma(g-\sum_r rm_r)} \prod_r \frac{1}{m_r!} \left(- \frac{(2\pi)^{2r}}{(2r+1)!} \right)^{m_r} \right] (-2\beta)^{g-1-\sum_r rm_r} 
   \end{split}
   \label{eq:AgJTGUE}
\end{equation}
where the sums over $m_r$ are now unconstrained. But note that the summand is zero whenever $\sum_r r m_r \geq g$, so for any given $g$ there are only finitely many terms. Also, all powers of $\beta$ are positive, so the expression is a finite polynomial. The term with $m_1 = m_2 = \ldots = 0$ corresponds to the Airy result. An alternative way of writing the result, which is evident from the construction but more efficient for explicit calculations, is:
\begin{equation}
    c_{2g}\big(\rho_0^\text{JT};\beta\big) = \frac{1}{(g-1)!} \,\left(\frac{d}{dy} \right)^{g-1}\left[ e^{-2\beta y} \left( \frac{2\pi \sqrt{y}}{\sinh(2\pi \sqrt{y})} \right)^{2g} \right]_{y=0} \,.
\end{equation}

For later reference, we record the first few coefficients:
\begin{equation}
\label{eq:c2gJTexplicit}
\begin{split}
 c_{2}\big(\rho_0^\text{JT};\beta\big)  &= 1 \,,\\
 c_{4}\big(\rho_0^\text{JT};\beta\big)  &= -\frac{2(2\pi)^2}{3} -2  \beta \,,\\
 c_{6}\big(\rho_0^\text{JT};\beta\big)  &= \frac{8(2\pi)^{4}}{15} +2(2\pi)^2 \beta +2  \beta^2 \,,\\
 c_{8}\big(\rho_0^\text{JT};\beta\big)  &= -\frac{16(2\pi)^{6}}{35} - \frac{28(2\pi)^{4}}{15} \, \beta - \frac{8(2\pi)^{2}}{3}\, \beta^2 - \frac{4}{3}\, \beta^3  \,.\\
\end{split}
\end{equation}
The $\uptau$-scaled SFF follows by plugging \eqref{eq:c2gJTexplicit} into \eqref{eq:GUESFF}:
\begin{equation}
\label{eq:ZJTgravity}
    {\cal K}_\beta^{\text{JT,GUE}}(\uptau)  = \frac{\uptau}{4 \pi \beta} - \frac{\uptau^3}{6 \pi}  + \frac{(3\beta+4\pi^2)\uptau^5}{30 \pi}  - \frac{(15 \beta^2 + 60 \pi^2 \beta + 64 \pi^2)\uptau^7}{315 \pi} - \frac{c_{8}\big(\rho_0^\text{JT};\beta\big)}{72\pi} \, \uptau^9 + \ldots 
\end{equation}
One can check by expanding to arbitrary orders in $\uptau$ that the above derivation is consistent with the closed-form expression given in eq.\ (2.12) of \cite{Saad:2022kfe}.

\subsection{The GOE $\uptau$-scaled SFF}
\label{sec:GOESFF}

We will now generalize the method from above to the GOE ensemble. Our task is to compute the Laplace transform of the universal sine kernel in the GOE universality class: 
\begin{equation}
\label{eq:GOElaplaceStart}
\begin{split}
    &{\cal K}_\beta^\text{GOE}(\uptau )=\int_0^\infty dE \, e^{-2\beta E} \; \text{min} \left\{ \frac{\uptau}{\pi} - \frac{\uptau}{2\pi} \, \log \left( 1 + \frac{\uptau}{\pi\rho_0(E)} \right) \,,\; 2\rho_0(E) - \frac{\uptau}{2\pi} \, \log \left( \frac{\uptau+\pi \rho_0(E)}{\uptau-\pi \rho_0(E)}\right) \right\}.
\end{split}
\end{equation}
This is equal to the following three contributions: 
\begin{equation}
\begin{split}
 {\cal K}_{\beta}^{{\text{GOE}}}(\uptau) &= 
 2 \times {\cal K}_{\beta}^{{\text{GUE}}}(\uptau)  \\
 &\quad -\,\frac{\uptau}{2\pi}\int_0^{E_*} dE \, e^{-2\beta E}  \, \log \left( \frac{ \frac{\uptau}{\pi} + \rho_0(E)}{\frac{\uptau}{\pi} - \rho_0(E)} \right)  -\frac{\uptau}{2\pi}\int_{E_*}^\infty dE \, e^{-2\beta E} \,  \log \left( 1 + \frac{\uptau}{\pi \rho_0(E)} \right) \,.
 \end{split}
 \label{eq:GOEdecompose}
\end{equation}
First we have (twice) the GUE result, second there is a low energy contribution associated with the microcanonical plateau (times $\uptau>2\pi \rho_0(E)$), and third is the high energy contribution associated with the microcanonical ramp (times $\uptau<2\pi \rho_0(E)$).\footnote{The three pieces are illustrated for the Airy model in figure \ref{fig:airyDecompose}.}

We will proceed by presenting the general result and discussing some examples. We leave the details of the derivation to Appendix \ref{app:derivations}.
For a general spectral density $\rho(E) = \rho_0(E) e^{S_0}$ we find the following $\uptau$-scaled spectral form factor:
\begin{equation}
\boxed{
\begin{aligned}
 {\cal K}_{\beta}^{\text{GOE}}(\uptau) = 
  \frac{\uptau}{2\pi \beta} &+ \sum_{g=1,2,...}^\infty \left[ A_{g}^\text{GOE}(\rho_0;\beta)+ B_{g}^\text{GOE}(\rho_0;\beta) \, \log(\uptau) \right] \, \uptau^{2g+1} \\
  & + \sum_{\gt=\frac{1}{2},\frac{3}{2},...}^\infty C_{\gt}^\text{GOE}(\rho_0;\beta) \, \uptau^{2\gt+1}
  \end{aligned}
 }
 \label{eq:GOEgeneral}
\end{equation}
where
\begin{equation}
\begin{split}
 A_{g}^\text{GOE}(\rho_0;\beta) 
 &= -\frac{2}{2\pi g(2g+1)} \left[1+\frac{4^{g-1}(2g+1)}{g} \big( g\,\log (4)-{}_2F_1(2g,2g,2g+1,-1) -1\big)\right] c_{2g}(\rho_0;\beta) \\
 &\quad + \frac{4^{g-1}}{\pi g} \,    d_{2g}(\rho_0;\beta)   \,,
  \\
  B_{g}^\text{GOE}(\rho_0;\beta) &
  = - \frac{4^{g-1}}{\pi g} \,  2c_{2g}(\rho_0;\beta)\,,
  \\
 C_{\gt}^\text{GOE}(\rho_0;\beta) &= - \frac{4^{\gt-1}}{\pi \gt} \, d_{2\gt}(\rho_0;\beta)\,.
 \end{split}
 \label{eq:generalA}
\end{equation}
The information about the specific choice of $\rho_0(E)$ is encoded in the coefficients
\begin{equation}
    \begin{split}
     c_{2g}(\rho_0;\beta) &\equiv \frac{1}{(2\pi)^{2g}} \oint \frac{dE}{2\pi i}  \, e^{-2\beta E} \, \rho_0(E)^{-2g} \,,\\
        d_{n}^{(\delta)}(\rho_0;\beta) &\equiv \frac{1}{(2\pi)^n} \int_\delta^\infty dE \, e^{-2\beta E} \, \rho_0(E)^{-n}\,.
    \end{split}
\end{equation}
and $d_n(\rho_0;\beta)$ is the {\it finite part} of $d_{n}^{(\delta)}(\rho_0;\beta)$ as $\delta \rightarrow 0$.

We observe three kinds of terms based on their $\uptau$-dependence: the $A$-type terms, which are odd powers of $\uptau^{2g+1}$ similar to the GUE case; the $B$-type terms, which multiply odd powers times logarithms, $\log(\uptau)\uptau^{2g+1}$; and the $C$-type coefficients corresponding to half-integer genus contributions, which produce even powers of $\uptau$. Note that the GUE piece and the low energy plateau contrbution $I^{\rm GOE}_{\rm low}$ contribute only to the A-type terms, while the high energy ramp part  $I^{\rm GOE}_{\rm high}$ is solely responsible for the B- and C-type terms and for part of the A-type terms. 

The dependence on $\beta$, encoded in $A_{g}^\text{GOE}(\rho_0;\beta)$, $B_{g}^\text{GOE}(\rho_0;\beta)$, $C_{g}^\text{GOE}(\rho_0;\beta)$, is very sensitive to the choice of spectral curve. This is a qualitatively new feature: while the GUE spectral form factor was only sensitive to the behavior of the spectral density near $E=0$ via $c_{2g}(\rho_0;\beta)$, the GOE case features new coefficients $d_n(\rho_0;\beta)$, which are sensitive to $\rho_0(E)$ at all energies. 
In the GUE, this meant that the genus $g$ term was only sensitive to the first $g-1$ terms in the low energy expansion of $\rho_0(E)$. For instance, in GUE JT gravity the $\uptau$-scaled SFF was not sensitive to the high energy exponential growth of  the density of states for any $g$. Instead in the GOE case, there is a non-trivial high-energy contribution to the Laplace transform \eqref{eq:GOElaplaceStart}. We discuss this contribution in detail in Appendix \ref{sec:goeHigh}. This shows how the GOE is more generic and qualitatively sensitive to new physics. 

A few more comments are in order: 
\begin{enumerate}
\item[$(i)$] In the above expressions we introduce the following notation: $g$ and $\gt$ are both interpreted as `genus'. However, $g\in \mathbb{Z}_+$ indicates integer genus, while $\gt \in \frac{1}{2} + \mathbb{Z}_+$ indicates half-integer genus contributions. Half-integer genus contributions are associated with crosscap geometries when a geometric calculation is available.
\item[$(ii)$] The coefficients $B_g^\text{GOE}$ and $C_\gt^\text{GOE}$ only exist in the non-orientable symmetry classes (GOE and -- as discussed later -- GSE). Further, the very first term in $A_g^\text{GOE}$ equals twice $A_g^\text{GUE}$, but it receives additional additive corrections. As will become clear, the coefficients always add up such that the logarithmic dependence is actually of the form $\log(2\beta \uptau^2)$.
\item[$(iii)$] The coefficients $d_{n}$ are infrared-divergent, both for $n=2g$ even and for $n=2\tilde g$ odd. This is parametrized by the cutoff $\delta \rightarrow 0$. There appear powerlaw divergences up to degree $\delta^{-g+1}$ and (for integer genus) a logarithmic divergence $\log \delta$. In the genus expansion, we only keep the finite piece in the limit of small $\delta$. We note that this prescription is not ad hoc, but follows from the detailed derivation in Appendix \ref{app:derivations}, where we show that any divergences indeed get removed. For a summary of this mechanism, see section \ref{sec:periodicorbits}.
\item[$(iv)$] We can also work with a different regulator, which is often more convenient in practice. In many examples the finite piece of $d_{n}^{(\delta)}$ can be extracted by {\it analytic continuation in genus}. As we will illustrate, for half-integer genus, $d_{2\tilde{g}}$ can often be obtained by using integral formulas valid for $\text{Re}(\tilde{g})<1$ and analytically continuing the result to the desired value of $\tilde{g}$. A similar strategy also works for $d_{2g}$ (i.e., even integer index), but with an additional subtlety: analytic continuation of $d_n$ proceeds from $\text{Re}(n)<1$ to $n=2(g-\varepsilon)$, shifted from an even integer by a small amount $\varepsilon$. In the limit $\varepsilon \rightarrow 0$, one obtains a divergent contribution scaling as $\frac{1}{\varepsilon}$, as well as a finite contribution; again we need to extract the finite one, which we can write as:
\begin{equation}
\label{eq:drelation}
    \begin{split}
    d_{2g}(\rho_0;\beta) &= 
    \text{coeff}_{\varepsilon^0} \left( d_{2(g-\varepsilon)}(\rho_0;\beta) \right) = \lim_{\varepsilon\rightarrow 0} \, \frac{d}{d\varepsilon} \left[ \varepsilon\, d_{2(g-\varepsilon)}(\rho_0;\beta)\right]
    \end{split}
\end{equation}
where the infrared cutoff is no longer important.
We note that the coefficient $c_{2g}$ is also encoded in $d_{2g-\varepsilon}$, but via the divergent part:
\begin{equation}
\label{eq:cdrelation}
    \begin{split}
    c_{2g}(\rho_0;\beta) &=  \text{coeff}_{\varepsilon^{-1}} \left( d_{2(g-\varepsilon)}(\rho_0;\beta) \right) =\; \lim_{\varepsilon\rightarrow 0} \left[ \varepsilon \, d_{2(g-\varepsilon)}(\rho_0;\beta) \right] \,.
    \end{split}
\end{equation}
We can therefore extract all required coefficients from a single calculation that obtains the finite and divergent pieces of $d_{2(g-\varepsilon)}$ and $d_{2(\tilde{g}-\varepsilon)}$. See section \ref{sec:periodicorbits} for more discussion of this.
\end{enumerate}

\subsubsection{Examples}

\subsubsection*{Example I: Airy model} 
\label{sec:ex1Airy}

We begin with the Airy model $\rho_0^\text{Airy}(E) = \tfrac{\sqrt{E}}{2\pi}$. The coefficients sensitive to low energies are the same as in the GUE:
\begin{equation}  c_{2g}\big(\rho_0^\text{Airy};\beta\big) = \frac{(-1)^{g-1}}{(g-1)!} \, (2\beta)^{g-1}\,.
\end{equation}
Next, we discuss the coefficients $d_n$, which are sensitive to all energies. For illustration we discuss different regularization schemes. 

\paragraph{{\it Infrared cutoff regularization:}}
With an infrared regulator $\delta$, we find
\begin{equation}
    \begin{split}   d_{2g}^{(\delta)}\big(\rho_0^\text{Airy};\beta\big) &= \frac{(-1)^{g-1}}{(g-1)!} \left[ \psi(g)-\log\left(2\beta\right) \right] (2\beta)^{g-1} + \sum_{m=1}^{g-1} \frac{(-2\beta)^{g-1-m}}{m(g-1-m)!} \, \frac{1}{\delta^{m}} - \frac{(-2\beta)^{g-1}}{(g-1)!} \, \log(\delta) + {\cal O}(\delta)\,,
    \\
d_{2\tilde{g}}^{(\delta)}\big(\rho_0^\text{Airy};\beta\big) &=  \Gamma(1-\tilde g) (2\beta)^{\tilde g-1} + \sum_{\tilde{m} = \frac{1}{2},\frac{3}{2},\ldots}^{\tilde{g}-1} \frac{(-2\beta)^{\tilde{g}-1-m}}{m \, \Gamma(\tilde g-m)} \, \frac{1}{\delta^m} \,.
    \end{split}
    \label{eq:AiryDeltaDiv}
\end{equation}
The powerlaw divergences are scheme-dependent and irrelevant for the SFF.
The relevant (and scheme-independent) terms, according to \eqref{eq:GOEgeneral}, are the finite ones:
\begin{equation}
\label{eq:cdAiry}
    \begin{split}   d_{2g}\big(\rho_0^\text{Airy};\beta\big) &\equiv \left[ d_{2g}^{(\delta)}\big(\rho_0^\text{Airy};\beta\big) \right]_\text{finite} = \frac{(-1)^{g-1}}{(g-1)!} \left[ \psi(g)-\log\left(2\beta\right) \right] (2\beta)^{g-1} \,,\\
    d_{2\tilde g}\big(\rho_0^\text{Airy};\beta\big) &\equiv \left[ d_{2\tilde g}^{(\delta)}\big(\rho_0^\text{Airy};\beta\big)\right]_\text{finite} = \Gamma(1-\tilde g) (2\beta)^{\tilde g-1}\,.
    \end{split}
\end{equation}
We also note that the coefficient of $\log(\delta)$ reproduces $c_{2g}$; we discuss this in more detail in section \ref{sec:periodicorbits}.
Plugging into \eqref{eq:generalA}, we infer the expansion coefficients in the $\uptau$-scaled SFF \eqref{eq:GOEgeneral}:
\begin{equation}
\label{eq:AresultAiry}
\begin{split}
 A_{g}^\text{GOE}\big(\rho_0^\text{Airy};\beta\big) 
 &= -\frac{(-2\beta)^{g-1}}{\pi (2g+1) g!} + \frac{(-8\beta)^{g-1}}{\pi  g!} \left[\frac{1}{g}\;{}_2F_1(2g,2g,2g+1,-1) +  \psi(g+1)  - \log\left(8\beta\right)  \right] 
  \\
  B_{g}^\text{GOE}\big(\rho_0^\text{Airy};\beta\big) &= -\frac{2(-4)^{g-1}}{\pi g!} \,(2\beta)^{g-1}\,,
  \\
 C_{\gt}^\text{GOE}\big(\rho_0^\text{Airy};\beta\big) &= -\frac{4^{\gt-1}}{\pi \gt} \Gamma(1-\gt) \, (2\beta)^{\gt-1}.
 \end{split}
\end{equation}
Explicitly, up to genus $\frac{5}{2}$, we get:
\begin{equation}\label{eq:AiryGOESFF}
    \begin{split}
    {\cal K}_\beta^\text{GOE,Airy}(\uptau) &= \frac{1}{2\pi \beta}\,\uptau -\frac{1}{\sqrt{2\pi\beta}}\, \uptau^2   - \frac{1}{\pi} \left(\frac{1}{3}+\gamma +\log(2\beta\uptau^2)\right) \uptau^3  +  \frac{8}{3\pi} \,(2\pi\beta)^{\frac{1}{2}} \, \uptau^4 \\
    &\quad + \frac{4\beta}{\pi} \left(-\frac{7}{60} + \gamma + \log(2\beta\uptau^2) \right)  \uptau^5  - \frac{64}{15\pi^2} \, (2\pi\beta)^{\frac{3}{2}}\,  \uptau^6  + \ldots 
    \end{split}
\end{equation}
This is consistent with the expressions obtained using a different method in \cite{Weber:2024ieq}. 
Note that, after dividing by an overall factor of $\uptau^3$, \eqref{eq:AiryGOESFF} only depends on $\uptau$ and $\beta$ through the combination $\beta\uptau^2$. This is a simplification of the Airy model, which can be useful to study the late-time behavior.

\paragraph{{\it Analytic continuation in genus:}}
For illustration, consider now the other regulator described in the comments after \eqref{eq:GOEgeneral}, i.e., analytic continuation in genus. We use the formula 
\begin{equation}
\label{PowerIntegral}
\int_0^{\infty} dE\, e^{-2 \beta E} \,E^{x} =  \Gamma(1+x) (2\beta)^{-x-1}\,,
\end{equation}
which converges for $\text{Re}(x)>-1$. For half-integer genus $\gt$, we immediately find the finite piece of $d_{2\tilde g}$ by analytically continuing this formula to $x=-\tilde{g}$, in agreement with \eqref{eq:cdAiry}. For integer genus $g$, we analytically continue to $x=-g+\varepsilon$ and expand in small $\varepsilon$. The result is finite in the IR regulator $\delta$ (so we can drop it), but divergent as $\varepsilon \rightarrow 0$:
\begin{equation}
    \begin{split}
    d_{2(g-\varepsilon)}\big(\rho_0^\text{Airy};\beta\big) &= \frac{(-1)^{g-1}}{(g-1)!} \left[ \frac{1}{\varepsilon} +  \psi(g)-\log\left(2\beta\right) + {\cal O}(\varepsilon)\right] (2\beta)^{g-1}\,.
    \end{split}
\end{equation}
Both the finite and the divergent pieces have physical meaning. As anticipated on general grounds, the finite piece agrees with \eqref{eq:cdAiry}, thus confirming \eqref{eq:drelation}. The divergent piece is $c_{2g}$, as expected from \eqref{eq:cdrelation}.

\subsubsection*{Example II: $(2,p)$ minimal string for $p=3$}

Let us again consider the density of states for the $(2,p=3)$ minimal string, \eqref{eq:rho023}.
The coefficients $c_{2g}$ were already computed for the GUE, using the binomial expansion \eqref{eq:23expansion}:
\begin{equation}
    \begin{split}
c_{2g}\big(\rho_0^{(2,p=3)};\beta\big) &= \left( -\frac{2}{3\kappa} \right)^{g-1} \frac{(3\kappa\beta)^{3g-1}}{ \Gamma(g)} \, U\left(2g,3g;3\kappa\beta \right)  \\
    &=c_{2g}\big(\rho_0^{\text{Airy}};\beta\big)  \left[ 1 + \frac{2g(g-1)}{3\kappa\beta} + \ldots + \frac{(3g-2)!}{(2g-1)!(3\kappa\beta)^{g-1}}
    \right]
    \,.
    \end{split}
\end{equation}

We compute the coefficients $d_n$ using analytic continuation in genus. We start with the formula
\begin{equation}
    \int_0^\infty dE \, e^{-2\beta E}\left[ \sqrt{E}\left(1+\frac{2}{3\kappa} E\right) \right]^x
 = \Gamma\left(1+\frac{x}{2}\right)\left(\frac{3\kappa}{2}\right)^{1+\frac{x}{2}} \, U\left( 1+\frac{x}{2} ,\, 2+\frac{3x}{2} ;\, 3\kappa\beta \right)\,,
 \label{eq:23formula}
\end{equation}
where $U(a,b;x)$ is the confluent hypergeometric function. The integral converges for $\text{Re}(x)>-2$. As illustrated before, we can extract the relevant pieces of $d_n$ by analytically continuing \eqref{eq:23formula} to $x=-2\tilde g$ and $x=-2(g-\varepsilon)$.
We find:\footnote{We rewrite the hypergeometric functions using the identity $U(a,b;z) = z^{1-b}\, U(1+a-b,2-b;z)$.}
\begin{equation}
    \begin{split}
    d_{2\tilde g}\big(\rho_0^{(2,p=3)};\beta\big) &= \Gamma\left( 1-\tilde g\right) \, (2\beta)^{\tilde{g}-1}  \, (3\kappa\beta)^{2\tilde{g}}\, U \left(2\tilde g, 3\tilde g ; 3\kappa\beta\right)\,,
     \\
 d_{2(g-\varepsilon)}\big(\rho_0^{(2,p=3)};\beta\big) &= \Gamma\left( 1-g+\varepsilon\right) \,  (2\beta)^{g-\varepsilon-1}  \, (3\kappa\beta)^{2(g-\varepsilon)}\, U \left(2 (g-\varepsilon), 3 (g - \varepsilon); 3\kappa\beta\right)\,.
\end{split}
\label{eq:cdGOEstring}
\end{equation}
As a consistency check, note that the $\kappa \rightarrow \infty$ limit of these coefficients reproduces the Airy expressions in \eqref{eq:cdAiry}. Equivalently, the leading terms in the large $\beta$ expansion yield the Airy results. It is interesting to observe that the Airy results in fact contribute in a {\it multiplicative} way and can naturally be factored out:
\begin{equation}
    \begin{split}
    d_{2\tilde g}\big(\rho_0^{(2,p=3)};\beta\big) 
    &= d_{2\tilde g}\big(\rho_0^{\text{Airy}};\beta\big) \,\times (3\kappa\beta)^{2\tilde{g}}\, U \left(2\tilde g, 3\tilde g ; 3\kappa\beta\right)
     \\
 d_{2(g-\varepsilon)}\big(\rho_0^{(2,p=3)};\beta\big) 
 &= d_{2(g-\varepsilon)}\big(\rho_0^\text{Airy};\beta\big) \bigg\{ \left[ 1 + \frac{2g(g-1)}{3\kappa\beta} 
 + \ldots + \frac{(3g-2)!}{(2g-1)!(3\kappa\beta)^{g-1}} \right] \\
 &\qquad\qquad\qquad\qquad + \varepsilon \left[ -\frac{2(2g-1)}{3\kappa\beta}-\frac{8g^3-15g^2+2g+2}{(3\kappa\beta)^2} + \ldots \right] +{\cal O}(\varepsilon^2)\bigg\}
\end{split}
\label{eq:cdGOEstringExpanded}
\end{equation}
These expressions imply that all coefficients in the topological expansion (with $g> 1$) receive corrections in negative powers of $(3\beta\kappa)$, relative to the Airy model:
\begin{equation}
\begin{split}
A_g^\text{GOE}\big(\rho_0^{(2,p=3)};\beta\big) 
&= A_g^\text{GOE}\big(\rho_0^{\text{Airy}};\beta\big) \left[ 1 + \frac{2g(g-1)}{3\kappa\beta} + \ldots + \frac{(3g-2)!}{(2g-1)!(3\kappa\beta)^{g-1}}
    \right] \\
    &\quad - B_g^\text{GOE}\big(\rho_0^{\text{Airy}};\beta\big)\left[ -\frac{(2g-1)}{3\kappa\beta} - \frac{8g^3-15g^2+2g+2}{2(3\kappa\beta)^2} + \ldots \right] \,,
    \\
B_g^\text{GOE}\big(\rho_0^{(2,p=3)};\beta\big) &= \frac{2(-1)^g}{\pi g!} \,(8\beta)^{g-1}\left(3\kappa\beta\right)^{2g}  U(2g,\,3g;\,3\kappa\beta) \\
 &= B_g^\text{GOE}\big(\rho_0^{\text{Airy}};\beta\big) \left[ 1 + \frac{2g(g-1)}{3\kappa\beta} +  
 \ldots + \frac{(3g-2)!}{(2g-1)! (3\kappa\beta)^{g-1}} \right]\,,
 \\
C_\gt^\text{GOE}\big(\rho_0^{(2,p=3)};\beta\big) &= - \frac{4^{\gt-1}\Gamma\left(1-\gt \right)}{\pi \gt} \,( 2\beta)^{\gt-1} (3\kappa\beta)^{2\gt} \, U\left(2\gt, \,3\gt;\,3\kappa\beta\right)\\
&= C_{\gt}^\text{GOE}\big(\rho_0^{\text{Airy}};\beta\big) \left[ 1 + \frac{2\gt(\gt-1)}{3\kappa\beta} + \frac{(2\gt+1)\gt(\gt-1)(\gt-2)}{(3\kappa\beta)^2} +  \ldots \right]\,,
 \end{split}
 \label{eq:23logeven}
\end{equation}
where the Airy coefficients are given in \eqref{eq:AresultAiry}.

Having analytic expressions, we can also work out the {\it small} $\beta$ expansion. For example, the expansion of $C_\gt^\text{GOE}$ does not truncate and is therefore non-trivial to expand in small $\beta$:
\begin{equation}
\begin{split}
&C_\gt^\text{GOE}\big(\rho_0^{(2,p=3)};\beta\big) \\
&\qquad = - 
\left( \frac{8}{3\kappa} \right)^{\gt-1}\,\frac{2\,\Gamma\left(1-\gt\right)\Gamma\left(3\gt-1\right)}{\pi\,\Gamma\left(2\gt+1\right)} \left[ 1 + \frac{3(2\gt-2)}{(6\gt-4)}(\kappa\beta ) + \frac{3(2\gt-4)}{2(6\gt-4)}(\kappa\beta )^2 + \ldots
    \right] \\
    &\qquad \quad +\frac{3}{4\pi} (3\kappa)^{2\gt} \Gamma\left(-3\gt\right)\left[ 1 + 2\kappa \beta + \frac{3(2 \gt+1)}{3\gt+1} (\kappa\beta)^2 + \ldots\big) \right](2\beta)^{3\gt-1} 
\end{split}
\end{equation}
Importantly, note that this expression has no negative powers of $\beta$ and thus a well-defined high-temperature limit (the same true for $A_g^\text{GOE}$ and $B_g^\text{GOE}$). This was not manifest in \eqref{eq:23logeven}.

\subsubsection*{Example III: JT gravity}
\label{sec:GOE-JT-Laplace}

Finally, we turn to computing the canonical SFF for non-orientable JT gravity with density of states \eqref{eq:rho0JT}. We will evaluate the general formula \eqref{eq:GOEgeneral} in different ways. It is instructive to consider expansions in both low and high temperatures separately.

\paragraph{Low temperature expansion.}
In the low temperature limit, we can again consider the expansion of inverse powers of $\rho_0^\text{JT}$ given in \eqref{eq:rho0JTexpanded}. That expansion holds for any $g$, not necessarily integer. The computation of the coefficients $d_n(\rho_0^\text{JT};\beta)$ then reduces to a sum over Laplace transforms of powers of $E$:
\begin{equation}
\begin{split}
 d_{n}\big(\rho_0^\text{JT};\beta\big) &= \sum_{p\geq 0} 
 \left[ (2\pi)^{2p} \sum_{\{m_r\}_p} \frac{\Gamma\left(n + \sum_r m_r\right)}{\Gamma(n)} \prod_r \frac{1}{m_r!} \left(- \frac{1}{(2r+1)!} \right)^{m_r} \right] \Gamma\left(p+1-\frac{n}{2}\right) (2\beta)^{\frac{n}{2}-p-1}
 \end{split}
 \label{eq:dnJT}
\end{equation}
For odd integer $n=2\tilde{g}$, this expansion is finite but does not truncate. It determines the coefficients of even powers $\uptau^{2\tilde{g}+1}$ in the $\uptau$-scaled SFF via \eqref{eq:generalA}. For example:
\begin{equation}
    \begin{split}
    d_{1}\big(\rho_0^\text{JT};\beta\big) 
    &= \pi\, (2\pi \beta)^{-\frac{1}{2}}\left[1 - \frac{\pi^2}{6\beta} + \frac{7\pi^4}{120\beta^2} - \frac{31\pi^6}{1008 \beta^3}  + \ldots \right] \,,\\
    d_3\big(\rho_0^\text{JT};\beta\big) 
    &= -2\, (2\pi\beta)^{\frac{1}{2}} \left[1 + \frac{\pi^2}{2\beta} - \frac{17\pi^4}{120\beta^2} +\frac{457\pi^6}{5040 \beta^3} + \ldots \right] \,,
    \\
    d_5\big(\rho_0^\text{JT};\beta\big) 
    &= \frac{4}{3\pi}\, (2\pi\beta)^{\frac{3}{2}}\left[ 1 + \frac{5\pi^2}{2\beta} + \frac{9\pi^4}{8\beta^2} -\frac{367\pi^6}{1008\beta^3} + \ldots \right] \,.
    \end{split}
\end{equation}

We can similarly use \eqref{eq:dnJT} to evaluate the coefficients of the remaining terms in the topological expansion. Of course, \eqref{eq:dnJT} diverges for $n=2g$ an even integer, so we employ our usual regularization scheme and instead evaluate at $n=2(g-\varepsilon)$. For instance:
{\small
\begin{equation}
\begin{split}
    d_{2-2\varepsilon}\big(\rho_0^\text{JT};\beta\big) 
    &=  \frac{1}{\varepsilon} + \left[ - (\gamma +\log(2\beta)) - \frac{2\pi^2}{3\beta} + \frac{4\pi^4}{15\beta^2} - \frac{32\pi^6}{189\beta^3} - \frac{512\pi^{10}}{3465\beta^5} + \ldots  \right] \,,\\
    d_{4-2\varepsilon}\big(\rho_0^\text{JT};\beta\big) 
    &= \left[-2\beta -\frac{8\pi^2}{3} \right] \frac{1}{\varepsilon} \\
    &\quad + \left[ 2\beta\, (\gamma +\log(2\beta)-1)  + \frac{8\pi^2}{3} \left(\gamma+ \log(2\beta) + \frac{1}{2}\right) + \frac{88\pi^4}{45\beta} - \frac{992\pi^6}{945\beta^2} + \ldots \right] \,,
    \\
    d_{6-2\varepsilon}\big(\rho_0^\text{JT};\beta\big) 
    &= \left[ 2\beta^2 + 8\pi^2\beta + \frac{128 \pi^4}{15} \right] \frac{1}{\varepsilon} \\
    &\quad + \bigg[ -2  \beta^2\, \left( \gamma + \log(2\beta) -\frac{3}{2} \right) - 8 \pi^2\beta \, \left( \gamma + \log(2\beta) -\frac{2}{3} \right)  \\
    &\qquad\;\;\, - \frac{128\pi^4}{15}\left( \gamma + \log(2\beta) + \frac{31}{48} \right)  - \frac{6112\pi^6}{945 \beta} + \frac{18496 \pi^8}{4725\beta^2} + \ldots \bigg]
    \,.
    \end{split}
    \label{eq:depsJT1}
\end{equation}
}\normalsize
We can again confirm that the divergent pieces are exactly $c_{2g}(\rho_0^\text{JT};\beta)$, c.f., \eqref{eq:c2gJTexplicit}.

From these coefficients (and \eqref{eq:c2gJTexplicit}) we can readily construct the coefficients in the genus expansion \eqref{eq:GOEgeneral}. We find for the SFF in non-orientable JT gravity up to genus $\frac{5}{2}$:
{\small
\begin{equation}
\label{eq:JTGOEexpandedResult}
    \begin{split}
    {\cal K}^{\text{GOE,JT}}_\beta(\uptau) 
    &= \frac{1}{2\pi \beta}\,\uptau \\
    &\quad -\frac{1}{\sqrt{2\pi\beta}}\left[1 - \frac{\pi^2}{6\beta} + \frac{7\pi^4}{120\beta^2} - \frac{31\pi^6}{1008 \beta^3}  + \ldots \right]  \uptau^2 \\
    &\quad + \left[ - \frac{1}{\pi} \left(\frac{1}{3}+\gamma +\log(2\beta\uptau^2)\right) -\frac{2\pi}{3\beta} + \frac{4\pi^3}{15\beta^2} - \frac{32\pi^5}{189\beta^3}  + \ldots \right] \uptau^3 \\
    &\quad +  \frac{8}{3\pi} \,(2\pi\beta)^{\frac{1}{2}} \left[1 + \frac{\pi^2}{2\beta} - \frac{17\pi^4}{120\beta^2} +\frac{457\pi^6}{5040 \beta^3} + \ldots \right] \uptau^4 \\
    &\quad + \left[ \frac{4\beta}{\pi} \left(-\frac{7}{60} + \gamma + \log(2\beta\uptau^2) \right)  + \frac{16\pi}{3} \left(  \frac{83}{60}+ \gamma + \log(2\beta\uptau^3)\right)  + \frac{176\pi^3}{45\beta}  + \ldots\right] \uptau^5 \\
    &\quad - \frac{64}{15\pi^2} \, (2\pi\beta)^{\frac{3}{2}}\left[ 1 + \frac{5\pi^2}{2\beta} + \frac{9\pi^4}{8\beta^2} -\frac{367\pi^6}{1008\beta^3} + \ldots \right]  \uptau^6 \\
    &\quad + \ldots 
    \end{split}
\end{equation}
}\normalsize
This quantity was also computed in \cite{Tall:2024hgo} up to order $\uptau^3$, see their eq.\ (4.19). Our result matches theirs up to that order.
We observe that just like the simpler GUE case, it is still true that for each power of $\uptau$, the leading term in the low temperature expansion corresponds to the Airy SFF in \eqref{eq:AiryGOESFF}. However, in addition there is not only an infinite series in powers of $\beta$ at each genus, but also new logarithmic terms, appearing at ${\cal O}(\uptau^5)$ for the first time (and proliferating at higher genus).

\paragraph{High temperature expansion.}
In Appendix \ref{app:JTdetails}, we derive recursive formulas for the coefficients $d_n$ in JT gravity, expanded in small $\beta$. The first few coefficients are (for half-integer genus):
{\small
\begin{equation}
    \begin{split}
    d_1\big(\rho_0^\text{JT};\beta\big) 
    &= \frac{\pi}{4} -\frac{\pi}{16}\, \beta + \frac{\pi}{32} \, \beta^2 -\frac{17\pi}{768} \, \beta^3 + \frac{31 \pi}{1536} \, \beta^4 + \ldots \\
    d_3\big(\rho_0^\text{JT};\beta\big) 
    &= -\frac{\pi^3}{2} + \frac{\pi^3-12\pi}{8} \, \beta -\frac{\pi^3-10\pi}{16} \, \beta^2 +\frac{17\pi^3-168 \pi}{384} \, \beta^3  \ldots 
    \\
    d_5\big(\rho_0^\text{JT};\beta\big) 
    &= \frac{3\pi^5}{2} - \frac{3 \pi^5-40 \pi^3}{8}\, \beta + \frac{9\pi^5-100\pi^3+120\pi }{48} \, \beta^2+ \ldots 
    \end{split}
    \label{eq:JThightemp1}
\end{equation}
}\normalsize
and (for integer genus):
{\small
\begin{equation}
    \begin{split}
    d_{2-2\epsilon}\big(\rho_0^\text{JT};\beta\big) 
    &= \frac{1}{\epsilon} +\left[ 2(1-\log(4\pi)) - \frac{3\zeta(3)}{2\pi^2} \, \beta + \frac{15\zeta(5)}{8\pi^4} \, \beta^2  -\frac{105\zeta(7)}{32\pi^6}\,\beta^3 + \ldots \right] +\ldots
    \\
    d_{4-2\epsilon}\big(\rho_0^\text{JT};\beta\big) 
    &= \left[ -\frac{8\pi^2}{3}-2\beta \right] \frac{1}{\epsilon}  +\bigg[ \frac{8\pi^2}{9}\left(6\log(4\pi)-5\right)+\left(4\log(4\pi)+4\zeta(3)-\frac{22}{3}\right)\beta \\
&\qquad  +\frac{5(\zeta(3)-\zeta(5))}{\pi^2}\,\beta^2 - \frac{35(\zeta(5)-\zeta(7))}{4\pi^2}\, \beta^3 + \frac{315(\zeta(7)-\zeta(9))}{16\pi^6} \, \beta^4 +\ldots
    \bigg] + \ldots \\
    d_{6-2\epsilon}\big(\rho_0^\text{JT};\beta\big) 
    &= \left[ \frac{128\pi^4}{15} +8\pi^2\beta+2\beta^2 \right] \frac{1}{\epsilon}  + \bigg[ -\frac{64\pi^4}{225}\left(60 \log(4\pi)-47\right) \\ 
    &\qquad + \frac{4\pi^2}{15} \left(101-60 \log(4\pi)-48 \zeta(3) \right)\beta  + \ldots
    \bigg] +\ldots
    \end{split}
    \label{eq:JThightemp2}
\end{equation}
}\normalsize
Note that the divergent terms match with the low temperature expansion, \eqref{eq:depsJT1}, because their series expansion (in $\beta$) truncates. However, the finite terms have been resummed in a non-trivial way, which we explore further in Appendix \ref{app:JTresummation}.

Collecting all the above results, we can write down the first few terms in the genus expansion of the JT SFF at high temperatures:
{\small
\begin{equation}
\label{eq:JTGOEexpandedResultHighT}
    \begin{split}
    {\cal K}^{\text{GOE,JT}}_\beta(\uptau) 
    &= \frac{1}{2\pi \beta}\,\uptau \\
    &\quad + \left[ -\frac{1}{4} + \frac{\beta}{16} - \frac{\beta^2}{32} + \frac{17\beta^3}{768} - \frac{31 \beta^4}{1536} + \ldots \right] \uptau^2 \\
    &\quad + \left[\frac{5}{3\pi}  - \frac{3\zeta(3)}{2\pi^3} \, \beta + \frac{15 \zeta(5)}{8\pi^5}\, \beta^2 - \frac{105 \zeta(7)}{32\pi^7} \, \beta^3  + \ldots 
 -\frac{2}{\pi}\, \log(4\pi\uptau)  \right] \uptau^3 \\
   &\quad +  \left[\frac{2\pi^2}{3}+\frac{12-\pi^2}{6}\,\beta - \frac{10-\pi^2}{12} \, \beta^2 + \frac{168-17\pi^2}{288} \, \beta^3    + \ldots \right] \uptau^4 \\
    &\quad + \left[-\frac{188\pi}{45} - \frac{167-120 \zeta(3)}{15\pi} \,\beta + \frac{10(\zeta(3)-\zeta(5))}{\pi^3} \, \beta^2 + \ldots 
    + \frac{8}{3\pi} (4\pi^2 + 3\beta) \, \log(4\pi \uptau)
    \right] \uptau^5 \\
    &\quad + \left[ -\frac{24\pi^4}{5} - \frac{80\pi^2-6\pi^4}{5}\,\beta - \frac{9\pi^4-100\pi^2+120}{15} \, \beta^2 + \ldots \right] \uptau^6 \\
    &\quad + \ldots 
    \end{split}
\end{equation}
}\normalsize
This should be compared to the low temperature expansion \eqref{eq:JTGOEexpandedResult}. Evidently, the two are related by a non-trivial resummation. In Appendix \ref{app:JTresummation} we confirm this resummation explicitly for low genus, using zeta-function regularization.

\subsection{The GSE $\uptau$-scaled SFF}

In the GSE ensemble, the universal expression we wish to compute is 
\begin{equation}
\label{eq:GSElaplaceStart}
\begin{split}
    {\cal K}_\beta^\text{GSE}(\uptau)  =\int_0^\infty dE \, e^{-2\beta E} \; \text{min} \left\{ \frac{\uptau}{4\pi} - \frac{\uptau}{8\pi} \, \log \left| 1 - \frac{\uptau}{2\pi\rho_0(E)} \right| \,,\; \rho_0(E) \right\} \,.
\end{split}
\end{equation}
The two terms exchange dominance at $\rho(E_*) = \frac{\uptau}{4\pi}$. In addition to the non-analyticity at the transition point, the expression for the ramp is also divergent at $\rho_0=\frac{\uptau}{2\pi}$. Using similar methods as before, we will find:
\begin{equation}
\begin{split}
  {\cal K}_\beta^\text{GSE}(\uptau) &= \frac{\uptau}{8\pi \beta} + \sum_{g=1,2,\ldots} \left[ A_g^\text{GSE}(\rho_0;\beta) + B_g^\text{GSE}(\rho_0;\beta) \, \log \left( \frac{\uptau}{2} \right) \right] \, \left( \frac{\uptau}{2} \right)^{2g+1}  \\
  &\qquad\quad\; + \sum_{\gt= \frac{1}{2},\frac{3}{2},\ldots} \, C_{\gt}^\text{GSE}(\rho_0;\beta) \, \left( \frac{\uptau}{2} \right)^{2\gt+1} 
\end{split}
\label{eq:GSEgeneralResult}
\end{equation}
with
\begin{equation}
\begin{split}
 A_g^\text{GSE}(\rho_0;\beta) 
 &= \frac{1}{2} \, A_g^\text{GOE}(\rho_0;\beta)\,,\\
 B_g^\text{GSE}(\rho_0;\beta) &= \frac{1}{2} \,  B_g^\text{GOE}(\rho_0;\beta)\,,\\
 C_\gt^\text{GSE}(\rho_0;\beta) &= -\frac{1}{2} \,  C_\gt^\text{GOE}(\rho_0;\beta)\,.
\end{split}
\end{equation}
Note that the coefficients encode no new information compared to GOE. The GSE expression is obtained from the corresponding GOE expression \eqref{eq:GOEgeneral} by the following simple rule:
\begin{equation}
\boxed{
    {\cal K}_\beta^\text{GSE} (\uptau) = \frac{1}{2} \, {\cal K}_\beta^\text{GOE} \left( \frac{\uptau}{2} \right) \bigg|_{C^\text{GOE}_{\gt}(\rho_0;\beta) \rightarrow - C^\text{GOE}_{\gt}(\rho_0;\beta) } 
}
\end{equation}
The similarity with GOE is to be expected because both the GOE and GSE have a time-reversal symmetry ${\cal T}$, but in the former it is idempotent (${\cal T}^2=1$) while in the latter it satisfies ${\cal T}^2 = -1$. The derivation of \eqref{eq:GSEgeneralResult} is given in Appendix \ref{app:GSEderivation}. We shall not discuss examples, as they follow immediately from the GOE expressions.

\subsection{Convergence of the $\uptau$-scaled topological expansion}

The explicit expressions for the topological expansion in the $\uptau$-scaling limit are \eqref{eq:GUESFF}, \eqref{eq:GOEgeneral}, \eqref{eq:GSEgeneralResult}. In this subsection we study the convergence properties of these expansions. The most important feature is that all of them have a finite radius of convergence. Its value, however, depends on the details. 

\begin{figure}
\begin{center}\includegraphics[width=.6\textwidth]{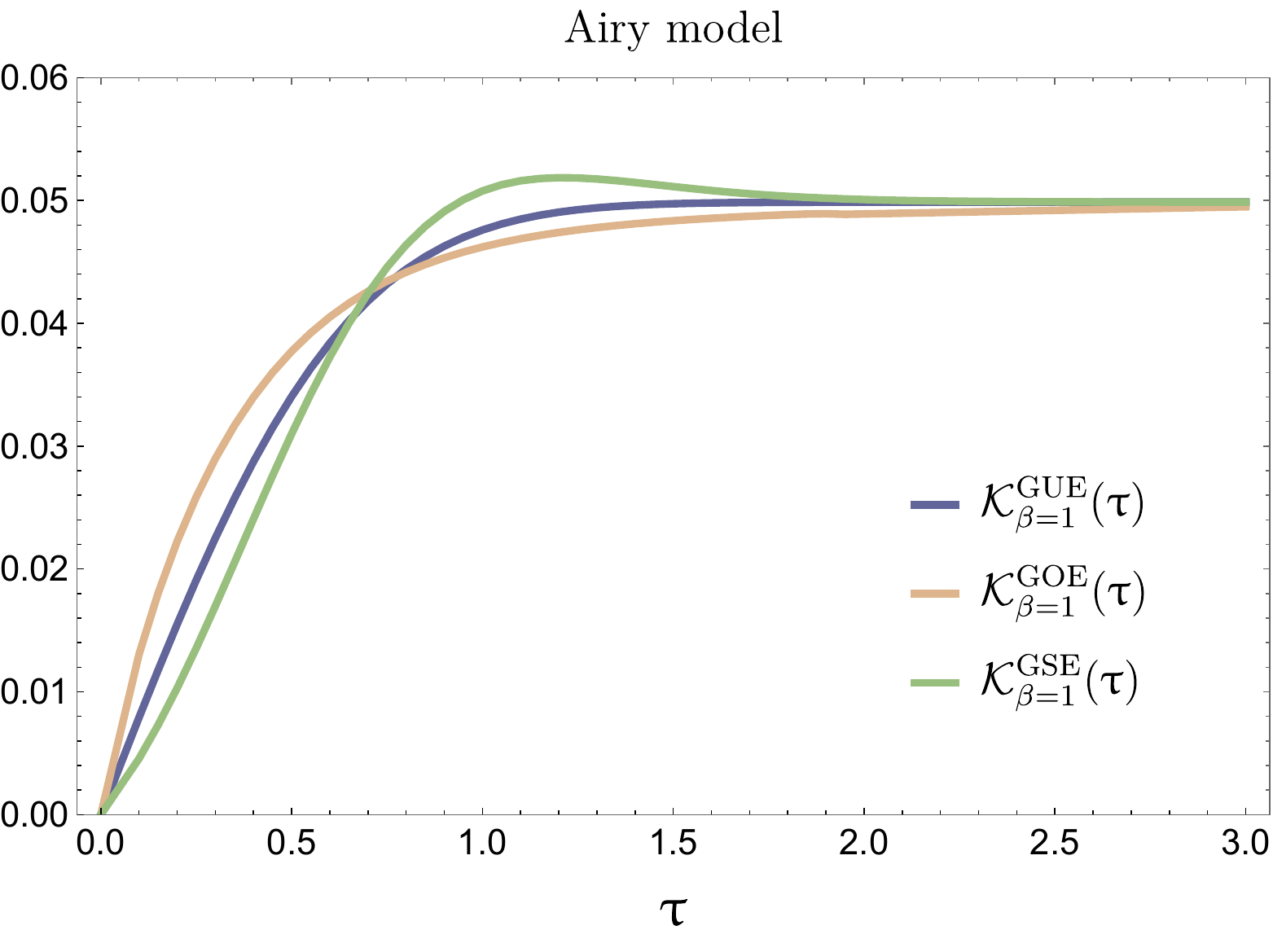}
    \end{center}
    \caption{The $\uptau$-scaled SFF ${\cal K}_\beta(\uptau)$ for the Airy model in the GUE, GOE, and GSE universality classes. In all three cases convergence to the plateau is manifest and the radius of convergence is infinite. The height of the asymptotic plateau is  $1/(8\sqrt{2\pi}\beta^{3/2})$. Curves were obtained for $\beta=1$ and using a truncated topological expansion including terms for $g=0,\ldots,100$.}
\label{fig:airyComparison}
\end{figure}

\paragraph{Example I: Airy model.} In the Airy model, the radius of convergence $\uptau_\text{max}$ of ${\cal K}_\beta(\uptau)$ is infinite in all three universality classes.\footnote{This is trivial to see for GUE (see \eqref{eq:AGUEairy}), as well as for the sums over $B_g$- and $C_\gt$-coefficients in GOE/GSE (see \eqref{eq:AresultAiry}). For the $A_g$-coefficients in GOE/GSE, the infinite radius of convergence follows from the very slow growth of the numbers appearing in \eqref{eq:AresultAiry}: $\frac{1}{g}\,{}_2F_1(2g,2g,2g+1;-1) + \psi(g+1) \sim \log(g)$ as $g\rightarrow \infty$. Another useful estimate is ${}_2F_1(2g,2g,2g+1;-1) \sim 2^{-2g+1}$ as $g\rightarrow \infty$.}
We can illustrate this numerically: in figure \ref{fig:airyComparison} we compare the different $\uptau$-scaled SFFs in the Airy model. Convergence is evident, where the transition to the plateau occurs at $\uptau \sim 1$. We can also see an interesting non-monotonicity in the GSE SFF. All three curves eventually converge to the same plateau value $\langle Z^\text{Airy}(2\beta)\rangle = 1/(8\sqrt{2\pi}\beta^{3/2})$, which provides a consistency check on our results.

\begin{figure}
\begin{center}\includegraphics[width=.6\textwidth]{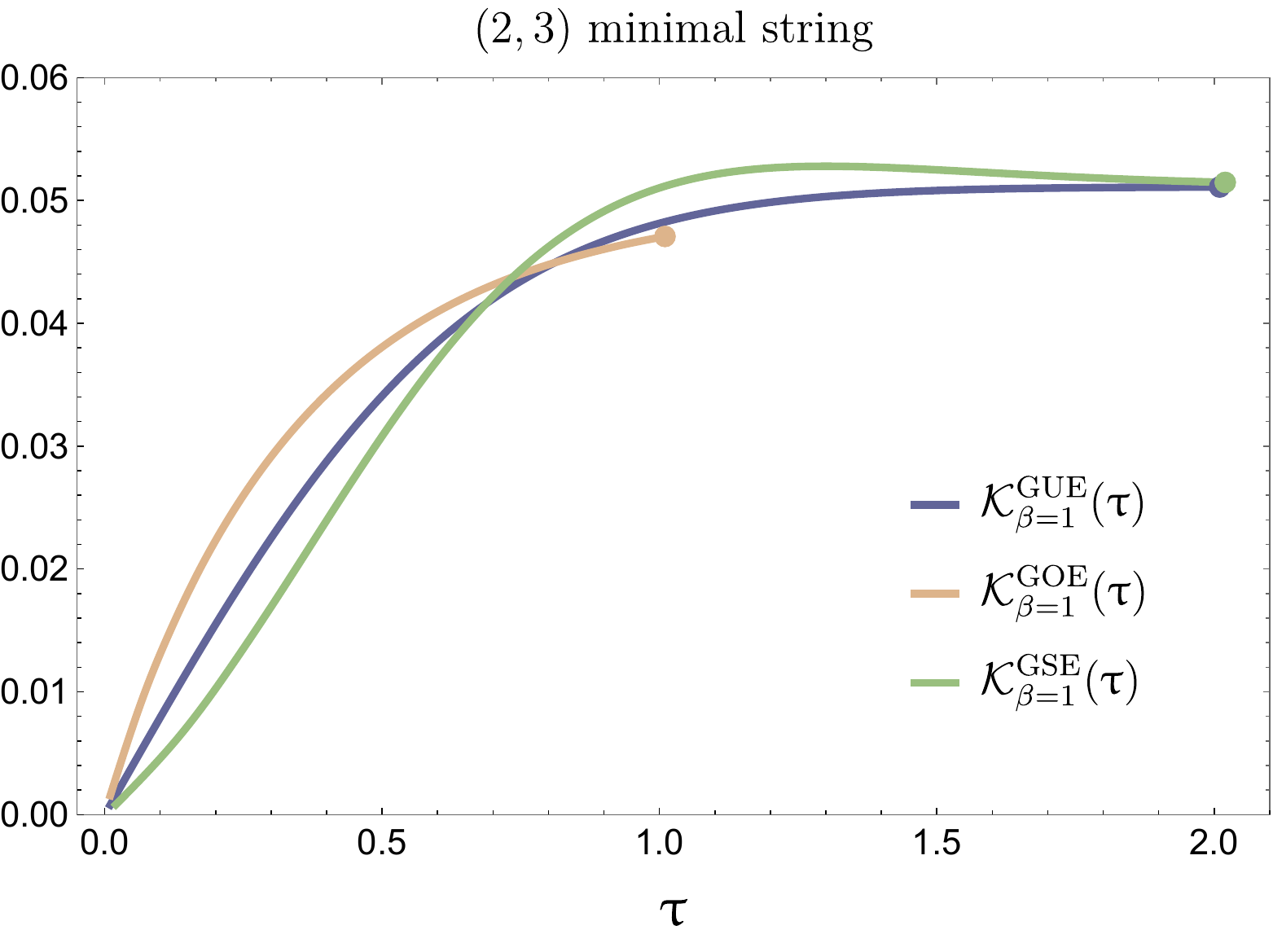}
    \end{center}
    \caption{The $\uptau$-scaled SFF ${\cal K}_\beta(\uptau)$ for the $(2,3)$ minimal string model in the GUE, GOE, and GSE universality classes. The genus expansion converges, but the radius of convergence is finite: solid dots indicate the largest value $\uptau_\text{max}$ for which the respective genus expansion converges. Curves were obtained for $\beta=1$, $\kappa=20$, and using a truncated topological expansion including terms for $g=0,\ldots,300$.}
\label{fig:21stringComparison}
\end{figure}

\paragraph{Example II: $(2,3)$ minimal string.} In the $(2,p)$ minimal string model for $p=3$ we also computed closed-form expressions for the coefficients in the topological expansion. While ${\cal K}_\beta(\uptau)$ is still given by a convergent genus expansion, the radius of convergence is no longer infinite (see also \cite{Blommaert:2022lbh} for the GUE case). For the three ensembles the radius of convergence is:\footnote{This can be confirmed by applying the ratio test to the three types of terms appearing in the topological expansion:
\begin{equation}
    (\uptau_\text{max})^2= 
    \text{min} \left\{ \lim_{g\rightarrow\infty} \left| \frac{A_g(\rho_0;\beta)}{A_{g+1}(\rho_0;\beta)}\right|
    \;,\;
    \lim_{g\rightarrow\infty} \left| \frac{B_g(\rho_0;\beta)}{B_{g+1}(\rho_0;\beta)}\right|
    \;,\;
    \lim_{\gt\rightarrow\infty} \left| \frac{C_\gt(\rho_0;\beta)}{C_{\gt+1}(\rho_0;\beta)}\right|
    \right\}\,.
\end{equation}
All three limits give the same radius of convergence for the $(2,p=3)$ minimal string, i.e., \eqref{eq:taumax21}.
}
\begin{equation}
\label{eq:taumax21}
    \uptau_\text{max}^\text{GUE} = \uptau_\text{max}^\text{GSE} = \frac{1}{3} \sqrt{2\kappa}\,,\qquad 
    \uptau_\text{max}^\text{GOE} = \frac{1}{6} \sqrt{2\kappa} \,.
\end{equation}
See figure \ref{fig:21stringComparison} for illustration. Since ${\cal K}_\beta(\uptau)$ is smooth and its series expansion has a finite radius of convergence, it can be analytically continued to $\uptau > \uptau_\text{max}$.

\paragraph{Example III: JT gravity.} In JT gravity the radius of convergence of the $\uptau$-scaled genus expansion is finite (thus allowing analytic continuation), but too small to visualize the convergence to the plateau (see also \cite{Blommaert:2022lbh}).


\newpage

\section{Topological recursion in non-orientable topological gravity}
\label{sec:toprec}

In this section, we investigate gravitational aspects of the $\uptau$-scaling limit. As described in section \ref{sec:gravitySummary}, the non-orientable case presents several novel subtleties. Our strategty will be to work in the Airy limit of JT gravity. This serves as a testing ground to explore some of the complications in non-orientable gravitational path integrals: in the Airy limit, crosscap divergences turn out to be absent, thus removing one technical issue and allowing us to focus on questions relating to the late-time limit. Due to the absence of crosscap divergences, we obtain a convergent topological recursion for the Weil-Petersson volumes of the non-orientable Airy model, $V^\text{Airy}_{g,n}$, which are manifestly finite. This recursion will be shown to be dual to the loop equations of the GOE matrix model, arising as a limit of the duality established in \cite{Stanford:2023dtm}. We present a table of explicit expressions for the non-orientable Airy Weil–Petersson volumes (table~\ref{tab:NOairyWP}).

\subsection{Review of non-orientable topological recursion}

We briefly review the calculation of  the Euclidean JT gravity path integral $Z_g\bigl(\beta_1,...,\beta_n\bigr)$ corresponding to non-orientable hyperbolic surfaces with $n$ Schwarzian boundaries and arbitrary numbers of handles $g$. This non-perturbative calculation relies on computing various Weil-Petersson (WP) volumes of moduli spaces of non-orientable bordered hyperbolic surfaces. Since we only discuss non-orientable ensembles from here onwards, we shall no longer use special labels to indicate this.

The decomposition into ``external'' trumpet factors glued to ``internal'' WP volumes was described in \eqref{eq:ZgGeneral}.
In order to compute such a path integral for given $(g,n)$, it is desirable to have a systematic way of calculating the WP volumes. This is achieved by Mirzakhani's topological recursion \cite{Mirzakhani:2006fta} and its non-orientable generalization \cite{Stanford:2023dtm}, which we review briefly.

\begin{figure}
\begin{center}\includegraphics[width=.78\textwidth]{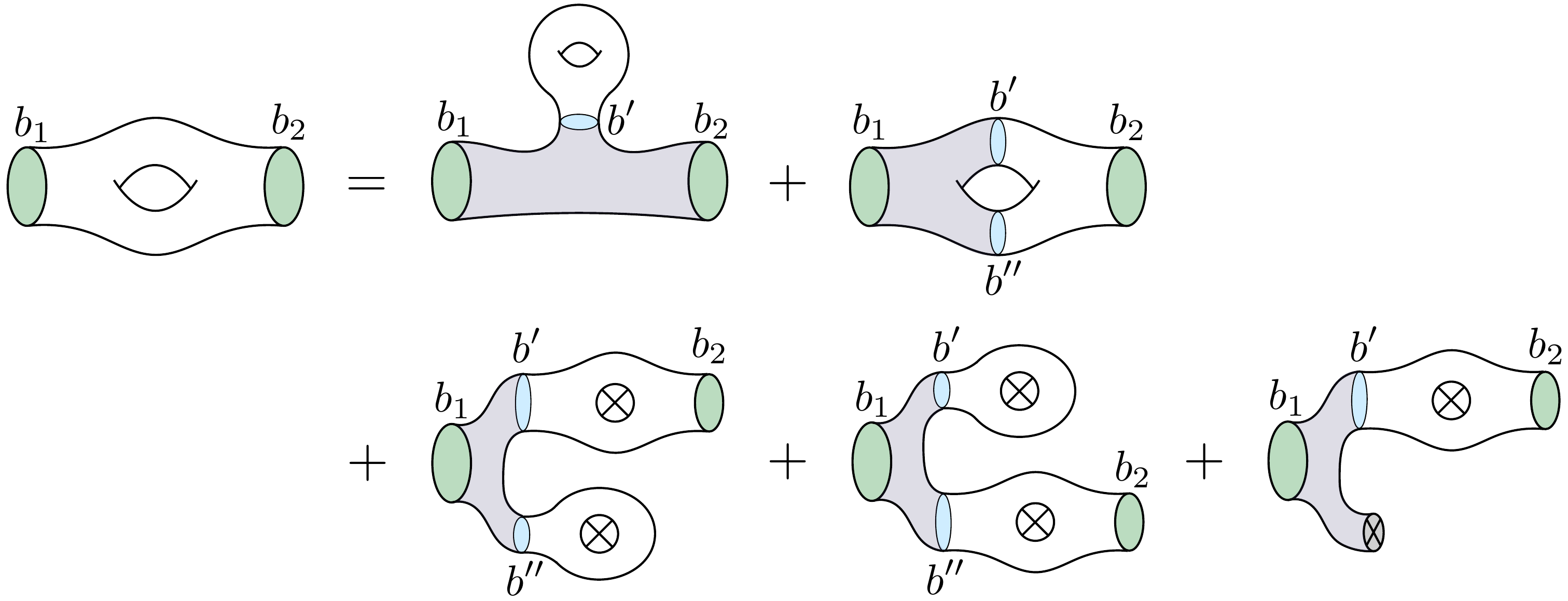}
    \end{center}
    \caption{Illustration of the topological recursion for non-orientable WP volumes. We show the different types of decompositions in the same order in which they appear in \eqref{eq:JTtoprecNonOrientable} for the case of $V_{1,2}(b_1,b_2)$. The three-holed sphere with boundary $b\equiv b_1$ is shaded in each case.}
\label{fig:genus1fates}
\end{figure}

The non-orientable extension of Mirzakhani's topological recursion \cite{Mirzakhani:2006fta} was developed in \cite{Stanford:2023dtm}. It expresses the WP volumes $V_{g,n+1}$ for the moduli space of non-orientable genus-$g$ surfaces with $n+1$ geodesic boundaries $b_1,\ldots,b_{n+1}$ in terms of the moduli space volumes for lower genus or lower number of boundaries: 
\begin{equation}
\label{eq:JTtoprecNonOrientable}
\begin{split}
   & \partial_b \Bigl[b V_{g,n+1} (b,\mathbf{b})\Bigr] = 
    \sum_{i=2}^{n+1} \int_0^{\infty} b' db' \  \Bigl( H( b',b+b_i ) + H( b',b-b_i  )  \Bigr) V_{g,n} (b',\mathbf{b} \backslash b_i)
    \\
    &\quad +
    \frac{1}{2} \int_0^{\infty} b' db' \ b'' db'' \  H( b'+b'',b ) \left( V_{g-1,n+2} (b',b'',\mathbf{b}) + \sum_{\substack{\text{stable}\\ \mathbf{b}=\mathbf{b}_1 \cup \mathbf{b}_2}} V_{g_1,|\mathbf{b}_1|+1} (b',\mathbf{b}_1) V_{g_2,|\mathbf{b}_2|+1} (b'',\mathbf{b}_2) \right)\\
    &\quad +\frac{1}{2} \int_0^{\infty} b' db' \left[ b' \, H(b',b) + 2 \log \left(1+e^{\frac{b-b'}{2}} \right) \left(1+e^{-\frac{b+b'}{2}} \right) \right]V_{g-\frac{1}{2},n+1}(b',\mathbf{b})
  \end{split}
\end{equation}
Amongst the boundaries $b_1,\ldots,b_{n+1}$, one singles out one boundary, w.l.o.g.\ $b\equiv b_1$. Then, $\mathbf{b} = \{b_2,...,b_{n+1}\}$ represents the lengths of the other fixed external boundaries. The recursion is obtained by removing from $\Sigma_{g,n+1}$ a three-holed sphere, one of whose boundaries is $b$. The different terms in \eqref{eq:JTtoprecNonOrientable} describe different possible fates for the remaining two boundaries of the three-holed sphere:
\begin{enumerate}
    \item[$(i)$] First line: the three-holed sphere is glued to the rest of the surface along one internal geodesic boundary $b'$. The third hole is an external geodesic boundary $b_i$.
    \item[$(ii)$] Second line: 
    the three-holed sphere is glued to the rest of the surface along two internal geodesic boundaries $b'$ and $b''$. The first term involving $V_{g-1,n+2}$ descibes the case where the rest of the surface is {\it connected}. The sum over ``stable'' decompositions accounts for the case where the rest of the surface is {\it disconnected}. In the latter case, the two internal gluings attach the three-holed sphere to two disconnected surfaces that obey $g_1+g_2 = g$ and $\mathbf{b}_1 \cup \mathbf{b}_2 = \mathbf{b}$. The sum over such stable decompositions excludes  cases where one component would involve the once- or twice-punctures spheres, $V_{0,1}$ or $V_{0,2}$.
    \item[$(iii)$] Third line: the three-holed sphere is glued to the rest of the surface along one internal geodesic boundary $b'$. The third hole is closed off with a crosscap.
\end{enumerate}
The integration kernels encode the details of the Riemannian geometries (i.e., the information about JT gravity). They are found to be:
\begin{equation}
    H(x,y) = \frac{1}{1+e^{\frac{x+y}{2}}} + \frac{1}{1+e^{\frac{x-y}{2}}}\,.
\end{equation}
We give an illustrative example in figure \ref{fig:genus1fates}. For more detailed explanations and derivations, we refer the reader to the literature \cite{Mirzakhani:2006fta,Dijkgraaf:2018vnm,Stanford:2023dtm}.

Note that the topological recursion for orientable geometries takes the same form as \eqref{eq:JTtoprecNonOrientable} with two small differences: the third line is absent in the orientable case, and the first line is multiplied by a factor $\frac{1}{2}$ (the inclusion of crosscaps leads to pairs of equivalent geometries with a single internal gluing surface). 

The initial conditions for the non-orientable recursion are given as follows:
\begin{equation}
    V_{0,2}(b_i,b_j) = \frac{2}{b_i} \,\delta ( b_i-b_j)\,, \ \ \ \ \ \ \  V_{0,3}(b_i,b_j,b_k) = 4\,, \ \ \ \ \ \ \  V_{\frac{1}{2},1} (b) = \frac{1}{2b  \tanh{\bigl(\frac{b}{4}\bigr)}}\,.
\end{equation}

The topological recursion for non-orientable WP volumes becomes ill-defined when crosscaps shrink to zero. A simple cure is to impose a minimum length $\epsilon$ on all one-sided geodesics, computing volumes on this ``$\epsilon$-regularized'' moduli space. For small $\epsilon$, the recursion still holds with only its initial conditions tweaked. In fact, these regularized volumes have been shown to coincide with those of the $(2,p)$ minimal string model, with $p=\epsilon^{-1}$ \cite{Stanford:2023dtm, Tall:2024hgo}. Equivalently, one can show that the non-orientable recursion reproduces the loop equations of a double-scaled orthogonal matrix integral whose spectral curve behaves like $y(z)\sim\sin(2\pi z)$. In contrast, Mirzakhani's original orientable recursion matches the unitary version of the same integral \cite{Eynard:2007kz, Eynard:2007fi}, which underpins its formal duality with JT gravity.

Here we adopt an alternative way to eliminate divergences by taking the Airy limit, i.e., sending the geodesic boundary lengths to infinity.\footnote{The Airy limit can be thought of as scaling towards the edge of the spectral curve. Since non-zero density acts as an order parameter for causal symmetry breaking, scaling towards the edge is akin to approaching a quantum critical point where the order parameter vanishes. For details see \cite{Altland:2024ubs,Altland:2025diq}.}

\subsection{Topological recursion in the Airy limit}

We will now derive topological recursion relations for the Airy model. The strategy is to employ a certain long-boundary limit of the JT gravity theory. We first recall this intuition in the orientable case, before generalizing to non-orientable WP volumes.

It was argued in \cite{Saad:2022kfe} that the WP volumes of the Airy (topological) model can be obtained from JT gravity in the limit of long geodesic boundary lengths. Let us briefly recall the derivation of this argument. Consider the JT gravity path integral with $n$ boundaries at inverse temperatures $\beta_1,\ldots,\beta_n$:
\begin{equation}
\begin{split}
  Z^{\text{JT}}(\beta_1,...,\beta_n) = \sum_{g\geq 0} e^{\chi S_0} \, Z_{g,n}^\text{JT} (\beta_1,...,\beta_n) \,, \qquad \chi = 2-2g-n \,.
  \end{split}
\end{equation}
It is known that the Airy limit of the $n$-boundary path integral is obtained from a homogeneous scaling of the boundary lengths and an infinite shift (renormalization) of the extremal entropy $S_0$ \cite{Saad:2022kfe}: 
\begin{equation}
\label{eq:airyfullscaling}
Z^{\text{Airy}}_{g,n}(\beta_1,...,\beta_n) \equiv \lim_{\Lambda \rightarrow \infty} \, e^{\chi(S_0+\frac{3}{2} \log \Lambda)} \ Z^{\text{JT}}_{g,n}(\Lambda\beta_1,...,\Lambda\beta_n) \,.
\end{equation}
This equation can be translated into a scaling limit for the WP volumes by writing both sides using the gluing prescription with trumpet wave functions:
\begin{equation}
 Z_{g,n}^\text{JT/Airy}(\beta_1,...,\beta_n) =  \left(\prod_{i=1}^n \int_{0}^{\infty} b_i db_i \ \frac{e^{-\frac{b_i^2}{4 \beta_i}}}{\sqrt{4 \pi \beta_i}} \right) V_{g,n}^\text{JT/Airy}(b_1,...,b_n)\,.
 \end{equation}
The large boundary limit of the full gravitational path integral \eqref{eq:airyfullscaling} thus translates directly into the scaling limit for WP volumes:
\begin{equation}
\label{eq:airyvolume}
    V^{\text{Airy}}_{g,n}(b_1,...,b_n) \equiv \lim_{\Lambda \rightarrow \infty} \Lambda^{3-3g-n} \ V_{g,n}^{\text{JT}}(\sqrt{\Lambda}\,b_1,...,\sqrt{\Lambda}\,b_n) \,.
\end{equation}
It is not obvious that \eqref{eq:airyvolume} should also apply to non-orientable WP volumes.
We will show that the analogous statement holds in the non-orientable case, where the long boundary limit of the WP volumes for non-orientable JT   is given by the  WP volumes for non-orientable Airy. We will take the limit directly at the level of the topological recursion and perform a number of consistency checks.

We can apply the scaling limit of WP volumes \eqref{eq:airyvolume} to Mirzakhani's recursion formula. The only input required for this is the large-boundary limit of the integration kernel $H(x,y)$:
\begin{equation}
 \lim_{\Lambda \rightarrow \infty} H \left( \sqrt{\Lambda} \, x, \sqrt{\Lambda} \, y\right) = 1+\theta(y-x) -\theta(y+x) \,,
\end{equation}
and similarly for the logarithm in the last line of \eqref{eq:JTtoprecNonOrientable}.

Applying the homogeneous scaling \eqref{eq:airyvolume} to the non-orientable topological recursion \eqref{eq:AiryTopRecNonOrientable}, we find the following recursion for the Airy model:
\begin{equation}
\label{eq:AiryTopRecNonOrientable}
\begin{split}
   &  \partial_b \Bigl[b\, V^{\text{Airy}}_{g,n+1} (b,\mathbf{b})\Bigr] = 
    \sum_{i=2}^{n+1}\left(\int_0^{b+b_i} b'db'  + \int_0^{|b-b_i|} b'db'  \right) V^{\text{Airy}}_{g,n} (b',\mathbf{b} \backslash b_i)
    \\
    &\qquad +
    \frac{1}{2} \int_0^{b} b'db' \int_{0}^{b-b'}b'' db''  \left( V^{\text{Airy}}_{g-1,n+2} (b',b'',\mathbf{b}) + \sum_{\substack{\text{stable}\\ \mathbf{b}=\mathbf{b}_1 \cup \mathbf{b}_2}} V^{\text{Airy}}_{g_1,|\mathbf{b}_1|+1} (b',\mathbf{b}_1) V^{\text{Airy}}_{g_2,|\mathbf{b}_2|+1} (b'',\mathbf{b}_2) \right)\\
    &\qquad +\frac{b}{2} \int_0^{b} b' db' \; V^{\text{Airy}}_{g-\frac{1}{2},n+1}\bigl(b',\mathbf{b}\bigr) \,.
  \end{split}
\end{equation}
\begin{figure}
\begin{center}\includegraphics[width=\textwidth]{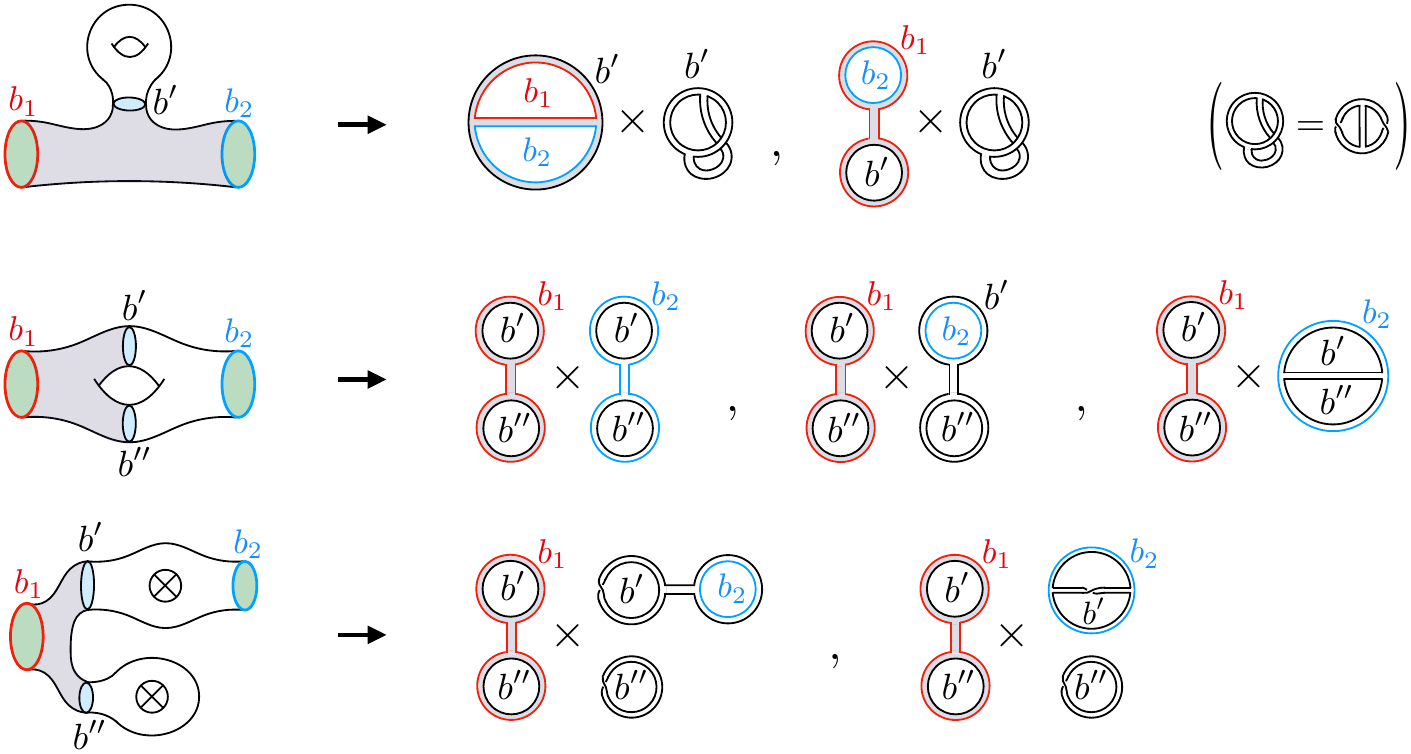}
    \end{center}
    \caption{Different terms in the topological recursion in the Airy model \eqref{eq:AiryTopRecNew} can be interpreted as ribbon graphs. We illustrate how the $g=1$ Riemann surfaces from figure \ref{fig:genus1fates} degenerate into inequivalent ribbon graphs in the Airy (long boundary) limit. Internal  boundaries $b'$ and $b''$ are drawn in black and should be glued together across the ``$\times$''. Cross caps degenerate into ``twists'', as shown in the third line. Handles can equivalently be obtained from planar graphs with twists (bracket in the first line). We do not show graphs obtained by exchanging $b_1 \leftrightarrow b_2$.}
\label{fig:genus1ribbon}
\end{figure}
Equivalently, the recursion can be written without the derivative:
{\small
\begin{equation}
\label{eq:AiryTopRecNew}
\boxed{
\begin{aligned}
   & b\, V^{\text{Airy}}_{g,n+1} (b,\mathbf{b}) \\
   &\;\;\; = 
    \sum_{i=2}^{n+1}\left(
     \int_0^{b+b_i} b'db' 
     (b+b_i-b') + \text{sgn}(b-b_i) \int_0^{|b-b_i|} b'db' (|b-b_i|-b')
     \right)
     V^{\text{Airy}}_{g,n} (b',\mathbf{b} \backslash b_i)
    \\
    &\;\;\; +
    \int_0^{b} b'db' \int_{0}^{b-b'}b'' db''\, \frac{b-b'-b''}{2}  \left( V^{\text{Airy}}_{g-1,n+2} (b',b'',\mathbf{b}) + \sum_{\substack{\text{stable}\\ \mathbf{b}=\mathbf{b}_1 \cup \mathbf{b}_2} }V^{\text{Airy}}_{g_1,|\mathbf{b}_1|+1} (b',\mathbf{b}_1) V^{\text{Airy}}_{g_2,|\mathbf{b}_2|+1} (b'',\mathbf{b}_2) \right)\!\!\!\\
    &\;\;\; + \int_0^{b} b' db' \,\frac{b^2-b'^2}{4}\,V^{\text{Airy}}_{g-\frac{1}{2},n+1}\bigl(b',\mathbf{b}\bigr)
  \end{aligned}
  }
\end{equation}
}\normalsize
This version of the non-orientable topological recursion is free of any crosscap divergences and therefore is an exact expression. With the initial conditions
\begin{equation}
     V^{\text{Airy}}_{0,2}(b_i,b_j) = \frac{2}{b_i}\, \delta(b_i-b_j)\,, \qquad V^{\text{Airy}}_{0,3}(b_i,b_j,b_k) = 4\,, \qquad V^{\text{Airy}}_{\frac{1}{2},1}(b) = \frac{1}{2b}\,,
\end{equation}
we can calculate non-orientable Airy volumes in a systematic way.\footnote{There is again an orientable version of the formula \eqref{eq:AiryTopRecNew}: one simply has to multiply the first line by $\frac{1}{2}$ and drop all terms involving half-integer genus geometries (in particular the last line). In the orientable case the initial conditions are: 
\begin{equation}
\label{eq:initialCondJT}
    V^{\text{GUE,Airy}}_{0,2}(b_i,b_j) = \frac{1}{b_i}\, \delta ( b_i-b_j)\,, \ \ \ \ \ \ \ V^{\text{GUE,Airy}}_{0,3}(b_i,b_j,b_k) = 1\,.
\end{equation}} 
Each term in the topological recursion can also be identified with a ribbon graph. This leads to a systematic geometrical enumeration of all terms appearing in the topological recursion. Some examples for $(g,n)=(1,2)$ are given in figure \ref{fig:genus1ribbon}.

\begin{table}
\begin{center}
\begin{tabular}{|l | l|} 
 \hline  
 $n=1$ & $V_{1,1}^{\text{Airy}} (b) = \frac{7\,b^2}{48} $
   \rule{0pt}{3ex} \\  [1ex]
   &  $V_{\frac{3}{2},1}^{\text{Airy}} (b) = \frac{b^5}{180}$ \\ [1ex]
   &
   $V_{2,1}^{\text{Airy}} (b) = \frac{37\,b^8}{442368}$ \\ [1ex]
   & 
   $V_{\frac{5}{2},1} ^{\text{Airy}} (b) =  \frac{b^{11}}{1496880}$ \\ [1ex]
   & 
   $V_{3,1} ^{\text{Airy}} (b) =  \frac{887\, b^{14}}{267544166400}$ \\ [1ex]
\hline
 $n=2$ & $V_{\frac{1}{2},2}^{\text{Airy}} (b_1,b_2)  = \theta(b_1-b_2) b_1 + \text{perm.} \equiv \text{max}(b_1,b_2)$ \rule{0pt}{3ex} \\ [1ex]
   &  $V_{1,2}^{\text{Airy}} (b_1,b_2) = \left(\frac{b_1^4}{32} + \frac{7 b_1^2 b_2^2}{96} \right) + \theta (b_1-b_2)\,\frac{ b_1^4 + 2 b_1 b_2^3}{24}   + \text{perm.} $
   \rule{0pt}{3ex} \\ [1ex]
   &  $V_{\frac{3}{2},2}^{\text{Airy}} (b_1,b_2) =
   \left( \frac{23b_1^7}{40320} + \frac{7 b_1^5 b_2^2}{1920} + \frac{7b_1^4 b_2^3}{1152}\right)$
   \rule{0pt}{3ex} \\ [1ex]
   &  $ \qquad\qquad\qquad\quad\;\; + \theta(b_1-b_2) \left(\frac{41b_1^7}{40320} + \frac{43 b_1^5 b_2^2}{5760} + \frac{b_1^3 b_2^4}{128} + \frac{5b_1 b_2^6}{1152}  \right) + \text{perm.}$ \rule{0pt}{3ex} \\ [1ex]
   &  $V_{2,2}^{\text{Airy}} (b_1,b_2) =
   \left(\frac{377 b_1^{10}}{46448640} + \frac{863 b_1^8 b_2^2}{5160960} + \frac{b_1^7 b_2^3}{7560} +\frac{97 b_1^6b_2^4}{122880} + \frac{b_1^5b_2^5}{7200} \right)$
   \rule{0pt}{3ex} \\ [1ex]
   & $\qquad\qquad\qquad\quad\;\; + \theta(b_1-b_2) \left(\frac{5b_1^{10}}{580608}+ \frac{3b_1^8b_2^2}{35840} + \frac{b_1^6 b_2^4}{69120} + \frac{b_1^3 b_2^7}{15120} + \frac{b_1 b_2^9}{22680} \right) + \text{perm.}
   $
   \rule{0pt}{3ex} \\ [1ex]
   &  $V_{\frac{5}{2},2}^{\text{Airy}} (b_1,b_2) = \frac{1}{45\cdot 2^{15}}\left( \frac{907b_1^{13}}{13860} + \frac{32743b_1^{11}b_2^2}{20790} + \frac{37b_1^{10}b_2^3}{18} + \frac{377b_1^9b_2^4}{36}+\frac{37 b_1^8b_2^5}{4} + \frac{143 b_1^7 b_2^6}{9} 
   \right)$
   \rule{0pt}{3ex} \\ [1ex]
   & $\qquad\qquad\qquad\quad\;\; + \theta(b_1-b_2)\, \frac{1}{225\cdot 2^{15}} \left(
   \frac{46547b_1^{13}}{108108}+\frac{49177b_1^{11}b_2^2}{4158} + \frac{50527b_1^9b_2^4}{756} + \frac{10483b_1^7b_2^6}{63} \right.$
   \rule{0pt}{3ex} \\ [1ex]
   & $\qquad\qquad\qquad\quad\;\; \left.
   + \frac{4351b_1^5b_2^8}{36} + \frac{551b_1^3b_2^{10}}{18} + \frac{13b_1b_2^{12}}{4}
   \right) + \text{perm.}
   $
   \rule{0pt}{3ex}
\\ [1ex] 
 \hline
$n=3$ & $ V_{0,3}^\text{Airy}(b_1,b_2,b_3) = 4 $ \rule{0pt}{3ex} \\ [1ex]
   & $ V_{\frac{1}{2},3}^\text{Airy}(b_1,b_2,b_3) = \left(\frac{b_1^3}{6}+ \frac{b_1b_2b_3}{3}\right) + \theta(b_1+b_2-b_3) \, \frac{(b_1+b_2-b_3)^3}{12} + \text{perm.} $ \rule{0pt}{3ex} \\ [1ex]
   &  $V_{1,3}^\text{Airy}(b_1,b_2,b_3) =
\left(\frac{3b_1^6}{320}-\frac{b_1^5 b_2}{80}+ \frac{5b_1^4 b_2^2}{48}- \frac{3 b_1^3 b_2^3}{72}+\frac{b_1^4 b_2 b_3}{16} + \frac{b_1^3 b_2^2 b_3}{24}  +\frac{7b_1^2 b_2^2 b_3^2}{144}  \right)  $ \rule{0pt}{2.6ex} \\ [1ex]
   & $\qquad\qquad\qquad\qquad\;\; + \theta(b_1-b_2) \,\frac{(b_1-b_2)^5b_3}{240} + \theta(b_1+b_2-b_3) \, \frac{(b_1+b_2-b_3)^5(b_1+b_2+2b_3)}{720} + \text{perm.}$ \rule{0pt}{2.6ex} \\ [1ex]
\hline
 $n=4$ & 
 $V_{0,4}^\text{Airy}(b_1,...,b_4) = \frac{2 b_1^2}{3} +\text{perm.}$ \rule{0pt}{3ex} \\ [1ex]
 & $V_{\frac{1}{2},4}^\text{Airy}(b_1,...,b_4) = \left( \frac{17 b_1^5}{720} + \frac{b_1^4 b_2}{48} + \frac{5b_1^3 b_2^2}{24} + \frac{b_1^3b_2b_3}{4}- \frac{b_1^2b_2^2 b_3}{8} + \frac{b_1^2b_2b_3b_4}{4} \right)+ \theta(b_1-b_2) \, \frac{(b_1-b_2)^5}{120}
 $
 \rule{0pt}{3ex} \\ [1ex]
 & $\qquad\qquad\qquad\qquad\;\; + \theta(b_1+b_2-b_3-b_4) \, \frac{(b_1+b_2-b_3-b_4)^5}{480} $
 \rule{0pt}{3ex} \\ [1ex]
 & $\qquad\qquad\qquad\qquad\;\; + \theta(b_1+b_2+b_3-b_4) \, \frac{(b_1+b_2+b_3-b_4)^5}{360}   + \text{perm.}$
 \rule{0pt}{3ex} \\ [1ex]
 \hline
 $n=5$ & 
 $V_{0,5}^\text{Airy}(b_1,...,b_5) = \left(\frac{b_1^4}{12} + \frac{2b_1^2 b_2^2}{3} \right)+ \text{perm.}
 $\rule{0pt}{3ex} \\ [1ex]
 \hline
 $n=6$ & 
 $V_{0,6}^\text{Airy}(b_1,...,b_6) = \left(\frac{b_1^6}{180} + \frac{b_1^4 b_2^2}{4}  + \frac{2 b_1^2 b_2^2 b_3^2}{3} \right)+ \text{perm.}
 $\rule{0pt}{3ex} \\ [1ex]
 \hline
\end{tabular}
\end{center}
\caption{Weil-Petersson volumes $V_{g,n}$ in the non-orientable Airy model. The notation ``$+\text{perm.}$'' means a sum of the given expression over all $n!$ permutations of the boundary lengths $b_i$ ($i=1,\ldots,n$).}
\label{tab:NOairyWP}
\end{table}

Applying the topological recursion \eqref{eq:AiryTopRecNonOrientable} repeatedly, we find the Weil-Petersson volumes for the moduli spaces of non-orientable geometries displayed in table \ref{tab:NOairyWP}. The WP volumes for $n=2$ boundaries are consistent with those obtained in \cite{Weber:2024ieq} using a different method.

In these examples, it is interesting to note how non-analyticities appear: the Weil-Petersson volumes have the form of a symmetric polynomial (as in the orientable case) plus non-symmetric polynomials multiplying step-functions that impose for the sum of some subset of boundary lengths to be larger than the sum of another subset. As the number of boundaries $n$ grows, an increasing number of polynomials is required to characterize the WP volumes. For example:
\begin{equation}
    \begin{split}
    V^\text{Airy}_{g,1} (b) &= P^{(1)}_{g,2}(b)  \,,\\
    V^\text{Airy}_{g,2} (b_1,b_2) &= P^{(1)}_{g,2}(b_1,b_2) + \theta(b_1-b_2) P^{(2)}_{g,2}(b_1,b_2)+ \text{perm.}  \,,\\
    V^\text{Airy}_{g,3} (b_1,b_2,b_3) &= P^{(1)}_{g,3}(b_1,b_2,b_3) + \theta(b_1-b_2) P^{(2)}_{g,3}(b_1,b_2,b_3) + \theta(b_1+b_2-b_3) P^{(3)}_{g,3}(b_1,b_2,b_3) + \text{perm.} 
    \end{split}
\end{equation}
where $P_{g,n}^{(i)}(\{b_i\})$ are polynomials of degree $(6g+2n-6)$, and ``$+\text{perm}$'' indicates a sum over all $n!$ permutations of $b_1,\ldots,b_n$.
The first term $P_{g,n}^{(1)}(\{b_i\})$ thus gives a symmetric polynomial contribution, while the non-analytic terms also get symmetrized. The volumes must, of course, be permutation symmetric and we have chosen to present them in a way that makes this property manifest.

\subsection{Loop equations in the  Airy GOE matrix model}
\label{sec:loopEqs}

The topological recursion for non-orientable surfaces in the Airy limit can be mapped to the loop equations of the GOE Airy matrix model. The derivation is analogous to the one for non-orientable JT gravity in \cite{Stanford:2023dtm}, which we refer to for more details. See also \cite{Weber:2024ieq} for a discussion of loop equations in the GOE Airy model.

The basic quantity of interest in the loop equations is the resolvent (Laplace transform of the WP volume):
\begin{equation}
\label{eq:resolventDef}
    R_{g,n}(-z_1^2,\dots,-z_n^2)\equiv (-1)^n \int V_{g,n}(b_1,\dots,b_n) \prod_{j=1}^n b_jdb_j \frac{e^{-b_jz_j}}{2z_j}\,.
\end{equation}
The loop equations for a GOE matrix model with spectral curve $y(-z^2)$ are then given by: 
\begin{equation}
    R_{g,n+1}(-z^2,{\bf X})=\frac{1}{2\pi i z} \int_{\delta+i\mathbb{R}}\frac{z'^2dz'}{z'^2-z^2} \frac{F_{g,n+1}(-z'^2,{\bf X})}{y(-z^2)}\,,
\end{equation}
where ${\bf X} = \{-z_2^2 ,\ldots,-z_{n+1}^2\}$ and the integrand satisfies the recursion
\begin{align}
    F_{g,n+1}(x,{\bf X})&= \sum_{i=2}^{n+1}\qty(\frac{1}{2\sqrt{-x}\sqrt{-x_i}(\sqrt{-x}+\sqrt{-x_i})^2}+\frac{1}{(x-x_i)^2})R_{g,n}(x,{\bf X}\backslash x_i)
   \nonumber
    \\&\quad 
   +R_{g-1,n+2}(x,x,{\bf X})+\sum_{\substack{\text{stable}\\ {\bf X}={\bf X}_1 \cup {\bf X}_2}}R_{h_1,|{\bf X}_1|+1}(x,{\bf X}_1)R_{h_2,|{\bf X}_2|+1}(x,{\bf X}_2)
     \nonumber\\&\quad
     -\partial_x R_{g-\frac{1}{2},n+1}(x,{\bf X})\,.
     \label{eq:loopRec}
\end{align}
Starting from the Airy topological recursion in \eqref{eq:AiryTopRecNew}, the first two lines are identical to the orientable case and are known to match with the Airy GUE loop equations \cite{Eynard:2007fi}. We are left with the task of matching the last line in \eqref{eq:AiryTopRecNew},
\begin{equation}
b\, V^{\text{Airy}}_{g,n+1} (b,\mathbf{b}) \supset \int_0^{b} b' db' \,\frac{b^2-b'^2}{4}\,V^{\text{Airy}}_{g-\frac{1}{2},n+1}\bigl(b',\mathbf{b}\bigr)\,,
\end{equation}
to the only remaining term in the GOE loop equations, namely the last term in \eqref{eq:loopRec}: 
\begin{equation}
R_{g,n+1}(-z^2,{\bf X})\supset \frac{1}{2\pi i z}\int_{\delta+i\mathbb{R}} \frac{{z'}^2dz'}{{z'}^2-z^2}\frac{1}{y(-{z'}^2)}\frac{1}{2z'}\, \partial_{z'}R_{g-\frac{1}{2},n+1}(-{z'}^2,{\bf X})\,.
\end{equation}
To show they match, we substitute $R_{g,n+1}$ and $R_{g-\frac{1}{2},n+1}$ in terms of WP volumes, using the relationship \eqref{eq:resolventDef}.
After inverting the Laplace transform of $V_{g,n+1}(b,\mathbf{b})$ \cite{Stanford:2023dtm}, we obtain: 
\begin{equation}
    \, V^{\text{Airy}}_{g,n+1} (b,\mathbf{b}) \supset \int_{0}^{\infty}b'db' \;V^{\text{Airy}}_{g-\frac{1}{2},n+1}(b',\mathbf{b}) \int_{\delta+i\mathbb{R}}\frac{dz}{i}\frac{\sinh(bz)}{y(-{z}^2)}\qty(-\partial_{z} \frac{e^{-b'z}}{z})\,.
\end{equation}
For this to match the corresponding term in the Airy topological recursion, the $z$ integral should be equal to: 
\begin{equation}
    \int_{\delta+i\mathbb{R}}\frac{dz}{i}\frac{\sinh(bz)}{y(-{z}^2)}\qty(-\partial_{z} \frac{e^{-b'z}}{z}) = \theta(b-b') \,\frac{b^2-b'^2}{4}\,,
\end{equation}
where the Airy matrix model spectral curve is $y(-{z}^2)=2\pi z$. 
This integral identity can be derived by using the residue theorem for the pole at $z=0$.  In particular, the following result is useful: 
\begin{equation}
\frac{1}{2\pi i}\int_{\delta-i\infty}^{\delta+i\infty}\frac{e^{-bz}}{z}\,dz
=\,\theta (- b\big)\,,
\end{equation}
where the $\theta(-b)$ follows from having to close the contour to the right if $b>0$, without enclosing any poles, and to the left if $b<0$, enclosing the $z=0$ pole. 
The identity establishes the match between the Airy limit of the non-orientable topological recursion and the loop equations of the GOE Airy matrix model. Note that in the loop equations we have not performed any scaling limits. Instead we have simply used the Airy curve $y(-{z}^2)=2\pi z$ and obtained the same result.

\section{Cancellations in WP volumes of non-orientable topological gravity}
\label{sec:cancellations}

Having observed the finiteness of WP volumes in the Airy model, we will now use this toy example to explore and resolve the complexities involved in the $\uptau$-scaling limit for non-orientable models of gravity. Recall the two-boundary spectral form factor in the analytically continued configuration $\beta_{1,2}=\beta \pm iT$:
\begin{equation}
\label{eq:Zg2Airy}
    Z_{g,2}^\text{Airy}(T,\beta) \equiv \frac{1}{4\pi \sqrt{T^2+\beta^2}} \int_0^\infty b_1 db_1 \, b_2 db_2 \; e^{-\frac{1}{4(T^2+\beta^2)} \left[ \beta(b_1^2+b_2^2) - iT (b_1^2-b_2^2) \right]} \, V_{g,2}^\text{Airy}(b_1,b_2) \,.
\end{equation}
For the orientable case, this produces the well-known result \eqref{eq:KGUEAiry} genus by genus.
The non-orientable case is significantly more complicated. The first few terms in the genus expansion in the non-orientable case are of the following form:
\begin{equation}
    \begin{split}
    Z^\text{Airy}(\uptau,\beta)  &\equiv \lim_{\substack{T,e^{S_0}\rightarrow\infty \\ \uptau \text{ fixed}}} \;\;
    \sum_{g=0,\frac{1}{2},1,\ldots} e^{-(2g+1)S_0} \, Z_{g,2}^\text{Airy}(T=\uptau \, e^{S_0},\beta)\\
    & = \lim_{\substack{T,e^{S_0}\rightarrow\infty \\ \uptau \text{ fixed}}} \bigg\{\left({\color{darkblue}\frac{1}{2\pi \beta}} + \ldots \right) \uptau  +  \left({\color{darkblue} - \frac{1}{\sqrt{2\pi \beta}}} + \ldots \right)\uptau^2  - \frac{1}{\pi}\left( \log \left( \frac{{\color{darkblue}\beta}}{2T} \right) +\frac{9+{\color{darkblue}1}}{3}   + \ldots\right) \uptau^3
    \\
    &\qquad\qquad\;\,  + \bigg(\!- \frac{\sqrt{2\pi}}{3\pi}\, T^{\frac{1}{2}} {\color{darkblue} \,+\frac{8\sqrt{2\pi\beta}}{3\pi} } + \ldots\bigg) \,\uptau^4 + \frac{4\beta}{\pi} \left(\log \left( \frac{{\color{darkblue}\beta}}{2 T} \right) + \frac{170{\color{darkblue}-7}}{60} + \ldots\right) \uptau^5  
    \\
    &\qquad\qquad\;\, +  \bigg( \frac{1}{15\sqrt{2\pi}} \, T^{\frac{3}{2}}+\frac{17 \beta}{6\sqrt{2\pi}} \, T^{\frac{1}{2}} {\color{darkblue}- \frac{64 (2\pi \beta)^{\frac{3}{2}}}{15\pi^2}} + \ldots\bigg) \uptau^6
    \\
    &\qquad\qquad\;\,
    -\frac{32\beta^2}{3\pi} \left( \frac{T^2}{240\beta^2} + \, \log\left(\frac{{\color{darkblue}\beta}}{2T}\right) 
+\frac{9800{\color{darkblue}-1503}}{3360} 
    + \ldots \right)\uptau^7
    + \ldots  \bigg\}
    \end{split}
    \label{eq:AiryGOEgravity}
\end{equation}
where, in each bracket, we drop terms that vanish as $T\rightarrow \infty$ with $\uptau$ held fixed. See also \cite{Weber:2024ieq}, where this expression was first discussed. For consistency, this expression should match the universal RMT result \eqref{eq:AiryGOESFF}:
\begin{equation}
    Z^\text{Airy}(\uptau,\beta)  \stackrel{?}{=} {\cal K}^\text{Airy}_\beta(\uptau) \,.
\end{equation}
Evidently, for such an equality to be true, a resummation of all terms with explicit $T$ dependence must occur to make them subleading. We highlighted in  ${\color{darkblue}\text{blue}}$ the terms that match corresponding terms in \eqref{eq:AiryGOESFF}. Crucially, the remaining terms in \eqref{eq:AiryGOEgravity} are divergent term by term as $T\rightarrow \infty$. A non-trivial resummation is required to obtain a complete match with the RMT result \eqref{eq:AiryGOESFF}.

Instead of attempting to explicitly resum the divergent expression,\footnote{See also \cite{Weber:2024ieq}, where similar observations were made and mathematical identities were given that may play a role in an explicit resummation.} we will offer a complementary perspective: we will give evidence for the finiteness of the $\uptau$-scaling limit of the gravitational SFF by working in microcanonical variables.
We will also put focus on the following, more hidden feature of \eqref{eq:AiryGOEgravity}: the naive degree of divergence of a genus $g$ term is actually much worse than \eqref{eq:AiryGOEgravity} shows. Many naively divergent terms are absent due to cancellations amongst the coefficients in the WP volumes $V^\text{Airy}_{g,2}$. We will discuss different perspectives on these cancellations and conjecture their general form.

\subsection{The microcanonical ramp from gravity}
\label{sec:microcanonical}

The divergences and required resummation that complicate the result of the canonical gravitational path integral \eqref{eq:AiryGOEgravity} can be avoided by working at fixed energy. This is achieved through an inverse Laplace transform of \eqref{eq:Zg2Airy}:
\begin{equation}
\begin{split}
    Z_{g,2}^\text{Airy}(T,E) 
    &\equiv \int \frac{d\beta}{\pi i} \; e^{2\beta E}  \, Z_{g,2}^\text{Airy}(T,\beta) \\
    &\approx \frac{2}{\pi T} \int_0^\infty b_1db_1 \, b_2db_2 \; e^{ \frac{i}{4T}(b_1^2 - b_2^2)} \, \delta\left( \frac{b_1^2+b_2^2}{T^2} - 8 E \right)
    \, V_{g,2}^\text{Airy}(b_1,b_2) \,,
\end{split}
\label{eq:invLaplace}
\end{equation}
where we used $T \gg \beta$.
In rescaled average/difference variables $b_{1,2} = \sqrt{T}\,(\bar b \pm \delta b )$, this becomes:
\begin{equation}
\begin{split}
    &Z_{g,2}^\text{Airy}(T,E) 
    \approx 
    \frac{2T^{3g+1}}{\pi}\int_{-\infty}^{\infty} d\delta b \int_{\delta b}^\infty d\bar b  \,\left( \bar b^2 - \delta b^2 \right)\, e^{ i\,  \delta b\,\bar b} \, \delta\!\left( \bar b^2 + \delta b^2 - 4 ET \right)
    \, V_{g,2}^\text{Airy}\left(\bar b+\delta b ,\, \bar b - \delta b \right) 
    \\
    &\qquad= \frac{2T^{3g+1}}{\pi} \int_{-\sqrt{4ET}}^{\sqrt{4ET}} d\delta b   \,\frac{2ET-\delta b^2}{\sqrt{4 ET-\delta b^2}} \, e^{i\, \delta b \sqrt{4ET-\delta b^2}} 
    \,V_{g,2}^\text{Airy}\left(\sqrt{4ET-\delta b^2}+\delta b ,\, \sqrt{4ET-\delta b^2} - \delta b \right)  
\end{split}
\label{eq:ZAiryAvDif}
\end{equation}
In the $\uptau$-scaling limit $ET=e^{S_0}E\uptau$ so for energies $E\gg \frac{e^{-S_0}}{\uptau}$  we can approximate the integral by: 
\begin{equation}
\label{eq:ZairyMicroCan}
\begin{split}
    Z_{g,2}^\text{Airy}(T,E) 
    &\approx \frac{\sqrt{4ET}\,T^{3g+1}}{\pi} \int_{-\infty}^{\infty} d\delta b   \; e^{i\sqrt{4ET}\, \delta b} 
    \; V_{g,2}^\text{Airy}\left(\sqrt{4ET}+\delta b ,\, \sqrt{4ET} - \delta b \right)  
\end{split}
\end{equation}
Corrections to every term in this expansion are ${\cal O}\left( \frac{1}{e^{2S_0}\uptau^2\rho_0^4} \right)$ and therefore negligible for energies  $E\gg \frac{e^{-S_0}}{\uptau}$.
We compare this to the expected  early time ``ramp part''  of the GOE microcanonical SFF in the $\uptau\ll \rho_0(E)$ expansion (i.e., the first term in the minimum function in \eqref{eq:KEtauGeneral}):
\begin{equation}
\label{KAiryseries}
    \begin{split}
    {\cal K}_E^\text{Airy}(\uptau) \big|_\text{ramp} & = \frac{\uptau}{\pi} - \frac{\uptau^2}{2\pi^2 \rho_0} +\frac{\uptau^3}{4\pi^3 \rho_0^2}-\frac{\uptau^4}{6\pi^4 \rho_0^3}   + \frac{\uptau^5}{8\pi^5\rho_0^4} +\mathcal{O}\qty(\frac{\uptau^{2g+1}}{\rho_0^{2g}})
    \end{split}
\end{equation}
where $\rho_0 \equiv \rho_0^\text{Airy}(E) = \frac{1}{2\pi} \sqrt{E}$. 

By plugging the WP volumes (table \ref{tab:NOairyWP}) into \eqref{eq:ZairyMicroCan}, it is possible to match \eqref{KAiryseries} genus by genus.\footnote{The required integrals are simple Fourier transforms of the form 
\begin{equation}
    \int_{-\infty}^\infty d\delta b \; e^{i\sqrt{4ET} \, \delta b} \, (\delta b)^n \, \theta(\delta b) = i^{n+1} \, n!\,(4ET)^{-\frac{n+1}{2}} \,. 
\end{equation}
} The fixed energy gravitational path integral of topological gravity then reproduces the ``ramp part'' of the  $\uptau$-scaled SFF in the GOE Airy matrix model: 
\begin{equation}
\label{eq:NOairyMicroCan}
\boxed{
     {\cal K}_E^\text{Airy}(\uptau)\big|_{\rm ramp}=\lim_{e^{S_0}\rightarrow\infty}\sum_{g=0,\frac{1}{2},1,\dots}e^{-(2g+1)S_0} \, Z^{\rm Airy}_{g,2}\big(\uptau \,e^{S_0},E \big)
     }
\end{equation}
where the left hand side is the {\it universal $\uptau$-scaled RMT result} and the right hand side is the {\it gravitational path integral} evaluated genus by genus for $E\uptau\gg e^{-S_0}$.

Crucially, \eqref{eq:NOairyMicroCan} is finite in the $\uptau$-scaling limit! No late time divergences occur, unlike in the canonical ensemble computation in \eqref{eq:AiryGOEgravity}. 
In fact, note how the terms in the series \eqref{KAiryseries} become more and more divergent near $E=0$, rendering the Laplace transform to the canonical ensemble divergent. The first divergent term occurs at $g=1$, given by $\sim\frac{\uptau^3}{E}$, and all subsequent terms are more divergent.\footnote{The only non-divergent term when passing to the canonical ensemble is $g=\frac{1}{2}$ and can thus be easily reproduced in JT gravity \cite{Saad:2022kfe}.}

Let us now comment on the subtle effects that give rise to the correctness and finiteness of \eqref{eq:NOairyMicroCan}. The WP volumes entering the integrals \eqref{eq:ZairyMicroCan} are generically piecewise polynomials of the form
\begin{equation}
\label{eq:Vg2structure}
\begin{split}
    V_{g,2}^\text{Airy} &= P^{(1)}_{g,2}(b_1,b_2) + \theta(b_1-b_2) \,P^{(2)}_{g,2}(b_1,b_2)  + \text{perm.}
\end{split}
\end{equation}
characterized by two polynomials of degree $6g-2$:
\begin{equation}
\label{eq:Vg2structure2}
\begin{split}
    P^{(1)}_{g,2}(b_1,b_2) &= \sum_{\gamma=0}^{ \lfloor 3g-1\rfloor} C_{6g-2-\gamma,\gamma}^\text{sym} \, b_1^{6g-2-\gamma}\, b_2^{\gamma} \,,\\
    P^{(2)}_{g,2}(b_1,b_2) &=  \;\sum_{\gamma=0}^{6g-2} \;C^{>}_{6g-2-\gamma,\gamma} \, b_1^{6g-2-\gamma}\, b_2^{\gamma}   \,.
\end{split}
\end{equation}
See table \ref{tab:NOairyWP} for examples. Two properties are crucial in the above derivation:

\noindent {\bf (1) Non-analyticities:} First, the integrals \eqref{eq:ZairyMicroCan} are only non-trivial because of the step-function terms in the non-orientable WP volumes, i.e., the polynomials $P_{g,2}^{(2)}$: only those terms contribute to \eqref{eq:NOairyMicroCan}. The analytic terms $P_{g,2}^{(1)}$ give vanishing contributions to the expression \eqref{eq:ZairyMicroCan} (except for $g=0$, trivially). This is consistent with the fact that in the GUE, where the WP volumes are always just symmetric polynomials, the expansion \eqref{eq:NOairyMicroCan} truncates after the first term, c.f., \eqref{eq:GUElaplaceStart}. That is, both in the orientable and in the non-orientable case, the symmetric polynomials in the WP volumes are only relevant for the ``plateau'' (determined by small energies).

\noindent {\bf (2) Cancellations in WP volumes:} Second, the step-function terms involve non-trivial coefficients $C^{>}_{\alpha,\gamma}$ ($\alpha+\gamma=6g-2$). For a generic piecewise polynomial \eqref{eq:Vg2structure}, the integral \eqref{eq:ZairyMicroCan} produces:
\begin{equation} 
\begin{split}
e^{-(2g+1)S_0}\, Z_{g,2}^\text{Airy}(T,E) \,&\stackrel{ET\gg 1}{\approx}  
 \,  
 \uptau^{2g+1} \times \frac{2}{\pi}  \sum_{k=0}^{2g-1} (-1)^{k+1}(2k+1)!\; {\cal C}_{g}^{(k)} \, (4\pi \rho_0)^{6g-4k-4}\, T^{4g-2k-2}   \,\\
 {\cal C}_{g}^{(k)}  &\;\;\, \equiv \sum_{\gamma=0}^{6g-2} \left[ \sum_{r=0}^k (-1)^r \binom{\gamma}{r} \binom{6g-2-2\gamma}{2k+1-2r}\right] C^{>}_{6g-2-\gamma,\gamma} 
 \end{split}
\end{equation}
For $k=0,\ldots,2g-2$ these are divergent as $T \rightarrow \infty$ with $\uptau$ held fixed.
The fact that these do not spoil \eqref{eq:NOairyMicroCan} is due to cancellations between the coefficients $C^{>}_{\alpha,\gamma}$  for each of the WP volumes individually. We claim that the constraints ${\cal C}_g^{(k)}=0$ for $k=0,\ldots,2g-2$. The final term ($k=2g-1$) gives the finite contributions collected in \eqref{eq:NOairyMicroCan}. Further contributions vanish upon $\uptau$-scaling, and we do not show them in the above formula.
For illustration, we give some examples of these cancellations and then formulate a conjecture about their generalization. 

\noindent {\it Genus $g=1$:} 
\begin{equation}
\label{eq:microCon1}
    \begin{split}
     e^{-3 S_0} Z_{1,2}^\text{Airy}(T,E) 
     &\approx -64 \pi \left({\color{darkgreen}2C^{>}_{4,0}+C^{>}_{3,1}-C^{>}_{1,3}-2 C^{>}_{0,4}} \right)  \rho_0^2 \,T^2 \uptau^3 \\
     &\quad\, + \frac{3}{2\pi^3}\left(2C^{>}_{4,0}-C^{>}_{3,1}+C^{>}_{1,3}-2 C^{>}_{0,4} \right)  \rho_0^{-2} \, \uptau^3 + \ldots
    \end{split}
\end{equation}
One can check that the coefficients given in table \ref{tab:NOairyWP} are such that the first line (green combination) vanishes and the second line produces the $\uptau^3$ term in \eqref{eq:NOairyMicroCan}. 

\noindent {\it Genus $g=\frac{3}{2}$:} 
\begin{equation}
\label{eq:microConThreeHalf}
    \begin{split}
     &e^{-4 S_0} Z_{\frac{3}{2},2}^\text{Airy}(T,E) \\
     &\qquad\approx -2^{11} \pi^4 \left({\color{darkgreen}7C^{>}_{7,0}+5C^{>}_{6,1}+3C^{>}_{5,2}+C^{>}_{4,3}-C^{>}_{3,4}-3C^{>}_{2,5}-5C^{>}_{1,6}-7C^>_{0,7}}\right)  \rho_0^5 T^4 \uptau^4 \\
     &\qquad\quad\, +48 \left({\color{darkgreen}35C^{>}_{7,0}+5C^{>}_{6,1}-5C^{>}_{5,2}-3C^{>}_{4,3}+3C^{>}_{3,4}+5C^{>}_{2,5}-5C^{>}_{1,6}-35C^>_{0,7}}\right)  \rho_0 T^2 \uptau^4 \\
     &\qquad\quad\, -\frac{15}{4\pi^4} \left(21C^{>}_{7,0}-9C^{>}_{6,1}+C^{>}_{5,2}+3C^{>}_{4,3}-3C^{>}_{3,4} -C^{>}_{2,5}+9C^{>}_{1,6}-21C^>_{0,7}\right) \rho_0^{-3}  \uptau^4  + \ldots 
    \end{split}
\end{equation}
Again, the coefficients are such that the divergent first two lines vanish and the third line yields the $\uptau^4$ term in \eqref{eq:NOairyMicroCan}.

\noindent {\it Genus $g=2$:} 
\begin{equation}
\label{eq:microCon2}
    \begin{split}
     &e^{-5 S_0} Z_{2,2}^\text{Airy}(T,E)\\ 
     &\;\;\approx -2^{18} \pi^7 \left({\color{darkgreen}5C^{>}_{10,0}+4C^{>}_{9,1}+3C^{>}_{8,2}+2C^{>}_{7,3}+C^{>}_{6,4}-C^{>}_{4,6}-2C^{>}_{3,7}-3C^{>}_{2,8}-4C^{>}_{1,9}-5C^{>}_{0,10}}\right)\rho_0^8  T^6\uptau^{5} \\
     &\quad\, +3\cdot 2^{13} \pi^3\left({\color{darkgreen}15C^{>}_{10,0}+6C^{>}_{9,1}+C^{>}_{8,2}-C^{>}_{7,3}-C^{>}_{6,4}+C^{>}_{4,6}+C^{>}_{3,7}-C^{>}_{2,8}-6C^{>}_{1,9}-15C^{>}_{0,10}}\right) \rho_0^4 T^4 \uptau^{5} \\
     &\quad\, -\frac{960}{\pi}\left({\color{darkgreen}63C^{>}_{10,0}-7C^{>}_{8,2}+3C^{>}_{6,4}-3C^{>}_{4,6}+7C^{>}_{2,8}-63C^{>}_{0,10}}\right) T^2  \uptau^{5}\\
     &\quad\, +\frac{315}{\pi^5}\left(15C^{>}_{10,0}-6C^{>}_{9,1}+C^{>}_{8,2}+C^{>}_{7,3}-C^{>}_{6,4}+C^{>}_{5,6}-C^{>}_{4,7}-C^{>}_{3,8}+6C^{>}_{2,9}-15C^{>}_{0,10}\right) \rho_0^{-4}  \uptau^{5}+ \ldots 
    \end{split}
\end{equation}
Again, the first three lines vanish, while the last line gives the correct universal $\uptau^5$ term in \eqref{eq:NOairyMicroCan}.

It would clearly be interesting to understand these cancellations better. In general we find that $(2g-1)$ linear combinations of $C^{>}_{\alpha,\gamma}$ must vanish to produce a finite contribution to the $\uptau$-scaled SFF at genus $g$. Such cancellations are reminiscent of similar cancellations in the GUE case, where an underlying mathematical structure (KdV hierarchy) can be employed to explain this conspiracy \cite{Blommaert:2022lbh}. A goal for a future investigation of such structures would be to prove the following statement (which we have confirmed up to $g=3$):
\mdfsetup{%
   middlelinewidth=.5pt,
   roundcorner=10pt}
\begin{mdframed}
\noindent{\bf Conjecture:} {\it For any fixed-genus $g$, the microcanonical gravitational path integral \eqref{eq:ZairyMicroCan} has a finite $\uptau$-scaling limit and reproduces the universal RMT coefficient multiplying $\rho_0^{-2g}\uptau^{2g+1}$. This is possible due to {\color{darkgreen}$(2g-1)$ cancellations} amongst coefficients $C^>_{\alpha,\gamma}$ of the non-analytic parts of the Weil-Petersson volumes:}
    \begin{equation}
    \label{eq:conjecture1}
        0\, \stackrel{!}{=} \, {\cal C}_{g}^{(k)}  \equiv \sum_{\gamma=0}^{6g-2} \left[ \sum_{r=0}^k (-1)^r \binom{\gamma}{r} \binom{6g-2-2\gamma}{2k+1-2r}\right] C^{>}_{6g-2-\gamma,\gamma} \qquad (k=0,\ldots,2g-2). 
    \end{equation}
\end{mdframed}

\subsection{Cancellations in the canonical SFF}

One can directly evaluate the canonical gravitational path integral \eqref{eq:Zg2Airy} genus by genus. An interesting class of terms is the $\log(\frac{\beta}{2T})$ terms visible in \eqref{eq:AiryGOEgravity}. These were partly discussed in \cite{Weber:2024ieq}, so we will be brief.

Consider again the generic form of a genus $g$ WP volume, \eqref{eq:Vg2structure}.
We can plug this ansatz into the trumpet path integral \eqref{eq:Zg2Airy} and find the following contribution:
\begin{equation}
\label{eq:Zg2CanonicalGeneral}
    \begin{split}
    Z_{g,2}^{\text{Airy}}(\beta_1,\beta_2)
    &=  \frac{2^{6g-2}}{\pi} \,\sum_{\gamma=0}^{\lfloor 3g-1\rfloor } C_{6g-2-\gamma,\gamma}^\text{sym} \, \Gamma\left(3g-\frac{\gamma}{2}\right) \Gamma\left(1+\frac{\gamma}{2} \right)\, \beta_1^{3g-\frac{1+\gamma}{2}} \, \beta_2^{\frac{1+\gamma}{2}}   \\
    &\quad + \frac{2^{6g-2}}{\pi}\sum_{\gamma=0}^{6g-2} C^>_{6g-2-\gamma,\gamma} \bigg[ \Gamma\left( 3g - \frac{\gamma}{2}\right) \Gamma\left( 1 +\frac{\gamma}{2} \right) \beta_1^{3g-\frac{1+\gamma}{2}} \beta_2^{\frac{1+\gamma}{2}}
    \\
    &\qquad\qquad\qquad\qquad\quad\;\; -\frac{ \Gamma(3g+1)}{3g-\frac{\gamma}{2}} \frac{\beta_1^{3g+1}}{\sqrt{\beta_1\beta_2}} \; {}_2F_1\left(3g+1,3g-\frac{\gamma}{2},3g+1-\frac{\gamma}{2}; \, -\frac{\beta_1}{\beta_2} \right) \bigg]
    \\
    &\quad +\, [\beta_1 \leftrightarrow \beta_2]
    \end{split}
\end{equation}
with $\beta_{1,2} = \beta\pm iT$.
The naive degree of divergence for $e^{-(2g+1)S_0} Z_{g,2}^{\text{Airy}}$ in the $\uptau$-scaling limit is $T^{4g-2}$ (multiplying $\uptau^{2g+1}$ for a total power of time $T^{6g-1}$). However, we find that these divergences largely cancel as they are proportional to special linear combinations of the coefficients in the WP volumes. We will now give some examples for these cancellations and then formulate a conjecture for arbitrary genus.\\

\noindent {\it Genus $g=1$:} 
\begin{equation}
    \begin{split}
    e^{-3S_0} Z_{1,2}^{\text{Airy}}(T,\beta)
    &= -\frac{4}{\pi} \left( {\color{darkgreen} 2C^>_{4,0} + C^>_{3,1}-C^>_{1,3}-2 C^>_{0,4}} \right) \, \frac{T^2 \, \uptau^3}{\beta^2} 
    \\
    &\quad\, + \frac{12}{\pi}\left( {\color{lightred}C^>_{3,1}-C^>_{1,3}} \right) \log\left(\frac{\beta}{2T}\right) \, \uptau^3
    \\
    &\quad\, -\frac{2}{\pi}
    \left( 42 C^>_{4,0}-C^>_{3,1}-8C^>_{2,2}+C^>_{1,3}-10C^>_{0,4} +32 C^\text{sym}_{4,0}-16 C^\text{sym}_{2,2}\right) \uptau^3 \,.
    \end{split}
    \label{eq:gOne}
\end{equation}
The first line diverges as $T^2$, but the linear combination of coefficients cancels for the actual genus 1 WP volume. This {\color{darkgreen}constraint} is the same as derived in \eqref{eq:microCon1}. The second line also diverges, but it does not cancel: $\frac{12}{\pi}(C_{3,1}^> - C_{1,3}^>) = -\frac{1}{\pi}$. A different mechanism (resummation of the genus expansion) is required to remove this {\color{lightred}logarithmic divergence}. The last line is finite upon $\uptau$-scaling and produces the corresponding ${\cal O}(\uptau^3)$ term in \eqref{eq:AiryGOEgravity}.

\noindent {\it Genus $g=\frac{3}{2}$:} 
\begin{equation}
\label{eq:gThreeHalf}
    \begin{split}
    &e^{-4S_0} Z_{\frac{3}{2},2}^{\text{Airy}}(T,\beta) \\
    &\qquad= -\frac{15}{\sqrt{2\pi}} \left( {\color{darkgreen} 7C^>_{7,0} +5 C^>_{6,1}+3 C^>_{5,2} + C^>_{4,3}- C^>_{3,4} -3 C^>_{2,5} - 5 C^>_{1,6} - 7 C^>_{0,7}} \right) \, \frac{T^4 \, \uptau^4}{\beta^{7/2}}
    \\
    &\qquad\quad\, -\frac{12}{\sqrt{2\pi}} \left({\color{darkgreen} 35 C^>_{7,0}+35 C^>_{6,1}+25 C^>_{5,2}+9C^>_{4,3}-9C^>_{3,4}-25C^>_{2,5}-35C^>_{1,6}-35 C^>_{0,7} }\right) \frac{T^2 \, \uptau^4}{\beta^{3/2}}
    \\
    &\qquad\quad\, +\frac{48}{\sqrt{2\pi}}
    \left({\color{lightred}35C^>_{7,0}-10C^>_{5,2}+8C^>_{3,4}-16C^>_{1,6} +35 C^\text{sym}_{7,0}-16C^\text{sym}_{6,1}-10C^\text{sym}_{5,2}+8C^\text{sym}_{4,3}}\right) \sqrt{T} \, \uptau^4 \\
    &\qquad\quad\, -\frac{30}{\sqrt{2\pi}}
    \left(21C^>_{7,0}+91 C^>_{6,1}+21 C^>_{5,2}-5 C^>_{4,3}+5C^>_{3,4}-21C^>_{2,5}-91 C^>_{1,6}-21C^>_{0,7}\right)  \sqrt{\beta} \,\uptau^4 \,.
    \end{split}
\end{equation}
The first two lines vanish, thus giving {\color{darkgreen} two constraints} on the WP volumes. One can check that these constraints are equivalent (but interestingly not identical) to those found in \eqref{eq:microConThreeHalf}. The third line does not cancel and thus leads to a {\color{lightred}powerlaw divergence}, which requires a different mechanism to remove. The last line is finite under $\uptau$-scaling, see ${\cal O}(\uptau^4)$ in \eqref{eq:AiryGOEgravity}.

\noindent {\it Genus $g=2$:} 
\begin{equation}
    \begin{split}
    &e^{-5S_0} Z_{2,2}^{\text{Airy}}(T,\beta) \\
    &\quad = -\frac{768 \,T^6 \uptau^5}{\pi \beta^5} \,\left({\color{darkgreen}5 C^>_{10,0}+4 C^>_{9,1}+3 C^>_{8,2}+2 C^>_{7,3}+ C^>_{6,4}\;-\; [C^>_{\alpha,\gamma}\rightarrow C^>_{\gamma,\alpha}]\, }\right)\\
    &\qquad
    - \frac{384 \,T^4 \uptau^5}{\pi \beta^3} \,\left({\color{darkgreen}55 C^>_{10,0}+50 C^>_{9,1}+41 C^>_{8,2}+29 C^>_{7,3}+15 C^>_{6,4} \; - \;[C^>_{\alpha,\gamma}\rightarrow C^>_{\gamma,\alpha}] \,}\right) \\
    &\qquad
    - \frac{96 \,T^2 \uptau^5}{\pi \beta} \,\left({\color{darkgreen} 495C^>_{10,0}+420C^>_{9,1}+441 C^>_{8,2}+394 C^>_{7,3}+235 C^>_{6,4} \; - \; [C^>_{\alpha,\gamma}\rightarrow C^>_{\gamma,\alpha}]\,}
    \right)
    \\
    &\qquad + 720 \, T \uptau^5 \,\left( {\color{orange} 21 \left(C^>_{9,1}+C^>_{1,9}+2 C^\text{sym}_{9,1}\right)-7\left(C^>_{7,3}+C^>_{3,7}+2 C^\text{sym}_{7,3}\right)+5\left(C^>_{5,5}+2 C^\text{sym}_{5,5}\right) }  \right)
    \\
    &\qquad
    + \frac{48\,\beta \uptau^5}{\pi}\,\log\left(\frac{2T}{\beta}\right)\left({\color{lightred}2520 C^>_{9,1} -420 C^>_{7,3} \; - \; [C^>_{\alpha,\gamma}\rightarrow C^>_{\gamma,\alpha}] } \,\right)
    \\
    &\qquad +\frac{48\,\beta \uptau^5}{\pi}\Big(
    24445 C^>_{10,0}-4914C^>_{9,1}-4269C^>_{8,2}-973C^>_{7,3}-419C^>_{6,4}+931C^>_{4,6}+973C^>_{3,7}  \\
    &\qquad\qquad\qquad +1197C^>_{2,8}+4914C^>_{1,9}+1155C^>_{0,10} + 25600 C^\text{sym}_{10,0} -3072 C^\text{sym}_{8,2} +512 C^\text{sym}_{6,4}
    \Big)
    \end{split}
    \label{eq:genus2cancellations}
\end{equation}
The notation ``$-[C^>_{\alpha,\gamma}\rightarrow C^>_{\gamma,\alpha}]$'' means that we subtract the same combination of terms with indices swapped (see \eqref{eq:gOne} and \eqref{eq:gThreeHalf} for illustration of this structure).
The vanishing of the first three  linear combinations provides {\color{darkgreen}three constraints} on the coefficients in the WP volumes equivalent to those found in \eqref{eq:microCon2}. The last highlighted linear combination also vanishes; this therefore provides {\color{orange}one constraint}, which is of a different type as it also involves the coefficients $C^\text{sym}_{\alpha,\gamma}$. The {\color{lightred}logarithmic divergence} survives and the last line remains finite in the $\uptau$-scaling limit, see ${\cal O}(\uptau^5)$ in \eqref{eq:AiryGOEgravity}.

\noindent {\it Further examples:} 

At $g=\frac{5}{2}$ we find {\color{darkgreen}four constraints} on $C^>_{13-\gamma,\gamma}$ from cancellation of divergences scaling as $T^8,T^6,T^4,T^2$. The cancellations are equivalent to the linear combinations ${\cal C}_{g=\frac{5}{2}}^{(k)}$ in \eqref{eq:conjecture1} for $k=0,\ldots,3$. {\color{lightred}Two divergences} scaling as $T^{3/2}$ and $T^{1/2}$ survive and their removal requires all-order resummation, see ${\cal O}(\uptau^6)$ in \eqref{eq:AiryGOEgravity}.

At $g=3$ we find {\color{darkgreen}five constraints} on $C^>_{16-\gamma,\gamma}$ from the cancellation of divergences $\sim T^{10},T^8,T^6,T^4$ and $T^2 \log(2T/\beta)$, which are equivalent to ${\cal C}_{g=3}^{(k)}$ in \eqref{eq:conjecture1} for $k=0,\ldots,4$. We also find {\color{orange}one constraint} involving both $C^>_{16-\gamma,\gamma}$ and $C^\text{sym}_{16-\gamma,\gamma}$ from the cancellation of divergences $\sim T$:
\begin{equation}
    \begin{split}
    e^{-7S_0} Z_{3,2}^{\text{Airy}}(T,\beta) 
    & \supset 
    -1814400 \, \beta T \uptau^7 \big( {\color{orange}1001\big( C^>_{15,1} + C^>_{1,15} + 2 C^\text{sym}_{15,1}\big) - 143 \big( C^>_{13,3} + C^>_{3,13}+ 2 C^\text{sym}_{13,3}\big)} \\
    &\qquad\qquad\qquad \qquad\;\;\; {\color{orange}+33\big( C^>_{11,5} + C^>_{5,11} + 2 C^\text{sym}_{11,5}\big)  -7\big( C^>_{9,7} + C^>_{7,9} + 2 C^\text{sym}_{9,7}\big)} \big)\,.
    \end{split}
    \label{eq:genus3cancellations}
\end{equation}
The {\color{lightred}two divergences} that survive the cancellations scale as $T^2$ and $\log(2T/\beta)$, see ${\cal O}(\uptau^7)$ of \eqref{eq:AiryGOEgravity}.

At $g=\frac{7}{2}$, we find {\color{darkgreen}six constraints} on $C^>_{19-\gamma,\gamma}$ from cancellation of divergences $\sim T^{12},T^{10},\ldots,T^2$. The surviving {\color{lightred}three divergences} scale as $T^{5/2},T^{3/2},T^{1/2}$.\\

\noindent {\it General structure:} 

Based on the above observations, we are in the position to classify the divergences into three types and conjecture their general properties. We conjecture that $e^{-(2g+1)S_0} Z^\text{Airy}_{g,2}(T,\beta)$ exhibits the following divergences and cancellations as $T\rightarrow\infty$ with $\uptau=Te^{-S_0}$ held fixed:
\begin{itemize}
    \item {\bf Type I:} The first type of divergences is associated with terms that would scale as
    \begin{equation}
        \text{[type I]} \, \sim \, \left\{ \begin{aligned}
        &\left\{ \frac{T^{2+2n}}{\pi \beta^{3-g+2n}} \, \log\left(\frac{2T}{\beta}\right)\right\}_{0\leq n \leq \lfloor\frac{g-3}{2} \rfloor} \;\; \text{and} \;\; 
        \left\{ \frac{T^{2+2n}}{\pi \beta^{3-g+2n}} \right\}_{\lfloor \frac{g-3}{2} \rfloor < n \leq 2g-2}
        \quad (g\in \mathbb{Z}_+)\\
        & \left\{ \frac{T^{2+2n}}{\sqrt{\pi} \beta^{3-g+2n}} \right\}_{0\leq n \leq 2g-2} \qquad\qquad\qquad\qquad\qquad\qquad\qquad\qquad \quad\;\;  (g\in \tfrac{1}{2}+\mathbb{Z}_+)
        \end{aligned} 
        \right. 
    \end{equation}
    These divergences multiply vanishing linear combinations of $C^>_{\alpha,\gamma}$, which always appear in the {\it antisymmetric} combination $C^>_{\alpha,\gamma}-C^>_{\gamma,\alpha}$. These cancellations thus provide {\color{darkgreen}$(2g-1)$ constraints} on $C^>_{\alpha,\gamma}$, which are equivalent to those formulated in \eqref{eq:conjecture1}.
    \item {\bf Type II:} For integer genus, there could be additional divergences which would scale as
    \begin{equation}
        \text{[type II]} \, \sim \, \left\{ \frac{T^{1+2n}}{\beta^{2-g+2n}} \right\}_{0\leq n \leq \lfloor\frac{g}{2}\rfloor-1}  \qquad\quad (g\in \mathbb{Z}_+)
    \end{equation}
    These divergences multiply linear combinations of $C^>_{\alpha,\gamma}$ and $C^\text{sym}_{\alpha,\gamma}$, always appearing in the {\it symmetric} combination $C^>_{\alpha,\gamma}+ C^>_{\gamma,\alpha} + 2 C^\text{sym}_{\alpha,\gamma}$. Their vanishing thus provides a further   {\color{orange}$\lfloor \frac{g}{2}\rfloor $ constraints} on the coefficients of WP volumes, which are manifestly independent of type I.
    \item {\bf Type III:} Finally, we observe terms, which scale as
    \begin{equation}
    \label{eq:typeIII}
        \text{[type III]} \, \sim \, 
        \left\{ \begin{aligned}
        &\frac{\beta^{g-1}}{\pi} \, \log\left(\frac{2T}{\beta}\right)  \;\text{ and }\; \left\{\frac{T^{2+2n}}{\pi \beta^{g-3+2n}}\right\}_{0\leq n \leq \lfloor\frac{g-3}{2}\rfloor} \quad\;\;\,  (g\in \mathbb{Z}_+)\\
        &\left\{\frac{T^{\frac{1}{2}+n}}{\sqrt{\pi} \beta^{\frac{3}{2}-g+n}} \right\}_{0\leq n \leq \lfloor g-1 \rfloor} \qquad\qquad\qquad\qquad\qquad\quad\;\, (g\in \tfrac{1}{2}+\mathbb{Z}_+)
        \end{aligned} 
        \right. 
    \end{equation}
    These {\color{lightred}divergences} do not cancel at any fixed-genus. Their removal, required for consistency of the $\uptau$-scaling, must be due to a non-trivial resummation of the genus expansion.
\end{itemize}

It would be very interesting to prove (or improve) these conjectures. We note that a subset of the above (namely, type II cancellations and the powerlaw divergences of type III) are analogous to the constraints appearing in the orientable case \cite{Weber:2022sov,Weber:2024ieq}. The remaining cancellations are new and exclusive to the non-orientable scenario: we emphasize again that type I divergences -- if they didn't cancel -- would be more severe than in the orientable case, where the highest degree of any potential divergence is only $T^{g-1}\, \uptau^{2g+1}$, rather than $T^{4g-2}\, \uptau^{2g+1}$. We expect that a proper understanding will require new mathematical insights, for example about a generalization of intersection numbers to non-orientable geometries.

\section{Lessons about the genus expansion to all orders}
\label{sec:periodicorbits}

In this section, we draw some lessons and formulate some challenges regarding the GOE genus expansion.

\subsection{Resumming the genus expansion}

Recall the following convenient split of the universal GOE sine kernel into a GUE piece, a low energy ``plateau'' piece and a high energy ``ramp'' contribution (c.f., \eqref{eq:GOEdecomposeApp}):
\begin{equation}
\begin{split}
 {\cal K}^{{\text{GOE}}}(\uptau) = 
 2 \times {\cal K}^{{\text{GUE}}}(\uptau)  + \Khighlow^{{\text{GOE}}}_{\lowElabel}(\uptau) + \Khighlow^{{\text{GOE}}}_{\highElabel}(\uptau) \,.
 \end{split}
 \label{eq:GOEdecompose5}
\end{equation}

\paragraph{$\Khighlow^{{\text{GOE}}}_{\highElabel,E}$ from gravity.}
We showed in \eqref{eq:NOairyMicroCan} that the fixed energy gravitational path integral of the Airy model reproduces genus by genus the prediction from the GOE Airy matrix integral in the $\uptau$-scaling limit:
\begin{equation}
\label{eq:NOairyMicroCanRepeat}
\lim_{e^{S_0}\rightarrow\infty}\sum_{g=0,\frac{1}{2},1,\dots}e^{-(2g+1)S_0} \, Z^{\rm Airy}_{g,2}(\uptau\, e^{S_0},E)= \frac{\uptau}{\pi} + \sum_{g= \frac{1}{2},1,\frac{3}{2},\ldots}  \frac{(-1)^{2g} \,\uptau^{2g+1}}{4g\pi^{2g+1} \rho_0^{2g}}\,.
\end{equation}
The genus expansion we have obtained in this way is now convergent and can be resummed to: 
\begin{equation}
\eqref{eq:NOairyMicroCanRepeat} =\frac{\uptau}{\pi} - \frac{\uptau}{2\pi} \, \log\left(1 + \frac{\uptau}{\pi\rho_0(E)} \right).
\label{eq:genusresum} 
\end{equation}
The expansion converges within a finite radius $|\uptau| <\pi \rho_0(E)$, set by the nearest singularity of the logarithm, but can be analytically continued to arbitrary $\uptau>0$.
The resummed expression is equal to the first argument in the minimum function in the GOE microcanonical SFF, \eqref{eq:KEtauGeneral}.
Working with the fixed energy gravitational path integral has thus allowed us to reproduce the prediction of RMT for times before the Heisenberg/plateau time $\uptau<\uptau_H=2\pi\rho_0$ via a convergent genus expansion. 
After resummation, we can pass to the canonical ensemble via a Laplace transform. The expression we obtained from gravity is trustworthy only before the plateau time, i.e., for high energies $E>E_*(\uptau)$. Therefore:
\begin{equation}
\int_{E_*}^{\infty} dE \, e^{-2\beta E}\qty(\frac{\uptau}{\pi} - \frac{\uptau}{2\pi} \, \log\left(1 + \frac{\uptau}{\pi\rho_0(E)} \right))= \frac{\uptau}{2\pi\beta} + \Khighlow_{\highElabel,\beta}^{{\text{GOE}}}(\uptau)\subset {\cal K}^{\rm GOE}_\beta(\uptau) \,.
\end{equation}
This is exactly the linear ramp plus the ``high energy'' piece $\Khighlow_{\highElabel,\beta}^{{\text{GOE}}}$ in the RMT canonical SFF, which we analyze and Laplace transform in section \ref{sec:GOESFF}. The explicit expression for $\Khighlow_{\highElabel,\beta}^{{\text{GOE}}}$ can be found in appendix \ref{app:derivations} and is analyzed in detail in section \ref{sec:goeHigh}. 
Its behavior is plotted in figure \ref{fig:airyDecompose}; we can see that it asymptotes to a constant plateau, although with the wrong height. 

Let us pause to comment on the significance of this equality. Recall that non-orientable two-dimensional gravity did not have a well-defined $\uptau$-scaling limit, since each genus $g$ two-boundary wormhole is divergent in this limit:  $Z^{\rm (unor.)}_{g,2}(\beta,T=e^{S_0}\uptau)\rightarrow \infty$, see \eqref{eq:AiryGOEgravity}. This was not the case in the simpler orientable/GUE case where each genus $g$ wormhole admitted a finite $\uptau$-scaling limit and individually produced a contribution $A_g^\text{GUE}(\beta)\uptau^{2g+1}$, which in turn can be resummed to the canonical RMT SFF $\mathcal{K}^{\rm GUE}_{\beta}(\uptau)$.

We concluded that for non-orientable two-dimensional gravity to have a well-defined $\uptau$-scaling limit, a non-trivial genus resummation must occur, which cancels the late time divergences and produces a finite result, matching the RMT prediction (see \eqref{eq:typeIII} and c.f.\ \cite{Tall:2024hgo}). We have now directly exhibited such a resummation and obtained a finite $\uptau$-scaled SFF from gravity which reproduces the high energy piece of the RMT prediction, 
$\Khighlow_{\highElabel,\beta}^{{\text{GOE}}}\subset {\cal K}^{\rm GOE}_\beta(\uptau)$. 
To perform the resummation and cure the divergences of the canonical SFF in gravity it was crucial to work with the fixed energy gravitational path integral, computing and resumming the genus $g$ contributions at fixed energy, and only transforming  back to fixed temperature  \textit{after} the resummation. Any potential divergences were cancelled due to {\color{darkgreen}$(2g-1)$} ``type I'' cancellations amongst WP volume coefficients.

\paragraph{Towards the plateau.}
Having reproduced from gravity the full microcanonical RMT SFF for times before the plateau time $\uptau<\uptau_H$, it is natural to pose the question of how to proceed beyond and access the plateau region.  
In the GUE, a microcanonical gravitational derivation of the plateau can be achieved fully. We exhibit this in Appendix \ref{sec:microcanonicalPlateau}. The upshot is that the microcanonical ${\cal K}^\text{GUE}_E(\uptau)$ admits a formal genus expansion, where the plateau is realized as an infinite sum of delta-functions. The orientable gravitational path integral reproduces this structure term-by-term. Terms that would naively diverge upon $\uptau$-scaling are {\it a priori} produced, but they occur with vanishing coefficient due to {\color{orange}$(g-1)$} cancellations between coefficients of WP volumes, c.f., \eqref{eq:GUEcancellationsGrav}. These cancellations were discussed using different methods in \cite{Weber:2022sov,Blommaert:2022lbh}.

The fact that the GUE Airy model admits a finite description in gravity, allows us to declare victory regarding the second of three pieces of the microcanonical GOE SFF \eqref{eq:GOEdecompose5}, i.e., the contribution $2\, {\cal K}_E^\text{GUE}(\uptau)$. We note that it seems to make physical sense to split off this contribution: the GUE WP volumes as well as the analytic part of the GOE volumes are symmetric positive polynomials. As one can check, the GOE WP volumes are such that after subtracting twice their GUE counterpart, the analytic piece remains a symmetric positive polynomial. The missing low-energy piece $\Khighlow^{{\text{GOE}}}_{\lowElabel,E}$ is therefore entirely determined by the non-analytic terms in the WP volumes, and its own ``reduced'' symmetric positive polynomial.

We end with another observation about $\Khighlow^{{\text{GOE}}}_{\lowElabel,E}$. Note that the cancellations in the GUE Airy model have an analog in the GOE: for integer genus, there are {\color{orange}$\lfloor \frac{g}{2} \rfloor$} ``type II'' cancellations of the same type, plus {\color{lightred}$\lfloor \frac{g-1}{2} \rfloor$} ``type III'' polynomial divergences: together they amount to $(g-1)$ combinations of WP volume coefficients analogous to those in the GUE, but only half of the combinations cancel. The other half needs to be resummed to combine with $\log(T)$ divergences and produce a finite SFF. At half-integer genus, the number of cancellations of the type analogous to the GUE is {\color{orange}zero}, and the full set of {\color{lightred}$\lfloor g-1\rfloor$} combinations survives in the form of ``type III'' powerlaw divergences.

The fact that the GOE exhibits fewer ``type II'' cancellations poses an obstruction to obtaining a finite result from the gravitational path integral. We leave a detailed investigation as an important challenge for the future.

\subsection{Periodic orbits and gravity}
\label{sec:perOrbitsGravity}

The series expansion \eqref{eq:NOairyMicroCanRepeat} has been explained previously (for general $\rho_0$) in terms of encounters in periodic orbit theory
\cite{PhysRevE.72.046207}. The basic idea is to write the spectral density as a sum over classical periodic orbits, weighted by an action. Spectral correlations therefore display strong oscillations weighted by differences of classical orbit actions. The dominant contributions thus come from pairs of classical orbits with almost equal action. The leading term in \eqref{eq:NOairyMicroCanRepeat} is due to orbits with equal action \cite{berry1985}, the $\uptau^2$ term is explained by orbit pairs with a single close encounter in phase space \cite{Sieber_2001}, and all subsequent terms are systematically due to multiple encounters \cite{Heusler_2004,PhysRevE.72.046207}. In the GUE, different encounters always cancel, except at the leading order $\uptau$. This does not happen in the GOE, thus giving rise to the series \eqref{eq:NOairyMicroCanRepeat}.

\paragraph{Logarithmic divergences in encounter theory and gravity.}
Having seen an exact match between the microcanonical SFF in encounter theory and the high-energy gravitational path integral computations in the Airy model, we can now ask about the canonical ensemble. As anticipated by \cite{Saad:2022kfe}, the naive contribution of the genus $g$ piece in \eqref{eq:NOairyMicroCanRepeat} in encounter theory should be
\begin{equation}
\label{eq:KAiryEncounter}
    e^{-S_0}\,{\cal K}^\text{Airy}_\beta(\uptau) \stackrel{?}{\supset} \int_{\frac{1}{T}}^\infty dE\, e^{-2\beta E} \, \left(\frac{(-1)^{2g} \,\uptau^{2g+1}}{4g\pi^{2g+1} \rho_0^{2g}}\right) \qquad\quad \text{(encounter theory)} \,.
\end{equation}
The low energy cutoff is estimated such that the regime $ET < 1$ is excluded from the integration: in this region the encounter picture breaks down because the orbit action becomes ${\cal O}(1)$. We recognize the integral as precisely the coefficient $d_n^{(\delta)}(\rho_0;\beta)$, which featured in the general formula for the $\uptau$-scaled SFF in the Airy model, with the cutoff  chosen as $\delta = \frac{1}{T}$. The above contribution is therefore precisely one of the terms in the general result \eqref{eq:GOEgeneral}, but with the specific cutoff $\delta=\frac{1}{T}$.
Indeed, for integer $g$, this physical choice of cutoff yields a particularly nice result:
\begin{equation}
\begin{split}
    g\in\mathbb{Z}_+:\quad\;\; \eqref{eq:KAiryEncounter}
    &= \frac{4^{g-1}}{g\pi} \, d_{2g}^{(\frac{1}{T})}(\rho_0;\beta) \,\uptau^{2g+1}
    \\
    &= \frac{4^{g-1}}{g\pi} \left[ \frac{(-1)^g}{(g-1)!} \, \log \left( \frac{1}{T} \right)+ 
    \frac{(-1)^{g-1}}{(g-1)!} \left(\psi(g)-\log(2\beta)\right) + \ldots\right](2\beta)^{g-1}\, \uptau^{2g+1}\,.
\end{split}
\label{eq:periodicOrbitsIntegerRes}
\end{equation}
The logarithmic term matches precisely the analogous term in the gravitational calculation \eqref{eq:AiryGOEgravity}. We recall that these terms are divergent in the $\uptau$-scaling limit and need to be resummed. It is remarkable that the natural periodic orbit cutoff in \eqref{eq:KAiryEncounter} produces exactly the same logarithmic divergences as the gravitational path integral for all $g$. Being cutoff-independent, the finite terms in \eqref{eq:periodicOrbitsIntegerRes} of course also match the corresponding terms in \eqref{eq:cdAiry}.

At half-integer genus we similarly find that the universal finite piece of \eqref{eq:KAiryEncounter} matches the one entering in the $C_{\gt}$ coefficient of the $\uptau$-scaled SFF \eqref{eq:generalA}. The divergent terms have the same structure (i.e., the same powers in $\beta$ and $T$) as type III divergences classified above; the numerical coefficients, however, are not universal (see below).

\paragraph{Finiteness of $\uptau$-scaling in RMT.} 
To summarize, we encountered different perspectives on IR divergences. For notational simplicity, consider integer genus $g$ (half-integer is analogous). The gravitational path integral, using WP volumes and trumpet wave functions, yields divergences of type III:
\begin{equation}
\label{eq:gravityDivergence}
    \text{gravity:} \quad e^{-(2g+1)S_0} Z_{g,2}^\text{Airy}(T,\beta)
    \sim \left[\frac{(-8\beta)^{g-1}}{\pi g!} \log \left(\frac{T}{2\beta}\right) + \sum_{n=0}^{\lfloor \frac{g-3}{2} \rfloor} c_n^\text{(grav)} \, \frac{T^{2+2n}}{\beta^{3-g+2n}} + \text{(finite)} \right] \uptau^{2g+1}.
\end{equation}
The simple periodic orbit estimate \eqref{eq:KAiryEncounter} yields similar divergences but with different coefficients:
\begin{equation}
\label{eq:encounterDivergence}
    \text{encounters:} \quad e^{-(2g+1)S_0}\, Z_{g,2}^\text{Airy}(T,\beta)
    \sim \left[\frac{(-8\beta)^{g-1}}{\pi g!} \log \left(\frac{T}{2\beta}\right) + \sum_{n=0}^{\lfloor \frac{g-3}{2} \rfloor} c_n^\text{(enc.)} \, \frac{T^{2+2n}}{ \beta^{3-g+2n}} + \text{(finite)} \right] \uptau^{2g+1}
\end{equation}
The exact Laplace transform of the universal RMT sine kernel yields a finite result:
\begin{equation}
    \text{RMT:} \quad e^{-(2g+1)S_0}\, Z_{g,2}^\text{Airy}(T,\beta) \equiv {\cal K}_\beta^\text{GOE,Airy}(\uptau)
    = \left[\frac{(-8\beta)^{g-1}}{\pi g!} \log \left(\frac{1}{2\beta\uptau^2}\right) + \text{(finite)} \right] \uptau^{2g+1}.
    \label{eq:RMTrepeat}
\end{equation}
The exact RMT expression is derived from the same coefficients $d_n^{(\delta)}$, but it extracts only their finite piece. The polynomially divergent contributions cancel exactly. The reason is laid out in detail in Appendix \ref{app:derivations}, but we wish to summarize it briefly: the Laplace transform is performed by a contour integral around the cut of a suitable discontinuous function, which extends along an interval of real energies $E$. The contour is then deformed and ultimately yields two contributions: an integral of a discontinuity {\it along the cut} (${\cal C}_1^{(\delta)}$ in figure \ref{fig:contours2}) and an integral along a small circle {\it around the branch point} $E=0$ (${\cal C}_2^{(\delta)}$ in figure \ref{fig:contours2}). The former is precisely the same integral that features in the periodic orbit calculation; the latter is an additional contribution that cancels all scheme-dependent divergences to produce a finite result.

The encounter result \eqref{eq:encounterDivergence} is only an effective estimate and there is no reason for it to be independent of the IR cutoff. However, the gravitational result \eqref{eq:gravityDivergence} should in principle match the finite RMT expression \eqref{eq:RMTrepeat} after summing over geometries.
This suggests that the divergences in the gravitational calculation have to resum into a finite expression. In particular, the resummation must turn $\log(T)$ into $\log(\uptau^{-2})$ \cite{Weber:2024ieq}. The finiteness of the Laplace transformed RMT expression thanks to contour deformation suggests that there might exist a prescription to improve the gravitational path integral calculation: there might exist a different contour in the (analytically continued) moduli space of two-boundary ribbon graphs along which the $\uptau$-scaled path integral is finite for any fixed $g$. It would be very interesting to find such a prescription, which would yield \eqref{eq:RMTrepeat} directly, genus by genus.

\paragraph{UV/IR relations.} We wish to give a different perspective on the structure of $\uptau$-scaling divergences and how they reflect properties of the spectral curve.
As we discussed in section \ref{sec:GOESFF}, all information about the spectral curve required to construct the $\uptau$-scaled genus expansion was encoded in the coefficients $c_{2g}$ and $d_{n}^{(\delta)}$, which we reproduce here:
\begin{equation}
    \begin{split}
     c_{2g}(\rho_0;\beta) &\equiv \frac{1}{(2\pi)^{2g}} \oint \frac{dE}{2\pi i}  \, e^{-2\beta E} \, \rho_0(E)^{-2g} \,,\\
        d_{n}^{(\delta)}(\rho_0;\beta) &\equiv \frac{1}{(2\pi)^n} \int_\delta^\infty dE \, e^{-2\beta E} \, \rho_0(E)^{-n}\,.
    \end{split}
\end{equation}
and $d_n(\rho_0;\beta)$ entering the SFF is the finite part of $d_{n}^{(\delta)}(\rho_0;\beta)$ as $\delta \rightarrow 0$.

We emphasize the following property of these coefficients, which can be observed in examples (e.g., \eqref{eq:AiryDeltaDiv}) as well as in the general analysis of divergences (Appendix \ref{app:derivations}): the coefficients are not independent! The integrals $d_{n}^{(\delta)}$ are divergent as $\delta\rightarrow 0$. They exhibit both powerlaw divergences (for any genus) and a logarithmic divergence (for integer genus), in addition to the finite piece that defines $d_n$. The powerlaw divergences are a non-universal artifact of working with the IR cutoff regulator. But the coefficient of the logarithmic divergence in $d_{2g}$ is scheme-independent and it captures precisely the low-energy coefficient $c_{2g}$:
\begin{equation}
\begin{split}
    c_{2g}(\rho_0;\beta) &= \,\text{coeff}_{\log(1/\delta)} \big( d_{2g}^{(\delta)}(\rho_0;\beta) \big) \,,\\
    d_{2g}(\rho_0;\beta) &= \,\text{coeff}_{\delta^0} \big( d_{2g}^{(\delta)}(\rho_0;\beta) \big) \,.
\end{split}
\end{equation}
This holds not only for the Airy model, but for any spectral curve.

A similar relation holds in terms of the analytic regulator $g \rightarrow g-\varepsilon$, see \eqref{eq:cdrelation}. This scheme is minimal (analogous to dimensional regularization) in the sense that it does not lead to any non-universal divergences and the universal $\frac{1}{\varepsilon}$ divergence produces the meaningful coefficient $c_{2g}$:
\begin{equation}
\begin{split}
    c_{2g}(\rho_0;\beta) &= \text{coeff}_{\varepsilon^{-1}} \big( d_{2(g-\varepsilon)}(\rho_0;\beta) \big) \,,\\
    d_{2g}(\rho_0;\beta) &= \,\text{coeff}_{\varepsilon^0} \big( d_{2(g-\varepsilon)}(\rho_0;\beta) \big) \,.
\end{split}
\end{equation}
We propose to think of these as ``UV/IR relation'' similar to those appearing in dispersion relations.\footnote{Indeed, the derivation in Appendix \ref{app:derivations} is reminiscent of techniques appearing in the study of dispersion relations and UV constraints on effective field theory, see, for example, \cite{Adams:2006sv} and recent studies such as \cite{Bellazzini:2021oaj,Herrero-Valea:2022lfd}.} The $c_{2g}$ coefficients encode universal information about the strict IR limit of the density of states; on the other hand, the $d_{2g}$ coefficients encode information from all energy scales (projected onto the IR). In the general $\uptau$-scaled SFF for the GOE, \eqref{eq:generalA}, we can view the $B_g$ coefficients multiplying $\log(\uptau)$ as containing only UV information -- these terms are entirely associated with the ``plateau''. The other coefficients ($A_g$ and $C_{\tilde g}$) contain both UV and IR information.


\section{Conclusion}
\label{sec:conclusion}

The $\uptau$-scaling limit has been proposed as a way to study universal RMT behavior within a convergent topological expansion. In the first part of this paper, we derived expressions for canonical $\uptau$-scaled spectral form factors in non-orientable matrix models with generic spectral curve. We found that these offer qualitatively new features compared to the orientable case, and ultimately also admit a convergent topological expansion. In the second part of the paper, we studied aspects of path integrals in non-orientable gravitational theories. We used the Airy model as a convenient example that does not exhibit moduli space divergences, thus allowing us to focus on implementing topological recursion in detail and studying its implications for $\uptau$-scaling.

We observed, and conjectured, an intricate pattern of cancellations among coefficients defining the Weil-Petersson volumes for non-orientable ribbon graphs (Airy model). These structures call for a detailed and mathematically rigorous understanding. We expect that this will require new insights, such as a non-orientable generalization of intersection numbers.

We reiterated a feature of the gravitational path integral computation of the late-time spectral form factor (c.f. \cite{Weber:2024ieq}): the $\uptau$-scaling limit of individual geometries with fixed-genus is divergent in the non-orientable setting. Some of these divergences survive the cancellations in the WP volumes and obstruct $\uptau$-scaling at fixed-genus. We gave an indirect argument that these divergences must cancel upon all-order resummation of the topological expansion. However, an explicit mechanism for this resummation within the canonical ensemble remains to be uncovered. It would also be interesting to generalize this analysis to JT gravity: our expression \eqref{eq:JTGOEexpandedResult} suggests that the fixed-genus $\uptau$-scaling divergences proliferate in JT gravity and more terms need to be resummed in order for the gravitational path integral to reproduce the universal RMT results.

A possible approach to this problem is as follows. In \cite{Blommaert:2023vbz}, the authors have provided a Lorentzian calculation of the microcanonical spectral form factor for the GUE universality class. The Lorentzian calculation seemingly circumvents the need for intricate cancellations inherent in the Euclidean calculation, though perhaps it is still interesting to understand the role played by the mathematical structure responsible for these cancellations \cite{Blommaert:2022lbh}. In moving to the more generic non-orientable cases, one alternative to the Euclidean approach would be to utilize a similar Lorentzian calculation to reproduce the microcanonical SFF. This would require placing the Lorentzian conical singularities of \cite{Louko:1995jw} on nonorientable surfaces (whereas Euclidean defects are related to changes in the dilaton potential, or equivalently to the spectral density). One could then expect a match between the Lorentzian spacetimes and periodic orbits in the boundary theory, term by term in  the $\uptau$-expansion.

In section \ref{sec:periodicorbits} we commented on the close relation between high- and low-energy spectral information and how they enter in the $\uptau$-scaled SFF. These observations resonate with the phenomenon known as Riemann-Siegel lookalike \cite{Berry_1990,Bogomolny_1992} (see \cite{Winer:2023btb} for a recent discussion). It would be interesting to connect this to our discussion in section \ref{sec:perOrbitsGravity} and see if the plateau can be obtained using only the input from the ramp that we already reproduced from gravity.

\vspace{20pt}

\paragraph{Acknowledgments:}
The authors are grateful to Andreas Blommaert, Jan Boruch, Fabian Haneder, Klaus Richter, Jarod Tall, Juan Diego Urbina, Torsten Weber, Cynthia Yan, Zhenbin Yang, and Shunyu Yao for helpful discussions.
AE and FMH are supported by UK Research and Innovation (UKRI) under the UK government’s Horizon Europe Funding Guarantee EP/X030334/1. GDU is supported by Japan Science and Technology Agency (JST) as part of Adopting Sustainable Partnerships for Innovative Research Ecosystem (ASPIRE), Grant Number JPMJAP2318. MR is supported by a Discovery grant from NSERC.


\newpage

\appendix

\section{Derivation of $\uptau$-scaled topological expansion}
\label{app:derivations}

In this appendix we give detailed derivations of the $\uptau$-scaled topological expansions for general spectral curves. The template is the derivation given in section \ref{sec:GUEreview} for GUE. Here we discuss the derivation of \eqref{eq:GOEgeneral} (for GOE) and \eqref{eq:GSEgeneralResult} (for GSE).


\subsection{GOE: Derivation of eq. \eqref{eq:GOEgeneral}} 

\begin{figure}[h]
\begin{center}\includegraphics[width=.6\textwidth]{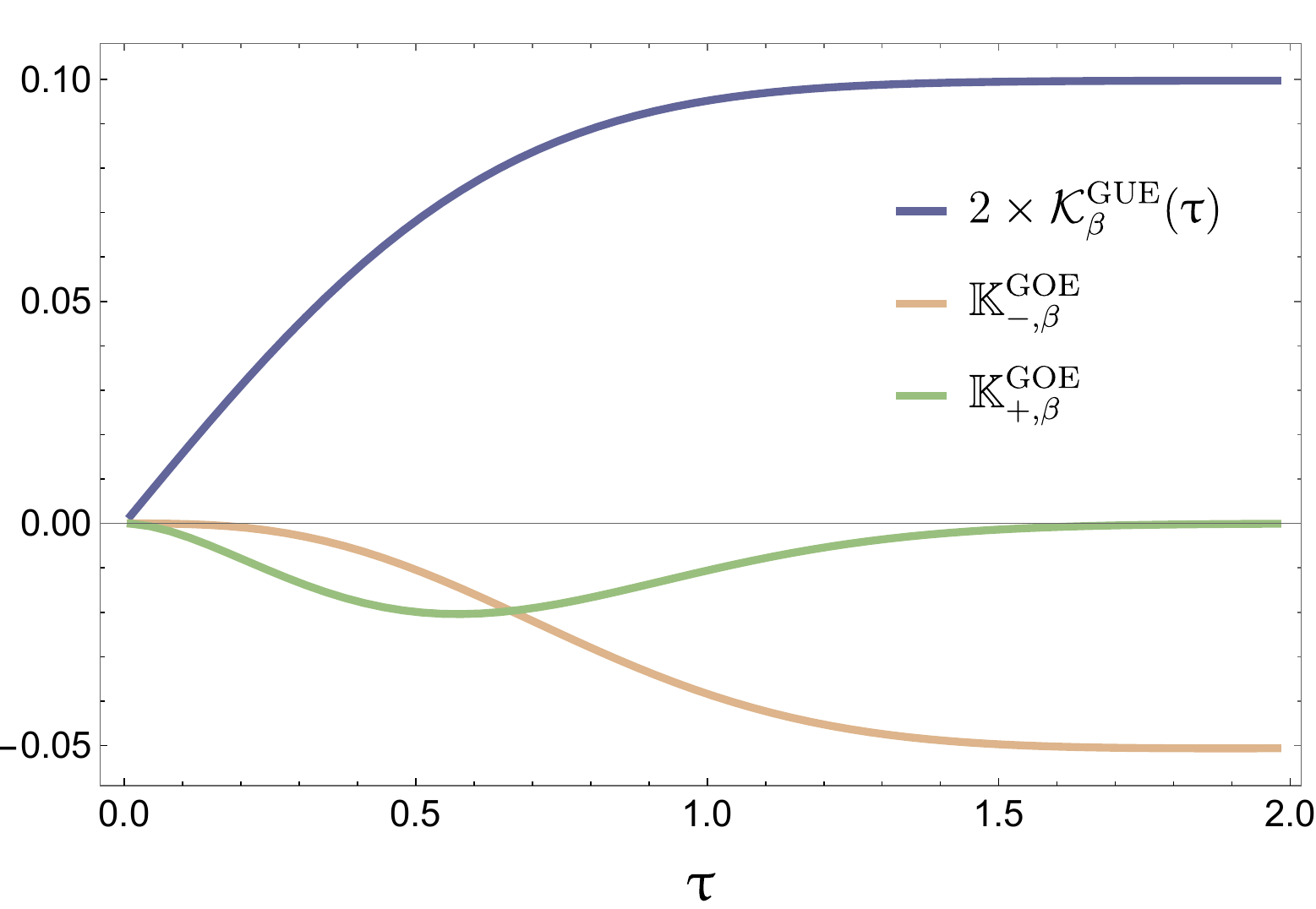}
    \end{center}
    \caption{Illustration of the physical significance of the three pieces in \eqref{eq:GOEdecomposeApp} for the GOE Airy model (for $\beta=1$). Each piece plateaus, but only their sum approaches the correct plateau height.}
\label{fig:airyDecompose}
\end{figure}

In the GOE universality class, there are non-trivial integrals for both the low- and the high-energy parts of the Laplace transform. We recall the split \eqref{eq:GOEdecompose}:
\begin{equation}
\begin{split}
 {\cal K}_{\beta}^{{\text{GOE}}}(\uptau) &= 
 2 \times {\cal K}_{\beta}^{{\text{GUE}}}(\uptau)  \\
 &\quad  \underbrace{-\,\frac{\uptau}{2\pi}\int_0^{E_*} dE \, e^{-2\beta E}  \, \log \left( \frac{ \frac{\uptau}{\pi} + \rho_0(E)}{\frac{\uptau}{\pi} - \rho_0(E)} \right) }_{ {\equiv \ \Khighlow^{{\text{GOE}}}_{\lowElabel,\beta}}} \,\underbrace{-\,\frac{\uptau}{2\pi}\int_{E_*}^\infty dE \, e^{-2\beta E} \,  \log \left( 1 + \frac{\uptau}{\pi \rho_0(E)} \right)}_{ {\equiv \ \Khighlow^{{\text{GOE}}}_{\highElabel,\beta}}} \,.
 \end{split}
 \label{eq:GOEdecomposeApp}
\end{equation}
The three pieces are illustrated for the Airy model in figure \ref{fig:airyDecompose}.
In the following, we bring the  integrals $\Khighlow^\text{GOE}_{\lowElabel,\beta}$ and $\Khighlow^\text{GOE}_{\highElabel,\beta}$ into a simple form to express the coefficients of the topological expansion, following a similar strategy as in the GUE case. The main physical novelty is that the coefficients in the topological expansion are no longer only sensitive to the analytic structure of $\rho_0(E)$ near $E=0$, but instead we encounter integrals along all $E\geq 0$.

\subsubsection{Low energy integral $\Khighlow^{{\text{GOE}}}_{\lowElabel,\beta}$}

We start with the low energy integral. The analysis is similar as for the GUE case and we will therefore be brief.
We again define a function that has a prescribed discontinuity along the branch cut $0\leq \rho_0 \leq \tfrac{\uptau}{2\pi}$:
\begin{equation}
\label{eq:f1def}
f_1(\rho_0)= \frac{\uptau}{2 \pi}\int_0^{\frac{\uptau}{2\pi}} \frac{d\rho'}{2\pi i} \, \frac{1 }{\rho'-\rho_0} \,\log (\frac{\frac{\uptau}{\pi}+\rho' }{\frac{\uptau}{\pi}-\rho'})\,.
\end{equation}
The discontinuity is by construction the integrand relevant for $\Khighlow^{{\text{GOE}}}_{\lowElabel,\beta}$. We find:
\begin{equation}
\begin{split}
    f_1(\rho_0) &= \frac{\uptau}{4\pi^2 i} \bigg[
    \text{Li}_2\left(\frac{z-1}{z-2}\right) - \text{Li}_2\left(\frac{z-1}{z+2} \right)+\text{Li}_2\left(\frac{z}{z+2} \right) - \text{Li}_2\left( \frac{z}{z-2}\right)\\
    &\qquad\qquad + \log \left( \frac{z-1}{z} \right) \log \left(\frac{2+z}{2-z} \right) 
    \bigg]_{z = \frac{2\pi \rho_0}{\uptau}} 
\end{split}
\end{equation}
This function allows us to turn the real integral into a contour integral which can be deformed away from the original interval. Taking $|\rho_0|$ large as in the GUE case, we can expand $f_1(\rho_0)$ in powers of $\uptau$. For illustration, note that these operations commute: the $\uptau$-expansion can also be obtained by first expanding the integrand in \eqref{eq:f1def} and then integrating term by term:  
\begin{equation}
f_1(\rho_0)=-\frac{\uptau}{2\pi }\sum_{n=0}^\infty \frac{1}{\rho_0^{n+1}} \int_0^{\frac{\uptau}{2\pi}} \frac{d\rho'}{2\pi i} \, (\rho')^n \,\log (\frac{\frac{\uptau}{\pi}+\rho' }{\frac{\uptau}{\pi}-\rho'})
=- \frac{1}{2\pi i} \sum_{n=1}^\infty a_n \, \frac{\left(\frac{\uptau}{2\pi}\right)^{n+1}}{\rho_0^n} \,,
\end{equation}
where the coefficients in the expansion are revealed to be the following `moments':
\begin{equation}
a_n = \int_0^1 dx \, x^{n-1} \log (\frac{2+x}{2-x})
   =  \frac{\log (3)}{n} - \frac{1}{n(n+1)} \, {}_2F_1 \left( 1 , \, \frac{n+1}{2} ,\, \frac{n+3}{2} ; \, \frac{1}{4} \right)\,.
\end{equation}
Finally we deform the contour again for each term to a small circle near $E=0$, as in the GUE case (see figure \ref{fig:contours}), and find
\begin{equation}
\label{eq:cnDef}
\begin{split}
\Khighlow^{{\text{GOE}}}_{\lowElabel,\beta} &= \oint_{[0,E_*]} dE \, e^{-2 \beta E}\,\big(f_1(\rho_0) + f_1(-\rho_0) \big)
\\
&= -\frac{1}{2\pi}\sum_{n=1}^\infty a_n \, \left(c_n(\rho_0;\beta)+c_n(-\rho_0;\beta)\right)\,  \uptau^{n+1}\,,
 \end{split}
\end{equation}
with $c_{n}(\rho_0;\beta)$ defined in \eqref{eq:PgDef}. As in the GUE case, the two terms in the sum cancel for odd $n$ and we are left with a sum over even $n=2g$:
\begin{equation}
\Khighlow^{{\text{GOE}}}_{\lowElabel,\beta} = -\sum_{g\geq 1} \frac{1}{2\pi g(2g+1)} \left[ \left(2g+1\right) \log(3)  - {}_2F_1 \left( 1 , \, g+\frac{1}{2} ,\, g+\frac{3}{2} ; \, \frac{1}{4} \right)\right]  c_{2g}(\rho_0;\beta)\; \uptau^{2g+1}\,.
\label{eq:GOEgeneralLowResult}
\end{equation}
Note that the coefficient outside of the square bracket is precisely two times the one appearing in the GUE case, see \eqref{eq:PgDef}.

\subsubsection{High energy integral $\Khighlow^{{\text{GOE}}}_{\highElabel,\beta}$}
\label{sec:goeHigh}

The new high energy integral is associated with the GOE ramp and features a non-compact integration region, making the manipulations more subtle. To work with a compact integration region, let us introduce a UV cutoff $\Lambda \rightarrow \infty$. 
To turn $\Khighlow^{{\text{GOE}}}_{\highElabel,\beta}$ into a contour integral we need to find a function $f_2^{(\Lambda)}(\rho_0)$
which has a cut for $E\in[E_*,\Lambda]$ with discontinuity 
$-\frac{\uptau}{2\pi}\log\big(1+\frac{\uptau}{\pi \rho_0}\big)$. 
Given such a function, we could then write:
\begin{equation}
\label{eq:IhighGeneral}
\Khighlow_{\highElabel,\beta}^{{\text{GOE}}} = \lim_{\Lambda\rightarrow\infty} \; \oint_{[E_*,\Lambda]} dE \, e^{-2\beta E} \left( f_2^{(\Lambda)}(\rho_0(E))+f_2^{(\Lambda)}(-\rho_0(E)) \right)\,,
\end{equation}
where the contour wraps the interval $[E_*,\Lambda]$ counterclockwise.
Following the same prescription as before, we find an integral expression for $f_2^{(\Lambda)}(\rho_0)$ that can be evaluated exactly:
\begin{equation}
    \begin{split}
        f_2^{(\Lambda)}(\rho_0)
        &=\frac{\uptau}{2\pi}\int_{\frac{\uptau}{2\pi}}^\Lambda \frac{d\rho'}{2\pi i} \, \frac{1 }{\rho'-\rho_0} \,\log \left(1 + \frac{\uptau}{\pi \rho'}\right) \\
        &= \frac{\uptau}{4\pi^2 i} \bigg[
        \text{Li}_2\left(\frac{z-1}{z+2}\right) - \text{Li}_2\left(\frac{z-\Lambda}{z+2} \right) - \text{Li}_2\left( \frac{z-1}{z} \right) + \text{Li}_2\left( \frac{z-\Lambda}{z} \right) \\
        &\qquad\qquad + \log \left( \frac{z+2}{z} \right) \log \left( \frac{z-\Lambda}{z-1} \right) \bigg]
    \end{split}
\end{equation}
One should keep the regulator and only take $\Lambda \rightarrow \infty$ in the end. However, with the benefit of hindsight, we note that this limit does not lead to any subtleties. We will therefore take the limit now, and -- in slight abuse of notation -- treat $E=\infty$ as a point which contours can wrap around. We can then consider the simplified function
\begin{equation}
\label{eq:fGOEcut}
\begin{split}
f_2\left(\rho_0\right)
& = \lim_{\Lambda \rightarrow \infty} \, f^{(\Lambda)}_2\left(\rho_0\right) =  \frac{\uptau}{4\pi^2 i} \left[
\text{Li}_2\left(\frac{2+z}{3}\right) -\text{Li}_2(z)-\log( 3) \log (1-z) 
+\frac{1}{2} (\log 3)^2\right]_{z = \frac{2 \pi \rho_{_0}}{\uptau}}
\end{split}
\end{equation}
This function has a cut along $z>1$ as expected. The discontinuity across the cut comes from the combined discontinuities of the dilogarithms and  the $\log(1-z)$. These combine to the correct discontinuity $-\frac{\uptau}{2\pi}\log(1+\frac{2}{z})$. 
The function $f_2(\rho_0)$ is only defined up to an additive constant, which does not contribute to the expansion below. The high-energy integral becomes:
\begin{equation}
\label{eq:IhighGeneralSimplified}
\Khighlow_{\highElabel,\beta}^{{\text{GOE}}} = \oint_{[E_*,\infty)} dE \, e^{-2\beta E} \left( f_2(\rho_0(E))+f_2(-\rho_0(E)) \right)\,,
\end{equation}
where the contour wraps around $[E_*,\infty)$ counterclockwise.

\begin{figure}
\begin{center}\includegraphics[width=.9\textwidth]{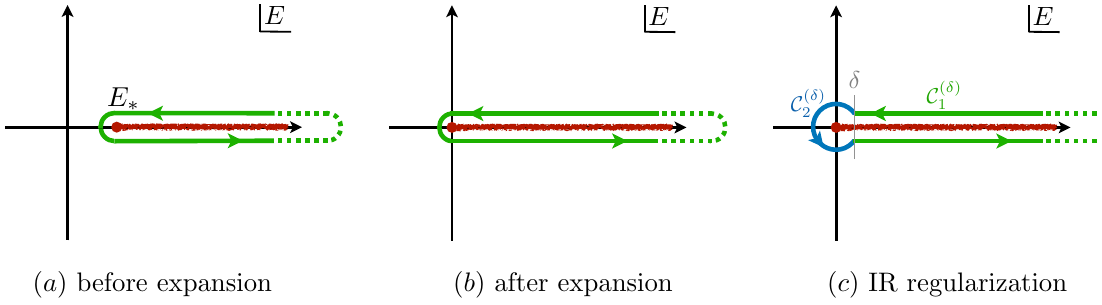}
    \end{center}
    \caption{Illustration of the contour deformation: $(a)$ we begin with a contour wrapping the interval $[E_*,\infty)$; $(b)$ we extend the contour around $E=0$ such as to avoid all singularities encountered after expanding the function $f_2\left(\rho_0\right)+f_2\left(-\rho_0\right)$; $(c)$ finally we treat the (divergent) integral around $E=0$ separately from the integral of the discontinuity.}
\label{fig:contours2}
\end{figure}

Our goal is to extract from this the topological expansion, i.e., a power series in $\uptau$. We would therefore like to expand the function that enters in the contour integral:\footnote{ To obtain this expansion, one needs to be careful about the branch cuts of $f_2(\rho_0)+f_2(-\rho_0)$. In practice, one can first write this function as indicated by \eqref{eq:fGOEcut}; then one uses the dilogarithm reflection identity to bring every term in a form where $z>0$ does not collide with the branch cut of $\text{Li}_2(\,\cdot\,)$. This avoids any branch cut ambiguities and leads to an expression that can be expanded in large $z$.}
\begin{equation}
\label{eq:GOEhighTemp}
\begin{split}
f_2(\rho_0)+f_2(-\rho_0) &= \frac{\uptau}{4\pi^2 i}\sum_{n\geq 1} \Bigg\{ \bigg[\frac{4^{-n}}{n} \, \Phi\left(-\frac{1}{2},1,2n\right)+\frac{1}{2n^2} -\frac{1}{2n}\left(\log(4)-
4^{-n} 
\log(9)\right)\\ 
&\qquad\qquad\quad + \frac{1}{2n}\left(\log\left(\frac{2 \pi \rho_0}{\uptau} \right) +\log\left(-\frac{2 \pi\rho_0}{\uptau}\right) \right) \bigg]  \left( \frac{\uptau}{\pi\rho_0} \right)^{2n} \\
&\qquad\qquad\quad +   
\frac{1}{2 n-1}\left(\log\left(\frac{2 \pi \rho_0}{\uptau}\right) - \log\left(-\frac{2 \pi\rho_0}{\uptau}\right) \right) \left( \frac{\uptau}{\pi \rho_0} \right)^{2n-1} \Bigg\}
\end{split}
\end{equation}
Note that performing this expansion inside the contour integral \eqref{eq:IhighGeneralSimplified} is subtle. The reason is that the term-by-term analytic structure of \eqref{eq:GOEhighTemp} is different than the analytic structure of the original function $f_2(\rho_0)+f_2(-\rho_0)$. Before expanding, the original contour for the energy integration is a loop wrapping the $E$-interval $[E_*,\infty)$, see figure \ref{fig:contours2}(a). This must be deformed such as to avoid all singularities that appear after expanding. It turns out that this makes the integration contour independent of $E_*$ (see figure \ref{fig:contours2}(b)), thus making all time dependence explicit. Since we are expanding in inverse powers of $\rho_0$, we have to choose a contour at large energy. We can then deform that contour at infinity, for each term separately. Each of the terms in the expansion has different analytic properties which determine the type of contour deformation needed to evaluate the energy integrals. While in the GUE case all terms in the expansion were even powers of $\rho$ leading to meromorphic functions of $E$, we now also have odd powers as well as logarithms which lead to functions with cuts. 

More conceptually, these observations mean that the analysis is now sensitive to features of the density of states at all energies, not only their expansion near $E=0$. This can be expected on general grounds. In the GUE case the semi-classical genus expansion at fixed (high) energy is trivial and all higher genus corrections in the canonical ensemble come from low energies. For the case of the GOE, every energy window in the Laplace transform already has a non-trivial genus expansion, and therefore all energy windows contribute to any fixed-genus term.

We proceed by considering the three types of terms in \eqref{eq:GOEhighTemp} separately and plugging each of them into \eqref{eq:IhighGeneralSimplified}.
First, let us choose the branch cut of $\rho_0(E)$ (see \eqref{eq:rhoGeneral}). It turns out to be most convenient to choose the branch cut of $\sqrt{E}$ along $E \in [0,\infty)$. The branch cut of $\log(z)$ is taken along $z\in (-\infty,0]$ as usual. With this choice, we have the following branch cut discontinuities:
\begin{equation}
    \begin{split}
        \text{Disc}_E \, \log\left(\frac{2\pi \rho_0(E)}{\uptau} \right) &= 2\pi i = \text{Disc}_E \, \log\left(-\frac{2\pi \rho_0(E)}{\uptau} \right)  \qquad \text{for } E \in [0,\infty)\,.
    \end{split}
\end{equation}

\paragraph{(1)\ Odd powers of $\uptau$.}
The first type of contribution in \eqref{eq:GOEhighTemp} is of the form
\begin{equation}
\label{eq:GOEhighTemp2}
\begin{split}
f_2(\rho_0)+f_2(-\rho_0) \supset \frac{\uptau}{4\pi^2 i}\sum_{g\geq 1} \left[\frac{4^{-g}}{g} \, \Phi\left(-\frac{1}{2},1,2g\right)+\frac{1}{2g^2} -\frac{1}{2g}\left(\log(4)-4^{-g}\log(9)\right) \right]  \left( \frac{\uptau}{\pi \rho_0} \right)^{2g}\,.
\end{split}
\end{equation}
This is of the same form as in the GUE case. As there, it necessitates a deformation of the contour wrapping $[E_*,\infty)$ into one that includes $E=0$. Subsequently, the expansion \eqref{eq:GOEhighTemp2} is allowed inside the integral and leads to the following contribution:
\begin{equation}
\label{eq:GOEhigh2b}
\Khighlow_{\highElabel,\beta}^{{\text{GOE}}} \supset 
\Khighlow_{\highElabel,\text{odd},\beta}^{{\text{GOE}}}\equiv
\frac{1}{2\pi}\sum_{g=1}^{\infty} \left[\frac{1}{g} \, \Phi\left(-\frac{1}{2},1,2g\right)+\frac{4^g}{2g^2} -\frac{4^g}{2g}\left(\log(4)-4^{-g}\log(9)\right) \right] c_{2g}(\rho_0;\beta) \, \uptau^{2g+1}
\end{equation}
where $c_n(\rho_0;\beta)$ was defined in \eqref{eq:cnDef} and captures the behavior of $\rho_0$ near $E=0$.

\paragraph{(2)\ Even powers of $\uptau$ multiplying logarithms.}
The even powers of $\uptau$ in the expansion are of the form
\begin{equation}
\label{eq:GOEtemp1}
    f_2\left(\rho_0\right)+f_2\left(-\rho_0\right) \supset \frac{\uptau}{4\pi^2 i}\sum_{n\geq 1} \frac{1}{2 n-1}\left[\log\left(\frac{2 \pi\rho_0}{\uptau}\right) -\log\left(-\frac{2 \pi\rho_0}{\uptau}\right) \right]\left(\frac{\uptau}{\pi \rho_0}\right)^{2n-1} \,.
\end{equation}
 Before implementing this expansion, we must deform the $E$-contour in \eqref{eq:IhighGeneralSimplified} such as to avoid the branch cut of \eqref{eq:GOEtemp1}. 
 Using the choices described above, the branch cut extends along $E \in [0,\infty)$. We are therefore required to extend the original contour wrapping $[E_*,\infty)$ to a new contour that wraps $[0,\infty)$, see figure \ref{fig:contours2}(b). With our choice of branch cuts, the difference of logarithms is constant everywhere along the contour:
 \begin{equation}
     \log\left(\frac{2 \pi\rho_0(E)}{\uptau}\right) -\log\left(-\frac{2 \pi\rho_0(E)}{\uptau}\right)
     = -\pi i\,.
 \end{equation}
The contour integral receives non-vanishing contributions due to the odd power of $\rho_0(E)$, which again has a branch cut along $[0,\infty)$. We split the integration contour into two pieces, ${\cal C}_1^{(\delta)} \cup {\cal C}_2^{(\delta)}$: the first piece consists of integrating the discontinuity across the branch cut along $[\delta,\infty)$, where $\delta$ is a small infrared regulator. The second piece consists of the  integration around $E=0$ along a circle of radius $\delta$. See figure \ref{fig:contours2}(c) for illustration.

\noindent $(i)$ The first contribution arises from integrating the discontinuity encountered in odd powers of $\rho_0(E)$ along ${\cal C}_1^{(\delta)}$:
\begin{equation}
    \text{Disc}_E \, \left(\frac{\uptau}{\pi \rho_0(E)}\right)^{2n-1} = 2\left(\frac{\uptau}{\pi \rho_0(E)}\right)^{2n-1}\,, \qquad  E \in [0,\infty) \,.
\end{equation}
Plugging this into the (IR-regulated) energy contour integral \eqref{eq:IhighGeneralSimplified}, we get a contribution:
\begin{equation}
\label{eq:GOEhigh1}
\begin{split}
\Khighlow_{\highElabel,\beta}^{{\text{GOE}}}\supset 
\Khighlow_{\highElabel,\text{even-A},\beta}^{{\text{GOE}}} &\equiv -\sum_{n=1}^{\infty} \frac{4^{n-1}}{\pi(2 n-1)} \,d_{2n-1}^{(\delta)}(\rho_0;\beta)\, \uptau^{2 n}\,,\\
\text{where}\quad d_{2n-1}^{(\delta)}(\rho_0;\beta) &\equiv  \int_\delta^{\infty} dE \, e^{- 2 \beta E} \,\frac{1}{(2\pi\rho_0(E))^{2n-1} }\,.
\end{split}
\end{equation}
Note that the integrals $d_{2n-1}^{(\delta)}$ are divergent as $\delta\rightarrow 0$. We return to this issue momentarily.

\noindent $(ii)$ The second contribution is due to integration along the small circle around $E=0$ at a regulated radius $|E|=\delta$, called ${\cal C}_2^{(\delta)}$ in figure \ref{fig:contours2}:
\begin{equation}
\label{eq:GOEhigh1b}
\begin{split}
\Khighlow_{\highElabel,\beta}^{{\text{GOE}}} \supset 
\Khighlow_{\highElabel,\text{even-B},\beta}^{{\text{GOE}}} &\equiv
-\sum_{n=1}^{\infty} \frac{4^{n-1}}{2\pi(2 n-1)} \, r_{2n-1}^{(\delta)} \, \uptau^{2n} \,,\\
\text{where} \quad r_{2n-1}^{(\delta)} &\equiv  \int_{{\cal C}_2^{(\delta)}} dE \, e^{-2\beta E}\, \frac{1}{(2\pi\rho_0(E))^{2n-1}}
\end{split}
\end{equation}
The integrals $r_{2n-1}^{(\delta)}$ are also divergent as $\delta \rightarrow 0$. We claim that the divergences cancel between \eqref{eq:GOEhigh1} and \eqref{eq:GOEhigh1b}. To see this, note that the integrand is the same in both cases and has a small energy expansion of a particular form:
\begin{equation}
   \frac{e^{-2\beta E}}{(2\pi\rho_0(E))^{2n-1}} =  E^{\frac{1}{2}-n} \left(1 + b_1 E+ b_2 E^2 + \ldots \right)\,.
\end{equation}
To extract the divergent pieces, this small-$E$ expansion is sufficient. The anti-derivative in this limit is:
\begin{equation}
  \int dE \,  \frac{e^{-2\beta E}}{\rho_0(E)^{2n-1}}
  = E^{\frac{3}{2}-n} \left( \frac{1}{\frac{3}{2}-n} + \frac{b_1 E}{\frac{5}{2}-n} + \frac{b_2 E^2}{\frac{7}{2}-n} + \ldots \right)
\end{equation}
This can be used to evaluate both \eqref{eq:GOEhigh1b} and the divergent part of \eqref{eq:GOEhigh1}. It is then obvious that both $d_{2n-1}^{(\delta)}$ and $r_{2n-1}^{(\delta)}$ have divergences of the form $\delta^{-a}$ for $a = n-\frac{3}{2}, \ldots , \frac{1}{2}$. It is also clear that these divergences cancel precisely between $d_{2n-1}^{(\delta)}$ and $r_{2n-1}^{(\delta)}$.\footnote{To confirm the cancellation of divergences, note two subteleties. First, $d_{2n-1}^{(\delta)}$ only has a single divergent boundary term, while $r_{2n-1}^{(\delta)}$ receives an identical divergence from both ends of the integration contour. This is consistent because the coefficient of the two integrals also differs by a factor 2, c.f., \eqref{eq:GOEhigh1} and \eqref{eq:GOEhigh1b}. Second, the divergence in $r_{2n-1}^{(\delta)}$ naively appears to have the wrong sign. However, the sign works out correctly if one carefully accounts for our choice of branch cut of $\sqrt{E}$ along $[0,\infty)$.} To find the finite piece of \eqref{eq:GOEhigh1}, one has to go beyond the small-$E$ expansion.

In summary, we see that all divergences cancel and we receive a finite contibution to the spectral form factor. The finite piece comes entirely from $d_{2n-1}^{(\delta)}$ and we can therefore give a simplified prescription:
\begin{equation}
\label{eq:GOEhighFinal1}
\begin{split}
\Khighlow_{\highElabel,\beta}^{{\text{GOE}}}\supset 
\Khighlow_{\highElabel,\text{even},\beta}^{{\text{GOE}}} &\equiv 
\Khighlow_{\highElabel,\text{even-A},\beta}^{{\text{GOE}}} + \Khighlow_{\highElabel,\text{even-B},\beta}^{{\text{GOE}}} \\
&= -\sum_{n=1}^{\infty} \frac{4^{n-1}}{\pi(2 n-1)} \,d_{2n-1}(\rho_0;\beta)\, \uptau^{2 n}\,,\\
\text{where}\quad d_{2n-1}(\rho_0;\beta) &\equiv \frac{1}{(2\pi)^{2n-1}} \left[\int_\delta^{\infty} dE \, e^{- 2 \beta E} \,\frac{1}{\rho_0(E)^{2n-1} } \right]_\text{finite}\,.
\end{split}
\end{equation}
The notation simply means that the integral is to be evaluated with an IR cutoff, but only the finite piece is kept. In the final result (second line of \eqref{eq:GOEgeneral}), we leave the prescription to pick the finite piece implicit.
There, we also use a half-integer index $\tilde{g} \equiv n-\frac{1}{2}$ to account for the terms \eqref{eq:GOEhighFinal1}.

\paragraph{(3)\ Odd powers of $\uptau$ multiplying logarithms.}
Finally, \eqref{eq:GOEhighTemp} contains terms of the form 
\begin{equation}
f_2(\rho_0)+f_2(-\rho_0) \supset \frac{\uptau}{4\pi^2 i}\sum_{n\geq 1} \frac{1}{2n}\left[\log\left(\frac{ 2\pi\rho_0}{\uptau} \right) +\log\left(-\frac{2 \pi\rho_0}{\uptau}\right) \right]\left( \frac{\pi \rho_0}{\uptau} \right)^{2n}\,.
\end{equation}
 This gives another series of terms with odd powers of time, namely
 \begin{equation}
\label{eq:GOEhigh3x}
\Khighlow_{\highElabel,\beta}^{{\text{GOE}}}\supset  \sum_{g\geq 1} \frac{4^{g-1}}{\pi g} \, d_{2g}^\text{log}(\rho_0;\beta) \, \uptau^{2g+1}
 \end{equation}
 where we defined 
  \begin{equation} 
 d_{2g}^\text{log}(\rho_0;\beta)= \frac{1}{(2\pi)^{2g}}\oint_{[0,\infty)} \frac{dE}{2 \pi i} \, e^{- 2 \beta E} \,\rho_0(E)^{-2 g} \left[\log\left(\frac{2 \pi\rho_0}{\uptau} \right) +\log\left(-\frac{2 \pi\rho_0}{\uptau}\right) \right] \,.
 \end{equation}
 Note that the integrand has a pole at $E=0$ and a branch cut along $[0,\infty)$.  It is convenient to split off a piece without branch cut containing $\log(\uptau)$:
   \begin{equation} 
 d_{2g}^\text{log}(\rho_0;\beta) =- 2 \, c_{2g}(\rho_0;\beta) \, \log(\uptau) + \frac{1}{(2\pi)^{2g}}\oint_{[0,\infty)} \frac{dE}{2 \pi i} \, e^{- 2 \beta E} \,\rho_0(E)^{-2 g} \left[\log\left(2 \pi\rho_0 \right) +\log\left(-2 \pi\rho_0\right) \right] \,,
 \end{equation}
 where we recognize the term proportional to $\log(\uptau)$ as a contour integral that can be contracted to a small circle around the pole at $E=0$, i.e., an expression involving the coefficient $c_{2g}$. It gives a contribution to the spectral form factor of the form
   \begin{equation}
\label{eq:GOEhigh00}
\begin{split}
\Khighlow_{\highElabel,\beta}^{{\text{GOE}}} \supset 
\Khighlow_{\highElabel,\text{log},\beta}^{{\text{GOE}}} &\equiv - \sum_{g=1}^{\infty} \frac{4^{g}}{2\pi g} \, c_{2g}(\rho_0;\beta)\, \log(\uptau) \, \uptau^{2 g+1}\,.
\end{split}
\end{equation}
The remaining integral has branch cuts and it will be convenient to turn the logarithms into powerlaw dependence via the identity $x^{-2(g-\varepsilon)} = x^{-2g}(1+2\varepsilon \, \log x + \ldots)$. This gives:
 \begin{equation} 
 d_{2g}^\text{log}(\rho_0;\beta) =- 2 \, c_{2g}(\rho_0;\beta) \, \log(\uptau) + \frac{1}{2}\,\lim_{\varepsilon\rightarrow 0} \, \frac{d}{d\varepsilon} \oint_{[0,\infty)} \frac{dE}{2 \pi i} \, e^{- 2 \beta E} \,\left[\left(2 \pi\rho_0 \right)^{-2(g-\varepsilon)} +\left(-2 \pi\rho_0\right)^{-2(g-\varepsilon)} \right] \,,
 \end{equation}
 As before, we split the integration into two piecewise contours ${\cal C}_{1,2}^{(\delta)}$, see figure \ref{fig:contours2}(c). The two contribute as follows:

\noindent $(i)$ The integral along ${\cal C}_1^{(\delta)}$ picks up the discontinuity across the cut, which exists for $\varepsilon>0$ and takes the following form:
 \begin{equation}
\text{Disc}_E\left[(2\pi\rho_0)^{-2(g-\varepsilon)}+(-2\pi\rho_0)^{-2(g-\varepsilon)} \right] 
= -2i \sin(2\pi (g-\varepsilon)) \,|2\pi \rho_0|^{-2(g-\varepsilon)} 
\qquad \text{for } E\in[\delta,\infty).
\end{equation}
 This gives the following contribution to the spectral form factor:
  \begin{equation}
\label{eq:GOEhigh3}
\begin{split}
\Khighlow_{\highElabel,\beta}^{{\text{GOE}}} \supset 
\Khighlow_{\highElabel,\text{odd-A},\beta}^{{\text{GOE}}} &\equiv \sum_{g=1}^{\infty} \frac{4^{g-1}}{\pi g} \, d_{2g}^{(\delta)}(\rho_0;\beta)\, \uptau^{2 g+1}\,,\\
\text{where}\quad d_{2g}^{(\delta)}(\rho_0;\beta) &\equiv -\frac{1}{2\pi}\,\lim_{\varepsilon \rightarrow 0} \, \frac{d}{d\varepsilon}\, \sin(2\pi(g-\varepsilon))\int_\delta^{\infty} dE \, e^{- 2 \beta E} \, |2\pi\rho_0|^{-2(g-\varepsilon)} \\
 &= \int_\delta^{\infty} dE \, e^{- 2 \beta E} \, (2\pi\rho_0)^{-2g} \,.
\end{split}
\end{equation}
As in the previous discussion, the integral is IR-divergent, featuring powers $\delta^{-a}$ for $a=1,\ldots,g-1$. Additionally, there is a $\log(\delta)$ divergence and a finite piece.

\noindent $(ii)$ The contour integral along the small circle around the origin ${\cal C}_2^{(\delta)}$ is
\begin{equation}
\label{eq:GOEhigh4}
\begin{split}
\Khighlow_{\highElabel,\beta}^{{\text{GOE}}} \supset 
\Khighlow_{\highElabel,\text{odd-B},\beta}^{{\text{GOE}}}&\equiv \sum_{n=1}^{\infty} \frac{4^{g-1}}{\pi g} \,  s_{2g}^{(\delta)}(\rho_0;\beta) \, \uptau^{2 g+1}\,,\\
\text{where}\quad s_{2g}^{(\delta)}(\rho_0;\beta) &\equiv \frac{1}{2} \, \lim_{\varepsilon \rightarrow 0 } \, \frac{d}{d\varepsilon}  \int_{{\cal C}_2^{(\delta)}}\frac{dE}{2\pi i} \, e^{- 2 \beta E} \,\left[\left( 2 \pi\rho_0\right)^{-2(g-\varepsilon)}  + \left(- 2 \pi\rho_0\right)^{-2(g-\varepsilon)}  \right] \,.
\end{split}
\end{equation}
As in the previous analysis, this integral produces divergences that precisely cancel those in \eqref{eq:GOEhigh3}. To see this, consider the small energy expansion
\begin{equation}
   \frac{e^{-2\beta E}}{(2\pi\rho_0(E))^{2g}} =  E^{-g} \left(1 + \tilde b_1 E+ \tilde b_2 E^2 + \ldots \right)\,.
\end{equation}
The IR divergences of \eqref{eq:GOEhigh3} evidently are of the form
\begin{equation}
    d_{2g}^{(\delta)}(\rho_0;\beta)
    = \frac{1}{(g-1)\delta^{g-1}} + \frac{\tilde b_1}{(g-2) \delta^{g-2}} + \ldots + \frac{\tilde b_{g-2}}{\delta} - \tilde b_{g-1} \, \log(\delta) + \text{[finite]}
    \label{eq:d2gDiv}
\end{equation}
In \eqref{eq:GOEhigh4} we need to keep the regulator until after integration:
\begin{equation}
    \begin{split}
    s_{2g}^{(\delta)}(\rho_0;\beta)
    & = -\frac{1}{2}\, \lim_{\varepsilon\rightarrow 0}\, \frac{d}{d\varepsilon} \Bigg[\frac{1}{2\pi i} \left(E^{\varepsilon} + (-E)^{\varepsilon} \right) \\
    &\qquad\;\; \times 
    \left(\frac{1}{(g-1)E^{g-1}} + \frac{\tilde b_1}{(g-2)E^{g-2}} 
    + \ldots + \frac{\tilde b_{g-2}}{E}- \frac{1}{\varepsilon}\,\tilde b_{g-1} + {\cal O}(E) \varepsilon^0 + {\cal O}(\varepsilon) \right)
    \Bigg]_{E=\delta+i0}^{E=\delta-i0}
    \end{split}
\end{equation}
As before, our convenient convention is to place the branch  cut of fractional powers along the positive real axis. We thus pick up the following discontinuity:
\begin{equation}
\begin{split}
    s_{2g}^{(\delta)}(\rho_0;\beta)
    &= -\lim_{\varepsilon\rightarrow 0}\, \frac{d}{d\varepsilon} \Bigg[  \left(\varepsilon +\varepsilon^2 \,\log(\delta) + {\cal O}(\varepsilon^3)\right) \\
    &\qquad\quad \times
    \left(\frac{1}{(g-1)\delta^{g-1}} + \frac{\tilde b_1}{(g-2)\delta^{g-2}} 
    + \ldots + \frac{\tilde b_{g-2}}{\delta}- \frac{1}{\varepsilon}\,\tilde b_{g-1} + {\cal O}(\delta) \varepsilon^0 + {\cal O}(\varepsilon) \right)
    \Bigg] \\
    & = -\frac{1}{(g-1)\delta^{g-1}} - \frac{\tilde b_1}{(g-2) \delta^{g-2}} - \ldots - \frac{\tilde b_{g-2}}{\delta} + \tilde b_{g-1} \, \log(\delta) \,,
\end{split}
\end{equation}
thus exactly canceling the divergent terms in \eqref{eq:d2gDiv}.

In summary, we can combine \eqref{eq:GOEhigh3} and \eqref{eq:GOEhigh4} to obtain a finite contribution to the $\uptau$-scaled spectral form factor:
\begin{equation}
\label{eq:GOEhighFinal2}
\begin{split}
\Khighlow_{\highElabel,\beta}^{{\text{GOE}}} \supset 
\Khighlow_{\highElabel,\text{odd},\beta}^{{\text{GOE}}}&\equiv \Khighlow_{\highElabel,\text{odd-A},\beta}^{{\text{GOE}}} + \Khighlow_{\highElabel,\text{odd-B},\beta}^{{\text{GOE}}} \\
&= \sum_{g=1}^{\infty} \frac{4^{g-1}}{\pi g} \,d_{2g}(\rho_0;\beta)\, \uptau^{2 g+1}\,,\\
\text{where}\quad d_{2g}(\rho_0;\beta) &\equiv \frac{1}{(2\pi)^{2g}} \left[\int_\delta^{\infty} dE \, e^{- 2 \beta E} \,\frac{1}{\rho_0(E)^{2g} } \right]_\text{finite}\,.
\end{split}
\end{equation}

\paragraph{Summary.} This completes the derivation. Let us collect all the pieces. The contributions \eqref{eq:GOEgeneralLowResult}, \eqref{eq:GOEhigh2b}, and \eqref{eq:GOEhigh00} combine into terms involving the ``low-energy'' coefficients $c_{2g}(\rho_0;\beta)$. When combining \eqref{eq:GOEgeneralLowResult}+\eqref{eq:GOEhigh2b}, the following identity is useful:
\begin{equation}
    \Phi\left(-\frac{1}{2},1,2g\right)+ \frac{1}{(2g+1)} \; {}_2F_1\left(1,g+\frac{1}{2},g+\frac{3}{2},\frac{1}{4}\right) = \frac{4^g}{2g} \; {}_2F_1(2g,2g,2g+1,-1)\,.
\end{equation}
Furthermore, \eqref{eq:GOEhighFinal1} and \eqref{eq:GOEhighFinal2} yield contributions determined by (the finite part of) the coefficients $d_{2g}(\rho_0;\beta)$ and $d_{2\tilde g}(\rho_0;\beta)$.
Adding up all these contributions, we get the final result \eqref{eq:GOEgeneral}.


\subsection{GSE: Derivation of eq. \eqref{eq:GSEgeneralResult}}
\label{app:GSEderivation}

The low-energy part of the Laplace transform in the GSE is identical to the GUE case, except that the integration limit is now defined by $\rho_0(E_*)= \frac{\uptau}{4\pi}$. Therefore:
\begin{equation}
{\cal K}_\beta^\text{GSE}(\uptau) ={\cal K}_\beta^\text{GUE}\left(\frac{\uptau}{2}\right) \underbrace{-\frac{\uptau}{8\pi} \int_{E_*}^\infty dE \, e^{-2\beta E} \, \log \left| 1 - \frac{\uptau}{2\pi \rho_0(E)} \right|}_{\Khighlow_{\highElabel,\beta}^{{\text{GSE}}}} \,.
\end{equation}
The new high energy (ramp) integral involves $-\frac{\uptau}{8\pi} \log |1-\frac{\uptau
}{2\pi\rho_0}|$. The function $f_3(\rho_0)$ which has this expression as its discontinuity across the cut for $\rho_0 > \frac{\uptau}{4\pi}$ is found using the Stieltjes transform:
\begin{equation}
\begin{split}
f_3\left(\rho_0\right) &= \frac{\uptau}{16 \pi^2 i} \bigg[
\text{Li}_2\left(\frac{1}{2z}\right)-\text{Li}_2\left(\frac{1}{2-2z}\right)+\frac{1}{2}\log( \frac{1}{z})^2+\frac{1}{2}\log(\frac{1}{z-1})^2 \\
&\qquad\qquad\quad +\log(z-1) \log \left(\frac{1}{z}\right) -\log(2-2z) \log\left(\frac{z-1}{z}\right)
\bigg]_{z=\frac{2\pi\rho_0}{\uptau
}}
\end{split}
\label{eq:GSEf3}
\end{equation}
One can indeed check that this function has a branch cut along $z>\frac{1}{2}$. The discontinuity is $\text{Disc}(f_3(\rho_0)) = -\frac{\uptau}{8\pi}\log(\frac{1}{z}-1)$ along $\frac{1}{2} < z < 1$, and $\text{Disc}(f_3(\rho_0)) = -\frac{\uptau}{8\pi}\log(1-\frac{1}{z})$ for $z>1$.

As for GUE and GOE, we symmetrize $f_3(\rho_0)$ and expand in large $z$. This gives:
\begin{equation}
\begin{split}
 f_3(\rho_0) + f_3(-\rho_0) &= \frac{\uptau}{16 \pi^2 i} \sum_{n\geq 1} \Bigg\{ \bigg[  {}_3F_2\left(1,1,1-2n;2,2;\frac{1}{2}\right) -\frac{H_{2n-1}}{n} + \frac{1}{2^{2n+1} n^2}  \\
 &\qquad\qquad\qquad + \frac{1}{2n} \left( \log\left(\frac{4\pi \rho_0}{\uptau}\right)+\log\left(-\frac{4\pi \rho_0}{\uptau}\right)\right) \bigg] \left( \frac{\uptau}{2\pi\rho_0}\right)^{2n}\\
 &\qquad\qquad\qquad - \frac{1}{2n-1} \left( \log\left(\frac{4\pi \rho_0}{\uptau}\right)-\log\left(-\frac{4\pi \rho_0}{\uptau}\right)\right)\left( \frac{\uptau}{2\pi\rho_0}\right)^{2n-1} \Bigg\} \,. 
\end{split}
\end{equation} 
Even though some numerical factors differ, the structure of this expression is the same as in the GOE case, \eqref{eq:GOEhighTemp}. For example, the second line is obtained from \eqref{eq:GOEhighTemp} by replacing $\uptau \rightarrow \frac{\uptau}{2}$ and multiplying the result by an overall factor $\frac{1}{2}$. Similarly for the third line, with an additional minus sign. The first line appears in the given form after analyzing \eqref{eq:GSEf3} term by term; it can be rewritten as follows:
\begin{equation}
    {}_3F_2\left(1,1,1-2n;2,2;\frac{1}{2}\right) -\frac{H_{2n-1}}{n}  + \frac{1}{2^{2n+1} n^2}
    = \frac{1}{2n^2} \big( {}_2F_1\left(2n,2n,2n+1;-1\right) +1 -n\,\log(4) \big)\,.
\end{equation}
We recognize the right hand side as the same combination that appears in $\frac{1}{2}\times A_g^\text{GOE}(\rho_0;\beta)$, see \eqref{eq:generalA}. 
Due to this non-trivial simplification, we can apply an analysis that is essentially identical to the GOE case. This yields the result \eqref{eq:GSEgeneralResult}.

\section{Details on $\uptau$-scaled GOE JT gravity}
\label{app:JT}

In this appendix we give further details on the $\uptau$-scaled JT gravity matrix model in the GOE universality class. Section \ref{app:JTdetails} contains a recursive construction of the high-temperature expansion. Section \ref{app:JTresummation} shows how to resum the high-temperature expansion into a low-temperature expansion.

\subsection{High-temperature expansion}
\label{app:JTdetails}

Here, we develop a method for computing the coefficients appearing in the $\uptau$-scaled SFF for non-orientable (GOE) JT gravity in a small-$\beta$ expansion. Consider the following generalization of the binomial formula ($x$ assumed real):
\begin{equation}
\label{eq:genBinom}
    (1+a)^x  =\sum_{k=0}^\infty
\frac{\Gamma(x+1)}{\Gamma(x-k+1)\Gamma(k+1)} \, a^k \,.
\end{equation}
Note that the expansion coefficients are the analytic continuation of the usual binomial coefficients. Furthermore, if $x$ is a positive integer then for every $k > x$ the numerator has a pole and therefore the series truncates and becomes the usual binomial formula. The series converges if $|a|<1$. Using this, we can write (for any $E$):
\begin{equation} \left[ \frac{1}{4\pi^2}\,\sinh(2\pi\sqrt{E}) \right]^{-n} 
=(8\pi^2)^n e^{-2\pi n\sqrt{E}} \, \sum_{k=0}^\infty \frac{(-1)^k\Gamma(1-n)}{\Gamma(1-n-k)\Gamma(k+1)}   e^{-4\pi  k  \sqrt{E}}
\end{equation}
We now Laplace transform each term in the sum,
\begin{equation}
\label{eq:genIntJT}
\begin{split}
    \int_0^\infty dE \, e^{- 2 \beta E } e^{-2\pi (n+2 k) \sqrt{E}} &= \frac{1}{2 \beta} -\frac{(n+2k) (2\pi)^{\frac{3}{2}}}{8 \beta^{3/2}} e^{\frac{\pi^2(n+2k)^2}{2 \beta}}\left[1 -\text{Erf}\left(\frac{\pi(n+2k)}{\sqrt{2 \beta}}\right)\right]\\
    &= \sum_{m=0}^\infty   \frac{\Gamma(2m+2)}{(2\pi^2)^{m+1}\,\Gamma(m+1)} \frac{\left(-\beta\right)^{m}  }{(n+2k)^{2(m+1)}}
\end{split}
\end{equation}
Note that it is important that $\beta$ and  $n+2k$ are positive, otherwise the integral has saddle point contributions which would not be analytic in $\beta$. For this range of parameters the expression is analytic around $\beta=0$, i.e. there is a series expansion in $\beta$ for each of the integrals appearing in the $t$ expansion. This gives then
\begin{equation}
\label{eq:JTsmallBetaExp}
   d_n\big(\rho_0^\text{JT};\beta\big)=(4\pi)^n \sum_{k=0}^\infty \frac{(-1)^k\Gamma(1-n)}{\Gamma(1-n-k)\Gamma(k+1)}  \sum_{m=0}^\infty  \frac{\Gamma(2m+2)}{(2\pi^2)^{m+1}\,\Gamma(m+1)} \frac{(-\beta)^{m}}{(n+2k)^{2(m+1)}}
\end{equation}
To simplify the coefficient of $\beta^m$ we need to perform the sum over the $k$ variable. Define
\begin{equation}
    \begin{split}
        I_r(a,n)&\equiv\sum_{k=0}^\infty \frac{(-1)^k\Gamma(1-n) \Gamma(r)}{\Gamma(1-n-k)\Gamma(k+1)} \, (-n)^{r-1} (a n + 2k )^{-r}\,.
\end{split}
\end{equation}

We can derive a recursive expression for $I_r(a,n)$. 
From the definition it follows that $\frac{dI_r}{da}=  I_{r+1}$. This allows an iterative evaluation of $I_r$ in terms of repeated derivatives of the seed function $I_1$:
\begin{equation}
    I_{r} (a,n) =  \partial_a^{r-1} I_1(a,n) \,,\qquad I_1(a,n) = \frac{\Gamma(1-n)\,\Gamma(\frac{a n}{2})}{2  \,\Gamma(1-n +\frac{a n }{2})}\,.
\end{equation}
Thus we obtain
\begin{equation}
 d_n\big(\rho_0^\text{JT};\beta\big)=(4\pi)^n \sum_{m=0}^\infty  \frac{I_{2m+2}(1,n)}{(-2n\pi^2)^{m+1}\,m!} \,\left(\frac{\beta}{n}\right)^{m}\,.
\end{equation}
Explicit examples can be found in \eqref{eq:JThightemp1} and \eqref{eq:JThightemp2}.

\subsection{Resummation of the high-temperature expansion}
\label{app:JTresummation}

To compare our results at high and low temperatures requires a non-trivial all-order resummation in $\beta$. Expressions to all orders in the low temperature regime were obtained in \cite{Tall:2024hgo} using gravitational calculations at genus $\frac{1}{2}$ and for genus 1. We will show how their expressions can be resummed into our all-order high-temperature results.

\paragraph{{\it Genus $g=\frac{1}{2}$:}}
We can use the high-temperature expressions \eqref{eq:genIntJT} to obtain the $\uptau$-scaled genus expansion in terms of infinite sums of error functions. For example, at order $\uptau^2$ (genus $\frac{1}{2}$) we find the following contribution to the SFF as a power series in small $\beta$:
\begin{equation}
\begin{split}
{\cal K}^{\text{GOE,JT}}_\beta(\uptau) &\supset  C_{\gt=\frac{1}{2}}^\text{GOE}\big(\rho_0^\text{JT};\beta\big) \, \uptau^{2}\\
 &= - 4\sum_{k=0}^\infty \left[ \frac{1}{2 \beta} -\frac{(2k+1) (2\pi)^{\frac{3}{2}}}{8 \beta^{3/2}}\, e^{\frac{\pi^2(2k+1)^2}{2 \beta}}\,\text{Erfc}\left(\frac{\pi(2k+1)}{\sqrt{2 \beta}}\right)\right] \uptau^{2}\\
 &= \sum_{k=0}^\infty \left[ -\frac{2}{((2k+1)\pi)^2} + \frac{6\beta}{((2k+1)\pi)^4} - \frac{30 \beta^2}{((2k+1)\pi)^6} + \ldots \right] \uptau^2\\
 &= \left[ -\frac{1}{4} + \frac{\beta}{16} - \frac{\beta^2}{32} + \frac{17\beta^3}{768} - \frac{31 \beta^4}{1536} + \ldots \right] \uptau^2
\end{split}
\label{eq:GOEJTgenushalf}
\end{equation}
This, of course, is nothing but the $\uptau^2$ term in \eqref{eq:JTGOEexpandedResultHighT}.
We would like to understand how this is consistent with our low temperature result \eqref{eq:JTGOEexpandedResult} (which is an expansion around $\beta=\infty$). In ref.\ \cite{Tall:2024hgo}, the complete large $\beta$ expansion was derived, thus generalizing \eqref{eq:JTGOEexpandedResult} to all orders; see their eq.\ (4.8). To resum, we formally expand their eq.\ (4.8) in {\it small} $\beta$ (i.e., outside the radius of convergence):
\begin{equation}
\begin{split}
\left[{\cal K}^{\text{GOE,JT}}_\beta(\uptau)\right]_{\text{\scriptsize\cite{Tall:2024hgo}}}&\supset  
-\frac{1}{\sqrt{2\pi\beta}}\left[1 + \sum_{k=1}^\infty (-1)^k\left( 2 - k \sqrt{2\pi\beta} \, e^{\frac{\beta k^2}{2}} \, \text{Erfc} \left( k \sqrt{\frac{\beta}{2}}\right) \right)\right]  \uptau^{2} \\
&= - \frac{1}{\sqrt{2\pi\beta}}\left[ 1 + \sum_{k=1}^\infty (-1)^k \left( 2 - k \sqrt{2\pi\beta} + 2 k^2 \beta - \frac{k^3}{4\pi} (2\pi\beta)^{\frac{3}{2}} + \ldots \right) \right]\uptau^{2}\,.
\end{split}
\label{eq:GOEJTgenushalfRichter}
\end{equation}
These sums can be regularized using $\zeta$-function regularization. This regularization renders the sums multiplying integer powers of $\beta$ finite and sets those with half-integer powers to zero. Using such a scheme, the power expansion in $\beta$ then exactly matches \eqref{eq:GOEJTgenushalf}.

\paragraph{{\it Genus $g=1$:}} A similar check can be performed for the SFF at order $\uptau^3$ (genus 1), where \cite{Tall:2024hgo} offer a similar resummed result to compare with. Our analysis yields the third line of \eqref{eq:JTGOEexpandedResultHighT}.
This should be compared with eq.\ (4.10) of \cite{Tall:2024hgo}. Again, their result is convergent for large $\beta$, but we can resum the small $\beta$ expansion using $\zeta$-function regularization:\footnote{ We use the following identities:
\begin{equation}
  \zeta'(s)= -\sum_{k\geq 1} k^{-s} \, \log(k)  \qquad \text{and} \qquad \zeta'(-2\ell) = \frac{(-1)^\ell(2\ell)!}{2^{2\ell+1} \, \pi^{2\ell}} \, \zeta(2\ell+1) \,,\quad \ell \in \mathbb{Z}_+ \,.
\end{equation}
}
\begin{equation}
\begin{split}
\left[{\cal K}^{\text{GOE,JT}}_\beta(\uptau)\right]_{\text{\scriptsize\cite{Tall:2024hgo}}}  &\supset 
\left[-\frac{\frac{1}{3}+\gamma+\log(2\beta\uptau^2)}{\pi} + \frac{4}{\pi} \sum_{k=1}^\infty \left( -1 + \frac{1}{2} (1+\beta k^2)\, e^{\frac{\beta k^2}{2}} \, E_1 \left( \frac{\beta k^2}{2} \right) \right) \right]\,\uptau^3 \\
&= \Bigg[ -\frac{\frac{1}{3}+\gamma+\log(2\beta\uptau^2)}{\pi} + \frac{4}{\pi} \sum_{k=1}^\infty\Bigg\{ \frac{ \log\big(\frac{2}{\beta}\big)-2-\gamma}{2}\, +
 \\
&\qquad\;\,
+ (\cdots) k^2 + (\cdots) k^4+ (\cdots) k^6 + \ldots - \sum_{\ell=0}^\infty\left( \frac{(2\ell+1)}{2^{\ell} \ell!} \, \beta^\ell \, \log(k) k^{2\ell} \right)  \Bigg\} \Bigg]\,\uptau^3  \\
&= \left[ -\frac{\frac{1}{3}+\gamma+\log(2\beta\uptau^2)}{\pi} 
+ \frac{4\zeta(0)}{\pi} \, \frac{ \log\big(\frac{2}{\beta}\big)-2-\gamma}{2}
+ \sum_{\ell=0}^\infty \frac{(2\ell+1)}{2^{\ell-2}\pi \ell!} \, \zeta'(-2\ell) \beta^\ell \right] \, \uptau^3\\
&= \left[ \frac{5}{3\pi}  - \frac{2}{\pi} \log(4\pi \uptau ) + \sum_{\ell=1}^\infty \frac{(-1)^\ell \,\Gamma\big( \frac{3}{2}+\ell \big)}{2^{\ell-2}\, \pi^{2\ell + \frac{3}{2}}} \, \zeta(2\ell+1) \, \beta^\ell\right] \, \uptau^3\,,
\end{split}
\label{eq:GOEJTgenusoneRichter}
\end{equation}
where the omitted terms in the second line involve even powers of $k$ that vanish after regularization. The last line precisely matches our corresponding term at ${\cal O}(\uptau^3)$ in \eqref{eq:JTGOEexpandedResultHighT} to arbitrarily high orders.\footnote{ We note one caveat:  the very first $k$-independent term in \eqref{eq:GOEJTgenusoneRichter} is not exactly what the authors of \cite{Tall:2024hgo} obtain directly from a gravity calculation. Their result from a purely gravitational calculation at genus $1$ is divergent in the $\uptau$-scaling limit. Writing it in the form of the first term in \eqref{eq:GOEJTgenusoneRichter} involves a non-trivial all-genus resummation, see \cite{Weber:2024ieq}. This is necessary in order to find a match with our manifestly finite $\uptau$-scaled analysis. This issue is separate from the resummation of the $\beta$-expansion and we comment on it in section \ref{sec:cancellations}.}


\section{Two-boundary partition function in non-orientable topological gravity}

For completeness, we give in table \ref{tab:twobdry} some exact results for the two-boundary gravitational path integral in the non-orientable Airy model. We present the results as functions of Euclidean boundary lengths $\beta_{1,2}$. These expressions can be obtained from \eqref{eq:Zg2CanonicalGeneral}.

\begin{table}[ht]
\begin{center}
\begin{tabular}{|l | l|} 
 \hline  
 $g=0$ & $Z^{\text{Airy}}_{0,2}(\beta_1,\beta_2) = \frac{\sqrt{\beta_1\beta_2}}{\pi (\beta_1+\beta_2)}$
   \rule{0pt}{3ex} \\  [1ex]
\hline
 $g=\tfrac{1}{2}$ & $Z^{\text{Airy}}_{\frac{1}{2},2}(\beta_1,\beta_2) = -\frac{1}{\sqrt{\pi}}\left\{ \frac{\beta_1 \beta_2}{\sqrt{\beta_1+\beta_2}} - \beta_1 \sqrt{\beta_2}- \beta_2 \sqrt{\beta_1}\right\}$ \rule{0pt}{3ex} \\ [1ex]
 \hline
$g=1$ & $Z^{\text{Airy}}_{1,2}(\beta_1,\beta_2) = \frac{1}{3\pi} \left\{\sqrt{\beta_1 \beta_2} \bigl(7 \beta_1^2 + 2\beta_1 \beta_2\bigr) + 3\beta_1 \beta_2^2 \arctan\left(\frac{\sqrt{\beta_1}}{\sqrt{\beta_2}}\right) + [\beta_1\leftrightarrow\beta_2] \right\}$ \rule{0pt}{3ex} \\ [1ex]
\hline
 $g=\tfrac{3}{2}$ & $Z^{\text{Airy}}_{\frac{3}{2},2}(\beta_1,\beta_2) = \frac{1}{48 \sqrt{\pi}} \Big\{ \sqrt{\beta_1} \bigl( 4 \beta_2^4 + 8 \beta_1 \beta_2^3 + 8 \beta_1^2 \beta_2^2 + 5 \beta_1^3 \beta_2 \bigr) $ \rule{0pt}{3ex} \\ [1ex]
& $\qquad\qquad\qquad\qquad\;\;+ \left(-5 \beta_1^3 \beta_2 -  \beta_1^2 \beta_2^2 \right) \sqrt{\beta_1+\beta_2} + [\beta_1\leftrightarrow\beta_2]\Big\}$ \rule{0pt}{3ex} \\ [1ex]
 \hline
 $g=2$ & $Z^{\text{Airy}}_{2,2}(\beta_1,\beta_2) = \frac{1}{90\pi}\Big\{
\sqrt{\beta_1\beta_2}\left(185\beta_1^5 + 435\beta_1^4\beta_2 + 60\pi\,\beta_1^{\frac{7}{2}}\beta_2^{\frac{3}{2}}  + 869\beta_1^3\beta_2^2 + 45\pi\,\beta_1^{\frac{5}{2}}\beta_2^{\frac{5}{2}} \right) $ \rule{0pt}{3ex} \\ [1ex]
& $\qquad\qquad\qquad\qquad\;\; +60\,\beta_1\beta_2^4(\beta_1+2\beta_2)
\arctan\!\left(\frac{\sqrt{\beta_1}}{\sqrt{\beta_2}}\right)
+ [\beta_1\leftrightarrow\beta_2]
\Big\}. $ \rule{0pt}{3ex} \\ [1ex]
 \hline
 $g=\tfrac{5}{2}$ & $Z_{\frac{5}{2},2}^\text{Airy}(\beta_1,\beta_2) = \frac{1}{90\sqrt{\pi}}
\Big\{
80\,\beta_{1}^{7}\sqrt{\beta_{2}}
+ 117\,\beta_{1}^{\frac{13}{2}}\beta_{2}
+ 368\,\beta_{1}^{\frac{11}{2}}\beta_{2}^{2}
$\rule{0pt}{3ex} \\ [1ex]
& $\qquad\qquad\qquad\qquad\;\; + 752\,\beta_{1}^{\frac{9}{2}}\beta_{2}^{3}+ 968\,\beta_{1}^{\frac{7}{2}}\beta_{2}^{4}+ 704\,\beta_{1}^{\frac{5}{2}}\beta_{2}^{5}
+ 320\,\beta_{1}^{\frac{3}{2}}\beta_{2}^{6}$\rule{0pt}{3ex} \\ [1ex]
& $\qquad\qquad\qquad\qquad\;\;
  - \left( 117 \beta_{1}^{6}\,\beta_{2} +217 \beta_1^5 \beta_2^2 -450 \beta_1^4 \beta_2^3\right)\sqrt{\beta_{1}+\beta_{2}} + [\beta_1\leftrightarrow\beta_2]
\Bigr\}$\rule{0pt}{3ex} \\ [1ex]
\hline
$g=3$ & $Z^{\text{Airy}}_{3,2}(\beta_1,\beta_2) = \frac{1}{5670\pi} \Big\{ \sqrt{\beta_1 \beta_2}\Big(6209\beta_1^8+26005\beta_1^7\beta_2+2520\pi\,\beta_1^{\frac{13}{2}}\beta_2^{\frac{3}{2}}+78533\beta_1^6\beta_2^2 $ \rule{0pt}{3ex} \\ [1ex]
& $\qquad\qquad\;\; +7560\pi\,\beta_1^{\frac{11}{2}}\beta_2^{\frac{5}{2}} +142378\beta_1^5\beta_2^3+15120\pi\,\beta_1^{\frac{9}{2}}\beta_2^{\frac{7}{2}}+170198\beta_1^4\beta_2^4 \Big) $ \rule{0pt}{3ex} \\ [1ex] 
& $ \qquad\qquad\;\; +5040\,\beta_1\beta_2^5(\beta_1+\beta_2)\left(2\beta_1^2+\beta_1\beta_2+\beta_2^2\right)
 \arctan\!\left(\frac{\sqrt{\beta_1}}{\sqrt{\beta_2}}\right)+ [\beta_1\leftrightarrow\beta_2]\Big\}$ \rule{0pt}{3ex} \\ [1ex]
 \hline
\end{tabular}
\end{center}
\caption{Two-boundary partition functions in non-orientable topological gravity (GOE Airy model).}
\label{tab:twobdry}
\end{table}

\section{The microcanonical plateau in orientable topological gravity}
\label{sec:microcanonicalPlateau}

The plateau in the universal SFF is due to spectral correlations at very small energies. The approximations used in section \ref{sec:microcanonical} to arrive at the general formula \eqref{eq:ZairyMicroCan} are therefore not valid. In this appendix we explore the structure of the $E\rightarrow 0$ correlations for the GUE Airy model.

\subsection{The microcanonical genus expansion}

In this section, we develop a formal microcanonical genus expansion of the GUE spectral form factor.
Consider the universal microcanonical SFF for the GUE:
\begin{equation}
\label{eq:GUEairyMicroUniversal}
    {\cal K}_E^\text{GUE}(\uptau) = \text{min} \left\{ \frac{\uptau}{2\pi} ,\, \rho_0(E) \right\} 
    = \frac{\uptau}{2\pi} + \left[ \theta\left( \rho_0(E)-\frac{\uptau}{2\pi}\right)-\theta(\rho_0(E))\right] \left(\frac{\uptau}{2\pi}-\rho_0(E)\right)\,.
\end{equation}
We expand the distributional contribution in a formal Taylor series:
\begin{equation}
\label{eq:thetaFormal}
    \theta\left( \rho_0-\frac{\uptau}{2\pi}\right)-\theta(\rho_0) = \sum_{n=1}^\infty \frac{(-\uptau)^n}{n!}\, \delta^{(n-1)}(2\pi\rho_0)
\end{equation}
We also note the following identity:
\begin{equation}
\label{eq:deltaIdentity}
    x^{k} \, \delta^{(n-1)}(x)
    = (-1)^k\,\frac{(n-1)!}{(n-1-k)!}\,\delta^{(n-1-k)}(x) \qquad (n-1-k\geq 0)\,.
\end{equation}
Using these identities, we can write:
\begin{equation}
    \begin{split}
     {\cal K}_E^\text{GUE}(\uptau)
     &=\frac{\uptau}{2\pi}+\left(\frac{\uptau}{2\pi}-\rho_0(E)\right)\sum_{n=1}^\infty \frac{(-\uptau)^n}{n!}\, \delta^{(n-1)}\big(2\pi\rho_0(E)\big)
     \\
     &=\frac{\uptau}{2\pi}-\sum_{n=0}^\infty \frac{(-\uptau)^{n+2}}{2\pi(n+2)!}\,\delta^{(n)}\big(2\pi\rho_0(E)\big)\,.
    \end{split}
    \label{eq:microGUEexpansion0}
\end{equation}
Due to the square root edge of $\rho_0(E)$, we can anticipate that only odd $n=2g-1$ will contribute to the sum. We write the resulting expression as follows:
\begin{equation}
\boxed{
\begin{split}
{\cal K}_E^\text{GUE}(\uptau)
    &=\frac{\uptau}{2\pi}-\sum_{g=1}^\infty \frac{1}{2\pi g(2g+1)} \, c_{2g}\big(\rho_0;E\big)  \, \uptau^{2g+1}\,,\\
    c_{2g}\big(\rho_0;E\big) &= -\frac{1}{2(2g-1)!} \; \delta^{(2g-1)}\big(2\pi \rho_0(E)\big) \,.
\end{split}
}
\label{eq:microGUEexpansion}
\end{equation}
The first term is the microcanonical ramp. The second term gives rise to the formal genus expansion of the plateau, which is singularly supported at $E=0$. Its $\rho_0$-dependent coefficient corresponds to the characteristic coefficients $c_{2g}$, now written in the microcanonical variables, c.f., \eqref{eq:GUESFF}.

\paragraph{Example: Airy model.} For the Airy spectral curve, the expression \eqref{eq:microGUEexpansion} can readily be written in terms of delta-functions localized at $E=0$, using 
\begin{equation}
\label{eq:deltaAiryExpand}
    \delta^{(2g-1)}\big( \sqrt{E} \big) 
    =  2\,(-1)^g \, \frac{(2g-1)!}{(g-1)!} \, \delta^{(g-1)}(E) \,.
\end{equation}
Plugging into \eqref{eq:microGUEexpansion}, we find:
\begin{equation}
    \begin{split}
     c_{2g}\big(\rho_0^\text{Airy};E\big) &= \frac{(-1)^{g-1}}{(g-1)!} \, \delta^{(g-1)}(E) \\
  \Rightarrow \qquad    {\cal K}_E^\text{GUE,Airy}(\uptau)
     &=\frac{\uptau}{2\pi}+\sum_{g=1}^\infty \frac{(-1)^g }{2\pi (2g+1)g!}\,\delta^{(g-1)}(E)\,\uptau^{2g+1}\,.
    \end{split}
    \label{eq:GUEAiryMicrocanonical}
\end{equation}
It is trivial to confirm that the Laplace transform of these expressions recovers the canonical results \eqref{eq:AGUEairy} and \eqref{eq:KGUEAiry}.

\paragraph{Example: $(2,3)$ minimal string.} For the spectral curve of the $(2,3)$ minimal string, we can similarly work out the microcanonical expansion coefficients. We observe the following generalization of \eqref{eq:deltaAiryExpand}, which can be confirmed by integrating against a test function:
\begin{equation}
\label{eq:delta23Expand}
\begin{split}
    \delta^{(2g-1)}\left( \sqrt{E} \left(1+\tfrac{2}{3\kappa} \, E \right) \right)
    &=  2\,(-1)^g \, \sum_{k=0}^{g-1}\frac{(2g+k-1)!}{k!(g-1-k)!} \, \left(\frac{2}{3\kappa}\right)^k \, \delta^{(g-1-k)}(E) \\
    &= \frac{2(-1)^{g}(2g-1)!}{(g-1)!} \, \left(\frac{3\kappa}{2} \right)^{2g}\; U\left(2g,3g; \frac{3\kappa}{2} \partial_E \right) \, \delta^{(3g-1)}(E) \,.
\end{split}
\end{equation}
This is, of course, directly related to the inverse Laplace transform of \eqref{eq:23ccoeff} and consistency with the canonical discussion thus follows.

\subsection{The microcanonical plateau from gravity}

Let us now reproduce the genus expansion of the plateau \eqref{eq:GUEAiryMicrocanonical} from the GUE topological gravity path integral. We could achieve this on a case-by-case basis by simply computing the inverse Laplace transform of the exact orientable two-boundary partition functions.\footnote{The first few ($g=1,2,3$) are given by:
\begin{equation}
    \begin{split}
    Z_{g,2}^\text{GUE,Airy}(\beta,T) \in \left\{ \frac{\sqrt{T^2 + \beta^2}}{4\pi\beta} \,,\; -\frac{(T^2 + \beta^2)^{\frac{3}{2}}}{6\pi} \,,\;\frac{\beta(T^2 + \beta^2)^{\frac{5}{2}}}{10\pi} \,, \ldots \right\}
    \end{split}
\end{equation}
}
Instead we wish to give a slightly more general perspective. We return to the starting point, the path microcanonical path integral \eqref{eq:Zg2Airy}, and systematically analyze the limit $E\rightarrow 0$. We expand the trumpet path integral in a power series in $\beta$:
\begin{equation}
    \begin{split}
    Z_{g,2}^\text{Airy}(T,\beta) = \frac{1}{4\pi \sqrt{T^2+\beta^2}} \int_0^\infty b_1 db_1 \, b_2db_2 \, \sum_{n=0}^\infty \frac{1}{n!}\left( - \frac{\beta(b_1^2 + b_2^2)}{4(T^2+\beta^2)}  \right)^n \, e^{\frac{i T}{4(T^2+\beta^2)}(b_1^2-b_2^2)}\; V_{g,2}^\text{Airy}(b_1,b_2) \,.
    \end{split}
\end{equation}
We can make the $T$-scaling explicit by rescaling 
$b_i \rightarrow  \sqrt{\frac{4(T^2+\beta^2)}{T}} \, b_i$:
\begin{equation}
    \begin{split}
    Z_{g,2}^\text{Airy}(T,\beta) = \frac{2^{6g}}{\pi} \sum_{n=0}^\infty \frac{(-\beta)^n}{n!}\, T^{3g-n} \left(1+\frac{\beta^2}{T^2}\right)^{3g+\frac{1}{2}} \int_0^\infty b_1 db_1 \, b_2db_2 \,  \left( b_1^2 + b_2^2 \right)^n \, e^{i(b_1^2-b_2^2)}\; V_{g,2}^\text{Airy}(b_1,b_2) \,.
    \end{split}
\end{equation}
The corresponding microcanonical expression is obtained by expanding the factor $(1+\frac{\beta^2}{T^2})^{3g+\frac{1}{2}}$ up to the required order and then replacing powers $\beta^n \rightarrow 2^{-n} \delta^{(n)}(E)$:
\begin{equation}
\label{eq:ZAiryMicroFinal}
    \begin{split}
    Z_{g,2}^\text{Airy}(T,E) &= \frac{2^{6g}}{\pi} \sum_{n=0}^\infty \frac{(-2)^{-n}}{n!}\,T^{3g-n}  \left[\delta^{(n)}(E) \, + \frac{(6g+1)}{8T^2}\, \delta^{(n+2)}(E) + \frac{(36g^2-1)}{128T^4} \, \delta^{(n+4)}(E) + \ldots \right] \\
    &\qquad\qquad\qquad \times  \int_0^\infty b_1 db_1 \, b_2db_2 \,  \left( b_1^2 + b_2^2 \right)^n \, e^{i(b_1^2-b_2^2)}\; V_{g,2}^\text{Airy}(b_1,b_2) \,.
    \end{split}
\end{equation}
Note that the second line is independent of $T$ or $E$ and simply corresponds to a certain universal symmetrized moment of the WP volume. 

So far, the expression is in principle valid for orientable as well as non-orientable geometries.\footnote{In the non-orientable case, the integrals in \eqref{eq:ZAiryMicroFinal} require regularization.} Recall now that in the orientable case, $V_{g,2}^\text{GUE,Airy}(b_1,b_2)$ is a symmetric polynomial. The integrals over $b_i$ for these types of terms can easily be performed, using the following Fresnel-type integrals:
\begin{equation}
\label{eq:symmpolyIntegral}
\begin{split}
    \int_0^\infty db \, b^m \, e^{\pm i b^2} = \frac{1}{2} \, e^{\pm \frac{i\pi}{4} (m+1)} \, \Gamma\left(\frac{m+1}{2} \right) \,.
\end{split}
\end{equation}
The WP volumes for the orientable model are well known (e.g., \cite{Blommaert:2022lbh}):
\begin{equation}
    \begin{split}
    V_{1,2}^{\text{GUE,Airy}}(b_1,b_2) &= \frac{b_1^4+2\,b_1^2b_2^2+b_2^4}{192}  \,,\\
    V_{2,2}^{\text{GUE,Airy}}(b_1,b_2) &= \frac{b_1^{10}+152\,b_1^8b_2^2+58\,b_1^6b_2^4}{4423680} + \text{perm.} \,,\\
    V_{3,2}^{\text{GUE,Airy}}(b_1,b_2) &= \frac{5\,b_1^{16}+200\, b_1^{14} b_2^2 +2156\, b_1^{12} b_2^4+8048\, b_1^{10} b_2^6 + 6070\, b_1^8 b_2^8}{4280706662400} + \text{perm.} \,,
    \end{split}
\end{equation}
and so on.
Explicit evaluation then yields:
\begin{equation}
\label{eq:GUEcancellationsGrav}
    \begin{split}
     e^{-3S_0}\,Z_{1,2}^\text{GUE,Airy}(T=\uptau e^{S_0},E) 
     &= - \frac{\uptau^3}{6\pi} \, \delta(E) + {\cal O}\big( e^{-S_0} \big)\,,
     \\
     e^{-5S_0}\,Z_{2,2}^\text{GUE,Airy}(T=\uptau e^{S_0},E) 
     &= {\color{orange}0} \times e^{S_0} \,\delta(E) + \frac{\uptau^5}{20\pi} \, \delta'(E) + {\cal O}\big( e^{-S_0} \big)\,,
     \\
     e^{-7S_0}\,Z_{3,2}^\text{GUE,Airy}(T=\uptau e^{S_0},E) 
     &= {\color{orange}0} \times e^{2S_0} \,\delta(E)+{\color{orange}0} \times e^{S_0} \,\delta'(E)   -\frac{\uptau^7}{84\pi} \, \delta''(E) + {\cal O}\big( e^{-S_0} \big)\,,
    \end{split}
\end{equation}
and so on. These terms build precisely the microcanonical distribution-valued expansion of the plateau, derived from RMT universality in \eqref{eq:GUEAiryMicrocanonical}. At genus $g$, the first $(g-1)$ terms would be divergent upon $\uptau$-scaling, but their coefficients turn out to be zero. This is due to {\color{orange}$(g-1)$ cancellations} among the coefficients of the WP volumes. For the GUE, these cancellations have been understood before \cite{Blommaert:2022lbh,Weber:2022sov}, so we will not discuss them in detail here. Our main observation is that the $\uptau$-scaled plateau has a formal microcanonical genus expansion in the GUE, which can be obtained from orientable topological gravity term-by-term.


\vspace{10pt}

\newpage
\bibliographystyle{JHEP}

\begin{thebibliography}{10}

\bibitem{Almheiri:2019psf}
A.~Almheiri, N.~Engelhardt, D.~Marolf and H.~Maxfield, \emph{{The entropy of
  bulk quantum fields and the entanglement wedge of an evaporating black
  hole}}, \href{https://doi.org/10.1007/JHEP12(2019)063}{\emph{JHEP} {\bfseries
  12} (2019) 063} [\href{https://arxiv.org/abs/1905.08762}{{\ttfamily
  1905.08762}}].

\bibitem{Penington:2019npb}
G.~Penington, \emph{{Entanglement Wedge Reconstruction and the Information
  Paradox}}, \href{https://doi.org/10.1007/JHEP09(2020)002}{\emph{JHEP}
  {\bfseries 09} (2020) 002}
  [\href{https://arxiv.org/abs/1905.08255}{{\ttfamily 1905.08255}}].

\bibitem{Saad:2018bqo}
P.~Saad, S.H.~Shenker and D.~Stanford, \emph{{A semiclassical ramp in SYK and
  in gravity}},  \href{https://arxiv.org/abs/1806.06840}{{\ttfamily
  1806.06840}}.

\bibitem{Saad:2019lba}
P.~Saad, S.H.~Shenker and D.~Stanford, \emph{{JT gravity as a matrix
  integral}},  \href{https://arxiv.org/abs/1903.11115}{{\ttfamily 1903.11115}}.

\bibitem{Cotler:2020ugk}
J.~Cotler and K.~Jensen, \emph{{AdS$_{3}$ gravity and random CFT}},
  \href{https://doi.org/10.1007/JHEP04(2021)033}{\emph{JHEP} {\bfseries 04}
  (2021) 033} [\href{https://arxiv.org/abs/2006.08648}{{\ttfamily
  2006.08648}}].

\bibitem{Harlow:2018tng}
D.~Harlow and H.~Ooguri, \emph{{Symmetries in quantum field theory and quantum
  gravity}}, \href{https://doi.org/10.1007/s00220-021-04040-y}{\emph{Commun.
  Math. Phys.} {\bfseries 383} (2021) 1669}
  [\href{https://arxiv.org/abs/1810.05338}{{\ttfamily 1810.05338}}].

\bibitem{Streater:1989vi}
R.F.~Streater and A.S.~Wightman, \emph{{PCT, spin and statistics, and all
  that}}, {Princeton University Press} (1989).

\bibitem{Harlow:2023hjb}
D.~Harlow and T.~Numasawa, \emph{{Gauging spacetime inversions in quantum
  gravity}},  \href{https://arxiv.org/abs/2311.09978}{{\ttfamily 2311.09978}}.

\bibitem{Grabovsky:2024vnb}
D.~Grabovsky and M.~Kolanowski, \emph{{Spin-refined partition functions and $
  \mathcal{CRT} $ black holes}},
  \href{https://doi.org/10.1007/JHEP12(2024)013}{\emph{JHEP} {\bfseries 12}
  (2024) 013} [\href{https://arxiv.org/abs/2406.07609}{{\ttfamily
  2406.07609}}].

\bibitem{Chen:2023mbc}
Y.~Chen and G.J.~Turiaci, \emph{{Spin-statistics for black hole microstates}},
  \href{https://doi.org/10.1007/JHEP04(2024)135}{\emph{JHEP} {\bfseries 04}
  (2024) 135} [\href{https://arxiv.org/abs/2309.03478}{{\ttfamily
  2309.03478}}].

\bibitem{Maloney:2007ud}
A.~Maloney and E.~Witten, \emph{{Quantum Gravity Partition Functions in Three
  Dimensions}}, \href{https://doi.org/10.1007/JHEP02(2010)029}{\emph{JHEP}
  {\bfseries 02} (2010) 029} [\href{https://arxiv.org/abs/0712.0155}{{\ttfamily
  0712.0155}}].

\bibitem{Keller:2014xba}
C.A.~Keller and A.~Maloney, \emph{{Poincare Series, 3D Gravity and CFT
  Spectroscopy}}, \href{https://doi.org/10.1007/JHEP02(2015)080}{\emph{JHEP}
  {\bfseries 02} (2015) 080} [\href{https://arxiv.org/abs/1407.6008}{{\ttfamily
  1407.6008}}].

\bibitem{Benjamin:2019stq}
N.~Benjamin, H.~Ooguri, S.-H.~Shao and Y.~Wang, \emph{{Light-cone modular
  bootstrap and pure gravity}},
  \href{https://doi.org/10.1103/PhysRevD.100.066029}{\emph{Phys. Rev. D}
  {\bfseries 100} (2019) 066029}
  [\href{https://arxiv.org/abs/1906.04184}{{\ttfamily 1906.04184}}].

\bibitem{Benjamin:2020mfz}
N.~Benjamin, S.~Collier and A.~Maloney, \emph{{Pure Gravity and Conical
  Defects}}, \href{https://doi.org/10.1007/JHEP09(2020)034}{\emph{JHEP}
  {\bfseries 09} (2020) 034}
  [\href{https://arxiv.org/abs/2004.14428}{{\ttfamily 2004.14428}}].

\bibitem{Alday:2019vdr}
L.F.~Alday and J.-B.~Bae, \emph{{Rademacher Expansions and the Spectrum of 2d
  CFT}}, \href{https://doi.org/10.1007/JHEP11(2020)134}{\emph{JHEP} {\bfseries
  11} (2020) 134} [\href{https://arxiv.org/abs/2001.00022}{{\ttfamily
  2001.00022}}].

\bibitem{DiUbaldo:2023hkc}
G.~Di~Ubaldo and E.~Perlmutter, \emph{{AdS3 Pure Gravity and Stringy
  Unitarity}},
  \href{https://doi.org/10.1103/PhysRevLett.132.041602}{\emph{Phys. Rev. Lett.}
  {\bfseries 132} (2024) 041602}
  [\href{https://arxiv.org/abs/2308.01787}{{\ttfamily 2308.01787}}].

\bibitem{Maxfield:2020ale}
H.~Maxfield and G.J.~Turiaci, \emph{{The path integral of 3D gravity near
  extremality; or, JT gravity with defects as a matrix integral}},
  \href{https://doi.org/10.1007/JHEP01(2021)118}{\emph{JHEP} {\bfseries 01}
  (2021) 118} [\href{https://arxiv.org/abs/2006.11317}{{\ttfamily
  2006.11317}}].

\bibitem{Hsin:2020mfa}
P.-S.~Hsin, L.V.~Iliesiu and Z.~Yang, \emph{{A violation of global symmetries
  from replica wormholes and the fate of black hole remnants}},
  \href{https://arxiv.org/abs/2011.09444}{{\ttfamily 2011.09444}}.

\bibitem{Stanford:2019vob}
D.~Stanford and E.~Witten, \emph{{JT gravity and the ensembles of random matrix
  theory}}, \href{https://doi.org/10.4310/ATMP.2020.v24.n6.a4}{\emph{Adv.
  Theor. Math. Phys.} {\bfseries 24} (2020) 1475}
  [\href{https://arxiv.org/abs/1907.03363}{{\ttfamily 1907.03363}}].

\bibitem{You_2017}
Y.-Z.~You, A.W.W.~Ludwig and C.~Xu, \emph{Sachdev-ye-kitaev model and
  thermalization on the boundary of many-body localized fermionic
  symmetry-protected topological states},
  \href{https://doi.org/10.1103/physrevb.95.115150}{\emph{Physical Review B}
  {\bfseries 95} (2017) }.

\bibitem{Cotler:2016fpe}
J.S.~Cotler, G.~Gur-Ari, M.~Hanada, J.~Polchinski, P.~Saad, S.H.~Shenker
  et~al., \emph{{Black Holes and Random Matrices}},
  \href{https://doi.org/10.1007/JHEP05(2017)118}{\emph{JHEP} {\bfseries 05}
  (2017) 118} [\href{https://arxiv.org/abs/1611.04650}{{\ttfamily
  1611.04650}}].

\bibitem{Yan:2022nod}
C.~Yan, \emph{{Crosscap contribution to late-time two-point correlators}},
  \href{https://doi.org/10.1007/JHEP12(2023)051}{\emph{JHEP} {\bfseries 12}
  (2023) 051} [\href{https://arxiv.org/abs/2203.14436}{{\ttfamily
  2203.14436}}].

\bibitem{Stanford:2023dtm}
D.~Stanford, \emph{{A Mirzakhani recursion for non-orientable surfaces}},
  \href{https://arxiv.org/abs/2303.04049}{{\ttfamily 2303.04049}}.

\bibitem{haake}
F.~Haake, \emph{Quantum Signatures of Chaos}, Physics and astronomy online
  library, Springer (2001).

\bibitem{Altland_1997}
A.~Altland and M.R.~Zirnbauer, \emph{Nonstandard symmetry classes in mesoscopic
  normal-superconducting hybrid structures},
  \href{https://doi.org/10.1103/physrevb.55.1142}{\emph{Physical Review B}
  {\bfseries 55} (1997) 1142–1161}.

\bibitem{zirnbauer2010symmetryclasses}
M.R.~Zirnbauer, \emph{Symmetry classes},  2010.

\bibitem{Simmons-Duffin:2016gjk}
D.~Simmons-Duffin, \emph{{The Conformal Bootstrap}},  in \emph{{Theoretical
  Advanced Study Institute in Elementary Particle Physics}: {New Frontiers in
  Fields and Strings}}, pp.~1--74, 2017,
  \href{https://doi.org/10.1142/9789813149441_0001}{DOI}
  [\href{https://arxiv.org/abs/1602.07982}{{\ttfamily 1602.07982}}].

\bibitem{Ginsparg:1988ui}
P.H.~Ginsparg, \emph{{APPLIED CONFORMAL FIELD THEORY}},  in \emph{{Les Houches
  Summer School in Theoretical Physics: Fields, Strings, Critical Phenomena}},
  9, 1988 [\href{https://arxiv.org/abs/hep-th/9108028}{{\ttfamily
  hep-th/9108028}}].

\bibitem{Yan:2023rjh}
C.~Yan, \emph{{More on torus wormholes in 3d gravity}},
  \href{https://doi.org/10.1007/JHEP11(2023)039}{\emph{JHEP} {\bfseries 11}
  (2023) 039} [\href{https://arxiv.org/abs/2305.10494}{{\ttfamily
  2305.10494}}].

\bibitem{Collier:2024mgv}
S.~Collier, L.~Eberhardt and M.~Zhang, \emph{{3d gravity from Virasoro TQFT:
  Holography, wormholes and knots}},
  \href{https://doi.org/10.21468/SciPostPhys.17.5.134}{\emph{SciPost Phys.}
  {\bfseries 17} (2024) 134}
  [\href{https://arxiv.org/abs/2401.13900}{{\ttfamily 2401.13900}}].

\bibitem{Collier:2023fwi}
S.~Collier, L.~Eberhardt and M.~Zhang, \emph{{Solving 3d Gravity with Virasoro
  TQFT}},  \href{https://arxiv.org/abs/2304.13650}{{\ttfamily 2304.13650}}.

\bibitem{Iliesiu:2020qvm}
L.V.~Iliesiu and G.J.~Turiaci, \emph{{The statistical mechanics of
  near-extremal black holes}},
  \href{https://arxiv.org/abs/2003.02860}{{\ttfamily 2003.02860}}.

\bibitem{Iliesiu:2021are}
L.V.~Iliesiu, M.~Kologlu and G.J.~Turiaci, \emph{{Supersymmetric indices
  factorize}},  \href{https://arxiv.org/abs/2107.09062}{{\ttfamily
  2107.09062}}.

\bibitem{Heydeman:2020hhw}
M.~Heydeman, L.V.~Iliesiu, G.J.~Turiaci and W.~Zhao, \emph{{The statistical
  mechanics of near-BPS black holes}},
  \href{https://arxiv.org/abs/2011.01953}{{\ttfamily 2011.01953}}.

\bibitem{DiUbaldo:2023qli}
G.~Di~Ubaldo and E.~Perlmutter, \emph{{AdS$_{3}$/RMT$_{2}$ duality}},
  \href{https://doi.org/10.1007/JHEP12(2023)179}{\emph{JHEP} {\bfseries 12}
  (2023) 179} [\href{https://arxiv.org/abs/2307.03707}{{\ttfamily
  2307.03707}}].

\bibitem{Boruch:2025ilr}
J.~Boruch, G.~Di~Ubaldo, F.M.~Haehl, E.~Perlmutter and M.~Rozali,
  \emph{{Modular-Invariant Random Matrix Theory and AdS3 Wormholes}},
  \href{https://doi.org/10.1103/4hhn-c6mp}{\emph{Phys. Rev. Lett.} {\bfseries
  135} (2025) 121602} [\href{https://arxiv.org/abs/2503.00101}{{\ttfamily
  2503.00101}}].

\bibitem{Haehl:2023tkr}
F.M.~Haehl, C.~Marteau, W.~Reeves and M.~Rozali, \emph{{Symmetries and spectral
  statistics in chaotic conformal field theories}},
  \href{https://doi.org/10.1007/JHEP07(2023)196}{\emph{JHEP} {\bfseries 07}
  (2023) 196} [\href{https://arxiv.org/abs/2302.14482}{{\ttfamily
  2302.14482}}].

\bibitem{Haehl:2023xys}
F.M.~Haehl, W.~Reeves and M.~Rozali, \emph{{Symmetries and spectral statistics
  in chaotic conformal field theories. Part II. Maass cusp forms and arithmetic
  chaos}}, \href{https://doi.org/10.1007/JHEP12(2023)161}{\emph{JHEP}
  {\bfseries 12} (2023) 161}
  [\href{https://arxiv.org/abs/2309.00611}{{\ttfamily 2309.00611}}].

\bibitem{Haehl:2023mhf}
F.M.~Haehl, W.~Reeves and M.~Rozali, \emph{{Euclidean wormholes in
  two-dimensional conformal field theories from quantum chaos and number
  theory}}, \href{https://doi.org/10.1103/PhysRevD.108.L101902}{\emph{Phys.
  Rev. D} {\bfseries 108} (2023) L101902}
  [\href{https://arxiv.org/abs/2309.02533}{{\ttfamily 2309.02533}}].

\bibitem{Jafferis:2025vyp}
D.L.~Jafferis, L.~Rozenberg and G.~Wong, \emph{{3d gravity as a random
  ensemble}}, \href{https://doi.org/10.1007/JHEP02(2025)208}{\emph{JHEP}
  {\bfseries 02} (2025) 208}
  [\href{https://arxiv.org/abs/2407.02649}{{\ttfamily 2407.02649}}].

\bibitem{deBoer:2025rct}
J.~de~Boer, J.~Kames-King and B.~Post, \emph{{Surgery and statistics in 3d
  gravity}},  \href{https://arxiv.org/abs/2506.04151}{{\ttfamily 2506.04151}}.

\bibitem{Saad:2022kfe}
P.~Saad, D.~Stanford, Z.~Yang and S.~Yao, \emph{{A convergent genus expansion
  for the plateau}},  \href{https://arxiv.org/abs/2210.11565}{{\ttfamily
  2210.11565}}.

\bibitem{Tall:2024hgo}
J.~Tall, T.~Weber, J.D.~Urbina and K.~Richter, \emph{{Chaos and moduli space
  volumes in unorientable JT gravity}},
  \href{https://arxiv.org/abs/2411.08129}{{\ttfamily 2411.08129}}.

\bibitem{Haake:2009scd}
F.~Haake, P.~Braun, A.~Altland, S.~Heusler and S.~M{\"u}ller,
  \emph{{Periodic-orbit theory of universal level correlations in quantum
  chaos}}, \href{https://doi.org/10.1088/1367-2630/11/10/103025}{\emph{New J.
  Phys.} {\bfseries 11} (2009) 103025}.

\bibitem{Okuyama:2020ncd}
K.~Okuyama and K.~Sakai, \emph{{Multi-boundary correlators in JT gravity}},
  \href{https://doi.org/10.1007/JHEP08(2020)126}{\emph{JHEP} {\bfseries 08}
  (2020) 126} [\href{https://arxiv.org/abs/2004.07555}{{\ttfamily
  2004.07555}}].

\bibitem{Muller2005}
S.~Müller, S.~Heusler, P.~Braun, F.~Haake and A.~Altland, \emph{Periodic-orbit
  theory of universality in quantum chaos},
  \href{https://doi.org/10.1103/physreve.72.046207}{\emph{Physical Review E}
  {\bfseries 72} (2005) }.

\bibitem{Blommaert:2022lbh}
A.~Blommaert, J.~Kruthoff and S.~Yao, \emph{{An integrable road to a
  perturbative plateau}},
  \href{https://doi.org/10.1007/JHEP04(2023)048}{\emph{JHEP} {\bfseries 04}
  (2023) 048} [\href{https://arxiv.org/abs/2208.13795}{{\ttfamily
  2208.13795}}].

\bibitem{Weber:2024ieq}
T.~Weber, J.~Tall, F.~Haneder, J.D.~Urbina and K.~Richter, \emph{{Unorientable
  topological gravity and orthogonal random matrix universality}},
  \href{https://doi.org/10.1007/JHEP07(2024)267}{\emph{JHEP} {\bfseries 07}
  (2024) 267} [\href{https://arxiv.org/abs/2405.17177}{{\ttfamily
  2405.17177}}].

\bibitem{Liu:2018hlr}
J.~Liu, \emph{{Spectral form factors and late time quantum chaos}},
  \href{https://doi.org/10.1103/PhysRevD.98.086026}{\emph{Phys. Rev. D}
  {\bfseries 98} (2018) 086026}
  [\href{https://arxiv.org/abs/1806.05316}{{\ttfamily 1806.05316}}].

\bibitem{Mirzakhani:2006fta}
M.~Mirzakhani, \emph{{Simple geodesics and Weil-Petersson volumes of moduli
  spaces of bordered Riemann surfaces}},
  \href{https://doi.org/10.1007/s00222-006-0013-2}{\emph{Invent. Math.}
  {\bfseries 167} (2006) 179}.

\bibitem{norbury2007lengthsgeodesicsnonorientablehyperbolic}
P.~Norbury, \emph{Lengths of geodesics on non-orientable hyperbolic surfaces},
  2007.

\bibitem{Eynard:2021zcj}
B.~Eynard and D.~Lewa{\'n}ski, \emph{{A natural basis for intersection
  numbers.}}, \href{https://doi.org/10.13137/2464-8728/35487}{\emph{Rend. Ist.
  Mat. Univ. Trieste} {\bfseries 55} (2023) 6}
  [\href{https://arxiv.org/abs/2108.00226}{{\ttfamily 2108.00226}}].

\bibitem{Eynard:2023qdr}
B.~Eynard, E.~Garcia-Failde, P.~Gregori, D.~Lewanski and R.~Schiappa,
  \emph{{Resurgent Asymptotics of Jackiw{\textendash}Teitelboim Gravity and the
  Nonperturbative Topological Recursion}},
  \href{https://doi.org/10.1007/s00023-023-01412-z}{\emph{Annales Henri
  Poincare} {\bfseries 25} (2024) 4121}
  [\href{https://arxiv.org/abs/2305.16940}{{\ttfamily 2305.16940}}].

\bibitem{Weber:2025mow}
T.~Weber, M.~Lents, J.~Dieplinger, J.D.~Urbina and K.~Richter,
  \emph{{Topological gravity for arbitrary Dyson index}},
  \href{https://arxiv.org/abs/2507.03172}{{\ttfamily 2507.03172}}.

\bibitem{Maldacena:2004sn}
J.M.~Maldacena, G.W.~Moore, N.~Seiberg and D.~Shih, \emph{{Exact vs.
  semiclassical target space of the minimal string}},
  \href{https://doi.org/10.1088/1126-6708/2004/10/020}{\emph{JHEP} {\bfseries
  10} (2004) 020} [\href{https://arxiv.org/abs/hep-th/0408039}{{\ttfamily
  hep-th/0408039}}].

\bibitem{Dijkgraaf:2018vnm}
R.~Dijkgraaf and E.~Witten, \emph{{Developments in Topological Gravity}},
  \href{https://doi.org/10.1142/S0217751X18300296}{\emph{Int. J. Mod. Phys. A}
  {\bfseries 33} (2018) 1830029}
  [\href{https://arxiv.org/abs/1804.03275}{{\ttfamily 1804.03275}}].

\bibitem{Eynard:2007kz}
B.~Eynard and N.~Orantin, \emph{{Invariants of algebraic curves and topological
  expansion}}, \href{https://doi.org/10.4310/CNTP.2007.v1.n2.a4}{\emph{Commun.
  Num. Theor. Phys.} {\bfseries 1} (2007) 347}
  [\href{https://arxiv.org/abs/math-ph/0702045}{{\ttfamily math-ph/0702045}}].

\bibitem{Eynard:2007fi}
B.~Eynard and N.~Orantin, \emph{{Weil-Petersson volume of moduli spaces,
  Mirzakhani's recursion and matrix models}},
  \href{https://arxiv.org/abs/0705.3600}{{\ttfamily 0705.3600}}.

\bibitem{Altland:2024ubs}
A.~Altland, K.W.~Kim, T.~Micklitz, M.~Rezaei, J.~Sonner and
  J.J.M.~Verbaarschot, \emph{{Quantum chaos on edge}},
  \href{https://doi.org/10.1103/PhysRevResearch.6.033286}{\emph{Phys. Rev.
  Res.} {\bfseries 6} (2024) 033286}
  [\href{https://arxiv.org/abs/2403.13516}{{\ttfamily 2403.13516}}].

\bibitem{Altland:2025diq}
A.~Altland, J.~van~der Heijden, T.~Micklitz, M.~Rozali and J.T.~de~Miranda,
  \emph{{The universality class of the first levels in low-dimensional
  gravity}},  \href{https://arxiv.org/abs/2505.18957}{{\ttfamily 2505.18957}}.

\bibitem{Weber:2022sov}
T.~Weber, F.~Haneder, K.~Richter and J.D.~Urbina, \emph{{Constraining
  Weil\textendash{}Petersson volumes by universal random matrix correlations in
  low-dimensional quantum gravity}},
  \href{https://doi.org/10.1088/1751-8121/acc8a5}{\emph{J. Phys. A} {\bfseries
  56} (2023) 205206} [\href{https://arxiv.org/abs/2208.13802}{{\ttfamily
  2208.13802}}].

\bibitem{PhysRevE.72.046207}
S.~M\"uller, S.~Heusler, P.~Braun, F.~Haake and A.~Altland,
  \emph{Periodic-orbit theory of universality in quantum chaos},
  \href{https://doi.org/10.1103/PhysRevE.72.046207}{\emph{Phys. Rev. E}
  {\bfseries 72} (2005) 046207}.

\bibitem{berry1985}
M.V.~{Berry}, \emph{{Semiclassical Theory of Spectral Rigidity}},
  \href{https://doi.org/10.1098/rspa.1985.0078}{\emph{Proceedings of the Royal
  Society of London Series A} {\bfseries 400} (1985) 229}.

\bibitem{Sieber_2001}
M.~Sieber and K.~Richter, \emph{Correlations between periodic orbits and their
  rôle in spectral statistics},
  \href{https://doi.org/10.1238/Physica.Topical.090a00128}{\emph{Physica
  Scripta} {\bfseries 2001} (2001) 128}.

\bibitem{Heusler_2004}
S.~Heusler, S.~M\"uller, P.~Braun and F.~Haake, \emph{Universal spectral form
  factor for chaotic dynamics},
  \href{https://doi.org/10.1088/0305-4470/37/3/l02}{\emph{Journal of Physics A:
  Mathematical and General} {\bfseries 37} (2004) L31}.

\bibitem{Adams:2006sv}
A.~Adams, N.~Arkani-Hamed, S.~Dubovsky, A.~Nicolis and R.~Rattazzi,
  \emph{{Causality, analyticity and an IR obstruction to UV completion}},
  \href{https://doi.org/10.1088/1126-6708/2006/10/014}{\emph{JHEP} {\bfseries
  10} (2006) 014} [\href{https://arxiv.org/abs/hep-th/0602178}{{\ttfamily
  hep-th/0602178}}].

\bibitem{Bellazzini:2021oaj}
B.~Bellazzini, M.~Riembau and F.~Riva, \emph{{IR side of positivity bounds}},
  \href{https://doi.org/10.1103/PhysRevD.106.105008}{\emph{Phys. Rev. D}
  {\bfseries 106} (2022) 105008}
  [\href{https://arxiv.org/abs/2112.12561}{{\ttfamily 2112.12561}}].

\bibitem{Herrero-Valea:2022lfd}
M.~Herrero-Valea, A.S.~Koshelev and A.~Tokareva, \emph{{UV graviton scattering
  and positivity bounds from IR dispersion relations}},
  \href{https://doi.org/10.1103/PhysRevD.106.105002}{\emph{Phys. Rev. D}
  {\bfseries 106} (2022) 105002}
  [\href{https://arxiv.org/abs/2205.13332}{{\ttfamily 2205.13332}}].

\bibitem{Blommaert:2023vbz}
A.~Blommaert, J.~Kruthoff and S.~Yao, \emph{{The power of Lorentzian
  wormholes}}, \href{https://doi.org/10.1007/JHEP10(2023)005}{\emph{JHEP}
  {\bfseries 10} (2023) 005}
  [\href{https://arxiv.org/abs/2302.01360}{{\ttfamily 2302.01360}}].

\bibitem{Louko:1995jw}
J.~Louko and R.D.~Sorkin, \emph{{Complex actions in two-dimensional topology
  change}}, \href{https://doi.org/10.1088/0264-9381/14/1/018}{\emph{Class.
  Quant. Grav.} {\bfseries 14} (1997) 179}
  [\href{https://arxiv.org/abs/gr-qc/9511023}{{\ttfamily gr-qc/9511023}}].

\bibitem{Berry_1990}
M.V.~Berry and J.P.~Keating, \emph{A rule for quantizing chaos?},
  \href{https://doi.org/10.1088/0305-4470/23/21/024}{\emph{Journal of Physics
  A: Mathematical and General} {\bfseries 23} (1990) 4839}.

\bibitem{Bogomolny_1992}
E.B.~Bogomolny, \emph{Semiclassical quantization of multidimensional systems},
  \href{https://doi.org/10.1088/0951-7715/5/4/001}{\emph{Nonlinearity}
  {\bfseries 5} (1992) 805}.

\bibitem{Winer:2023btb}
M.~Winer and B.~Swingle, \emph{{Reappearance of Thermalization Dynamics in the
  Late-Time Spectral Form Factor}},
  \href{https://arxiv.org/abs/2307.14415}{{\ttfamily 2307.14415}}.

\end{thebibliography}

\providecommand{\href}[2]{#2}\begingroup\raggedright\endgroup

\end{document}